\pdfoutput=1

\documentclass[11pt,twoside,a4paper,cmspaper,final,collab]{cms-tdr}
\def\svnVersion{df4c95d-D}\def\svnDate{2019/12/13}\def\cmsCernNoTag{CERN-EP-2019-175}\def\cmsCernDate{\today}\def\cmsMessage{"Published in the Journal of High Energy Physics as \href{http://dx.doi.org/10.1007/JHEP12(2019)061}{\doi{10.1007/JHEP12(2019)061}.}"}

\begin{document}\cmsNoteHeader{SMP-17-010}

\hyphenation{had-ron-i-za-tion}
\hyphenation{cal-or-i-me-ter}
\hyphenation{de-vices}
\newcommand{\rapidity}{\ensuremath{y}}
\newcommand{\phiStar}{\ensuremath{\phi^{\scriptscriptstyle *}_\eta}}
\newcommand{\mll}{\ensuremath{m_{\ell\ell}}}
\newcommand{\FxFx}{{\textsc{FxFx}}\xspace}
\newlength\cmsTabSkip\setlength{\cmsTabSkip}{1ex}
\providecommand{\NA}{\ensuremath{\text{---}}}

\cmsNoteHeader{SMP-17-010}

\title{Measurements of differential $\PZ$ boson production cross sections in proton-proton collisions at $\sqrt{s}=13\TeV$}

\date{\today}

\abstract{Measurements are presented of the differential cross sections for $\cPZ$ bosons produced in proton-proton collisions at $\sqrt{s} = 13\TeV$ and decaying to muons and electrons. The data analyzed were collected in 2016 with the CMS detector at the LHC and correspond to an integrated luminosity of $35.9\fbinv$. The measured fiducial inclusive product of cross section and branching fraction agrees with
next-to-next-to-leading order quantum chromodynamics calculations. Differential cross sections of the transverse
momentum \pt, the optimized angular variable $\phiStar$, and the rapidity of lepton pairs are
measured. The data are corrected for detector effects and compared to theoretical predictions using fixed order, resummed, and parton shower calculations. The uncertainties of the measured normalized cross sections are smaller than 0.5\% for $\phiStar < 0.5$ and for $\pt^{\cPZ} < 50\GeV$.}

\hypersetup{%
pdfauthor={CMS Collaboration},%
pdftitle={Measurements of differential Z boson production cross sections in pp collisions with CMS at sqrt{s} = 13 TeV},%
pdfsubject={CMS},%
pdfkeywords={CMS, physics, Z boson}}

\maketitle

\section{Introduction}
The measurement of the production of lepton pairs via the $\cPZ$ boson is important for the physics program of the CERN LHC.
The large cross section and clean experimental signature allow precision
tests of the standard model (SM), as well as constraints on the
parton distribution functions (PDFs) of the proton. In addition, a measurement of
the $\cPZ$ production process can set stringent constraints on physics beyond the standard
model. Moreover, dilepton events are valuable for calibrating the
detector and monitoring the LHC luminosity. The $\cPZ/\gamma^{*} \to \ell^+\ell^-$ process, where $\ell$ is a muon or an electron, is referred to as the $\cPZ$ boson process in this paper.

The $\cPZ$ boson production, identified via its decays into pairs of muons and electrons, can have nonzero transverse momentum, \pt, to the beam direction. This is due to the intrinsic \pt of the initial-state partons inside the proton, as well as initial-state radiation of gluons and quarks.
Measurements of the \pt distribution of the $\cPZ$ boson probe
various aspects of the strong interaction. In addition, an accurate theoretical
prediction of the \pt distribution is a key ingredient for a precise
measurement of the $\PW$ boson mass at the Tevatron and LHC.

Theoretical predictions of both the total and the differential $\cPZ$ boson production
cross section are available at next-to-next-to-leading order (NNLO) accuracy
in perturbative quantum chromodynamics
(QCD)~\cite{Melnikov:2006kv,Catani:2009sm}. Complete NNLO calculations of
vector boson production in association with a jet in hadronic collisions have
recently become available at $\mathcal{O}(\alpS^3)$ accuracy in the strong
coupling~\cite{Ridder:2015dxa,Boughezal:2015ded,Boughezal:2015dva}.
These calculations significantly reduce the factorization ($\mu_{\mathrm{F}}$) and
renormalization ($\mu_{\mathrm{R}}$) scale uncertainties,
which in turn reduce theoretical uncertainties in the prediction
of the \pt distribution in the high \pt region to the order of one
percent. Electroweak corrections are known at next-to-leading order (NLO) and
play an important role at high \pt~\cite{Dittmaier:2014qza,Lindert:2017olm}.

However, the fixed-order calculations are unreliable at low \pt due to soft
and collinear gluon radiation, resulting in large logarithmic
corrections~\cite{Collins:1984kg}. Resummation of the logarithmically divergent terms at
next-to-next-to-leading logarithmic (NNLL) accuracy has been matched with the
fixed-order predictions to achieve accurate predictions for the entire \pt
range~\cite{Balazs:1995nz,Catani:2015vma}. Fixed-order perturbative
calculations can also be combined with parton shower models~\cite{Sjostrand:2014zea,Gleisberg:2008ta,Bahr:2008pv} to obtain
fully exclusive predictions~\cite{Nason:2004rx,Frixione:2002ik,Alioli:2010xd,Alwall:2014hca}.
Transverse momentum dependent (TMD) PDFs~\cite{Angeles-Martinez:2015sea} can also
be used to incorporate resummation and nonperturbative effects.

The $\cPZ$ boson \pt and rapidity $\rapidity^{\cPZ}$ distributions were previously
measured, using $\Pe^+\Pe^-$ and $\mu^+\mu^-$ pairs, by the ATLAS, CMS, and
LHCb Collaborations in proton-proton ($\Pp\Pp$) collisions at $\sqrt{s}=7$, $8$,
and 13\TeV at the LHC~\cite{ATLAS_ZpT7TeV,ATLAS_ZptEta7TeV,Aad:2015auj,Aaboud:2016btc,Sirunyan:2018owv,CMS_ZpT7TeV,CMS_ZpT8TeV,CMS:2014jea,Khachatryan:2016nbe,LHCb_WZ7TeV,LHCb_Zee7TeV,LHCb_ZpT7TeV,LHCb_WZ8TeV,Aaij:2016mgv},
and in $\Pp\Pap$ at $\sqrt{s} = 1.8$ and 1.96\TeV by the CDF and D0 Collaborations at the Fermilab Tevatron~\cite{Affolder:1999jh,Abbott:1999yd,TevatronWZ:D0PhysRevLett2008_100,TevatronWZ:D0PhysLettB2010_693,TevatronWZ:D0PhysRevLett2011_106}.
The $\rapidity^{\cPZ}$ distribution in $\Pp\Pp$ collisions is strongly correlated
with the longitudinal momentum fraction $x$ of the initial partons and provides constraints on the PDFs
of proton. The precision of the $\cPZ$ boson \pt measurements is
limited by the uncertainties in the \pt measurements of
charged leptons from $\cPZ$ boson decays. The observable $\phiStar$~\cite{Banfi:2010cf,Banfi:2012du} is defined by the expression
\begin{linenomath}
\begin{align}
\label{eq0}
\phiStar = \tan \left( \frac{\pi -\Delta\phi}{2} \right) \, \sin(\theta^*_\eta), \quad  \cos(\theta^*_\eta) =\tanh\left(\frac{\Delta \eta}{2}\right),
\end{align}
\end{linenomath}
where  $\Delta \eta$ and $\Delta\phi$ are the differences in pseudorapidity and azimuthal angle, respectively, between the two leptons. In the limit of negligible lepton mass rapidity and pseudorapidity are identical.
The variable $\theta^*_\eta$ indicates the scattering angle of the lepton pairs with respect to the beam in the boosted frame where the leptons are aligned.
The observable $\phiStar$ follows an approximate relationship
$\phiStar \sim \pt^{\cPZ} / m_{\ell\ell}$, so the range $\phiStar \le 1$
corresponds to $\pt^{\cPZ}$ up to about 100\GeV for a lepton pair mass close
to the nominal $\cPZ$ boson mass. The measurement resolution of $\phiStar$ is
better than that of \pt since it depends only on the angular direction of
the leptons and benefits from the  excellent spatial resolution of the CMS
inner tracking system. The $\cPZ$ boson $\phiStar$ distribution was
previously measured by the D0~\cite{TevatronWZ:D0PhysRevLett2011_106}, ATLAS~\cite{Aad:2015auj},
CMS~\cite{Sirunyan:2017igm}, and LHCb~\cite{Aaij:2016mgv} Collaborations.

We present inclusive fiducial and differential production cross sections for
the $\cPZ$ boson as a function of \pt, $\phiStar$, and $\abs{\rapidity^{\cPZ}}$. The data
sample corresponds to an integrated luminosity of $35.9 \pm 0.9\fbinv$
collected with the CMS detector~\cite{Chatrchyan:2008zzk} at the LHC in 2016.

\section{The CMS detector}
\label{sec:CMSdetector}
The central feature of the CMS apparatus is a superconducting solenoid
of 6\unit{m} internal diameter, providing a magnetic field of
3.8\unit{T}. Within the solenoid volume there are a silicon pixel and strip
tracker, a lead tungstate crystal electromagnetic calorimeter (ECAL),
and a brass and scintillator hadron calorimeter, each composed
of a barrel and two endcap sections. Forward calorimeters extend the $\eta$
coverage provided by the barrel and endcap detectors. Muons are detected in gas-ionization
detectors embedded in the steel flux-return yoke outside the solenoid. A more detailed
description of the CMS detector, together with a definition of the coordinate system used
and the relevant kinematic variables, can be found in Ref.~\cite{Chatrchyan:2008zzk}.

The first level of the CMS trigger system, composed of custom hardware processors, uses
information from the calorimeters and muon detectors to select events of interest in a
fixed time interval of less than 4\mus. The second level, known as the high-level trigger,
consists of a farm of processors running a version of the full event reconstruction
software optimized for fast processing, and reduces the event rate to $\mathcal{O}$(1\unit{kHz})
before data storage~\cite{Khachatryan:2016bia}.

\section{Signal and background simulation}
\label{sec:Samples}
Monte Carlo event generators are used to simulate the signal and
background processes. The detector response is simulated using a detailed
specification of the CMS detector, based on the \GEANTfour
package~\cite{Agostinelli:2002hh}, and event reconstruction is performed with
the same algorithms used for data.

The simulated samples include the effect of additional pp interactions in the same or nearby
bunch crossings (pileup), with the distribution matching that observed in
data, with an average of about 23 interactions per crossing.

$\PW\cPZ$ and $\cPZ\cPZ$ production, via $\Pq\Paq$ annihilation,
are generated at NLO with
$\POWHEG$ 2.0~\cite{Alioli:2008gx,Nason:2004rx,Frixione:2002ik,Alioli:2010xd}.
The $\Pg\Pg \to \cPZ\cPZ$ process is simulated with MCFM 8.0~\cite{MCFM} at leading order.
The $\cPZ\gamma$, $\ttbar \cPZ$, $\PW\PW\cPZ$, $\PW\cPZ\cPZ$, and $\cPZ\cPZ\cPZ$
processes are generated with \MGvATNLO~2.3.3~\cite{Alwall:2014hca}.
The signal samples are simulated using \MGvATNLO and $\POWHEG$ at NLO.
The \MGvATNLO generator is used to compute the response matrix in the data unfolding procedure.
The $\PYTHIA$ 8.226~\cite{Sjostrand:2014zea} package is used
for parton showering, hadronization, and the underlying-event simulation,
with tune CUETP8M1~\cite{Skands:2014pea,Khachatryan:2015pea}.
The NNPDF 3.0~\cite{Ball:2014uwa} set of PDF, with the
perturbative order matching used in the matrix element calculations, is used in the simulated samples.

\section{Event selection and reconstruction}
The CMS particle-flow event algorithm~\cite{Sirunyan:2017ulk} aims to
reconstruct and identify each individual particle in an event, with an optimized
combination of all subdetector information. Particles are identified as charged
and neutral hadrons, leptons, and photons.

The reconstructed vertex with the largest value of summed physics-object
$\pt^2$ is the primary $\Pp\Pp$ interaction vertex.
The physics objects are the objects returned by a jet finding
algorithm~\cite{Cacciari:2008gp,Cacciari:2011ma} applied to
all charged particle tracks associated with the vertex plus the corresponding
associated missing transverse momentum, which is the negative vector sum of
the \pt of those jets.

Muons are reconstructed by associating a track reconstructed in the inner
silicon detectors with a track in the muon system. The selected muon
candidates must satisfy a set of requirements based on the number of spatial
measurements in the silicon tracker and in the muon system, and the fit quality
of the combined muon track~\cite{Chatrchyan:2012xi,Sirunyan:2018fpa}.
Matching muons to tracks  measured in the silicon tracker results in a relative
\pt resolution of 1\% for muons in the barrel and better than 3\% in the endcaps,
for \pt ranging from 20--100\GeV.
The \pt resolution in the barrel is less than 10\% for muons with \pt up to 1\TeV.

Electrons are reconstructed by associating a track reconstructed in the inner
silicon detectors with a cluster of energy in the ECAL~\cite{Khachatryan:2015hwa}.
The selected electron candidates cannot originate from photon conversions in the detector material, and they must satisfy a set of requirements based on the shower shape of the energy deposit in the
ECAL. The momentum resolution for electrons from
$\cPZ\to\Pep\Pem$ decays ranges from 1.7\% in the barrel
region to 4.5\% in the endcaps~\cite{Khachatryan:2015hwa}.

The lepton candidate tracks are required to be consistent with the primary
vertex of the event~\cite{TRK-11-001}. This requirement suppresses the
background of electron candidates from photon conversion, and lepton
candidates originating from in-flight decays of heavy quarks. The lepton
candidates are required to be isolated from other particles in the event. The
relative isolation for the lepton candidates with transverse momentum $\pt^{\ell}$ is defined as
\begin{linenomath}
\begin{equation}
R_\text{iso} = \bigg[ \sum_{\substack{\text{charged} \\ \text{hadrons}}} \!\! \pt \, + \,
\max\big(0, \sum_{\substack{\text{neutral} \\ \text{hadrons}}} \!\! \pt
\, + \, \sum_{\text{photons}} \!\! \pt \, - \, 0.5 \, \pt^\mathrm{PU}
\big)\bigg] \bigg/ \pt^{\ell},
\label{eq:iso}
\end{equation}
\end{linenomath}
where the sums run over the charged and neutral hadrons, and photons, in a
cone defined by
$\Delta R \equiv \sqrt{\smash[b]{\left(\Delta\eta\right)^2 + \left(\Delta\phi\right)^2}} = 0.4$ (0.3)
around the muon (electron) trajectory. The $\pt^\mathrm{PU}$ denotes the contribution of charged
particles from pileup, and the factor 0.5 corresponds to an approximate average ratio of neutral to
charged particles~\cite{Chatrchyan:2012xi,Khachatryan:2015hwa}. Only charged hadrons
originating from the primary vertex are included in the sum.

 Collision events are collected using single-electron and single-muon triggers that require the presence of an isolated lepton with \pt larger than 24\GeV, ensuring a trigger efficiency above 96\% for events passing the offline selection. The event selection aims to identify either $\mu^+\mu^-$ or $\Pe^+\Pe^-$ pairs compatible with a $\cPZ$ boson decay. Therefore, the selected $\cPZ$ boson
candidates are required to have two oppositely charged
same-flavor leptons, muons or electrons, with a reconstructed invariant mass
within 15\GeV the nominal $\cPZ$ boson mass~\cite{Tanabashi:2018oca}.
In addition, both leptons are required to have $\abs{\eta}<2.4$ and $\pt > 25\GeV$.
To reduce the background from multiboson events with a third lepton, events are rejected
if an additional loosely identified lepton is found with $\pt > 10\GeV$.

\section{Background estimation}\label{sec:backgrounds}
The contribution of background processes in the data sample is small relative
to the signal. The background processes can be split into two components,
one resonant and the other nonresonant. Resonant multiboson background processes stem
from events with genuine $\cPZ$ bosons, \eg, $\PW\cPZ$ diboson
production, and their contributions are estimated from simulation.

Nonresonant background stems from processes without $\cPZ$ bosons, mainly from leptonic decays of $\PW$ boson in $\ttbar$, $\cPqt\PW$, and $\PW\PW$ events.
Small contributions from single top quark events produced via s- and
t-channel processes, and $\cPZ \to \Pgt\Pgt$ events are also present.
The contribution of these nonresonant flavor-symmetric backgrounds is estimated
from events with two oppositely charged leptons of different flavor, $\Pe^{\pm}\Pgm^{\mp}$, that
pass all other analysis requirements. The method assumes lepton flavor symmetry
in the final states of these processes~\cite{Chatrchyan:2014tja}.
Since the $\PW$ boson leptonic decay branching fractions are well-known, the number of $\Pe\Pgm$ events selected inside
the $\cPZ$ boson mass window can be used to predict the nonresonant background
in the $\Pgm\Pgm$ and $\Pe\Pe$ channels.

A summary of the data, signal, and background yields after the full selection for the dimuon and dielectron
final states is shown in Table~\ref{tab:yields}. The contribution of the background
processes is below 1\%.

\begin{table}[hbtp]
  \begin{center}
\caption{Summary of data, expected signal, and background yields after the full selection. The predicted signal yields are quoted using \MGvATNLO. The statistical uncertainties in the simulated samples are below 0.1\%.\label{tab:yields}}
\begin{tabular}{lcccc}
\hline
Final state      & Data & $\cPZ\to\ell\ell$ & Resonant background & Nonresonant background \\[\cmsTabSkip]

$\mu\mu$         & $20.4 \times 10^{6}$ & $20.7 \times 10^{6}$ & $30 \times 10^{3}$ & $41 \times 10^{3}$ \\
$\Pe\Pe$         & $12.1 \times 10^{6}$ & $12.0 \times 10^{6}$ & $19 \times 10^{3}$ & $26 \times 10^{3}$ \\
\hline
  \end{tabular}
  \end{center}
\end{table}

\section{Analysis methods}\label{sec:fiducial}
The fiducial region is defined by a common set of kinematic selections
applied to both the $\mu^+\mu^-$ and $\Pe^+\Pe^-$ final states at generator level, emulating
the selection performed at the reconstruction level. Leptons are
required to have $\pt > 25\GeV$ and $\abs{\eta} < 2.4$, and a dilepton invariant
mass $\abs{m_{\ell\ell}-91.1876\,\GeV} < 15\GeV$. 
A small fraction (3\%) of selected signal events do not 
originate from the fiducial region because of detector effects. 
This contribution is treated as background and subtracted from the data yield.
The measured distributions, after
subtracting the contributions from the background processes, are corrected for
detector resolution effects and inefficiencies due to so-called dressed lepton
kinematics. The dressed leptons at generator level are defined by combining the four-momentum of
each lepton after the final-state photon radiation (FSR) with that of photons
found within a cone of $\Delta R = 0.1$ around the lepton. By using this definition, the measured kinematic distributions for $\cPZ$ boson decays to the muon final state and to the electron final state agree to better than 0.1\%. The rapidity
measurement is restricted to $\abs{\rapidity^{\cPZ}} < 2.4$. The \pt and $\phiStar$
measurements are restricted to $\pt < 1500\GeV$ and $\phiStar < 50$, respectively.
There are less than 0.001\% of events with $\pt > 1500\GeV$ and less than
0.02\% with $\phiStar > 50$.

The efficiencies for the reconstruction, identification, and isolation
requirements on the leptons are obtained in bins of \pt and $\eta$ using the
``tag-and-probe'' technique~\cite{CMS:2011aa}. Scale factors are applied as
event weights on the simulated samples to correct for the differences in the
efficiencies measured in the data and the simulation. The combined scale factor
for the reconstruction, identification, and isolation efficiencies for leptons ranges from 0.9 to 1.0, with an uncertainty of about 0.4 (0.7)\% for muons (electrons).
Momentum scale corrections are applied to the muons and electrons in
both data and simulated events~\cite{Bodek:2012id}.

The detector effects are expressed through a response matrix, calculated from
the simulated \MGvATNLO $\cPZ$ boson sample by associating
dressed and reconstructed objects for each observable independently.
To account for selection efficiencies and bin migrations, an unfolding procedure
based on a least squares minimization with Tikhonov regularization, as implemented in the
\textsc{TUnfold} framework~\cite{Schmitt:2012kp}, is applied.
The regularization reduces the effect of the statistical fluctuations present in the measured
distribution on the high-frequency content of the unfolded spectrum. The regularization strength
is chosen to minimize the global correlation coefficient~\cite{Schmitt:2016orm}.

\section{Systematic uncertainties}
\begin{figure}
	\centering
	\includegraphics[width=0.45\textwidth]{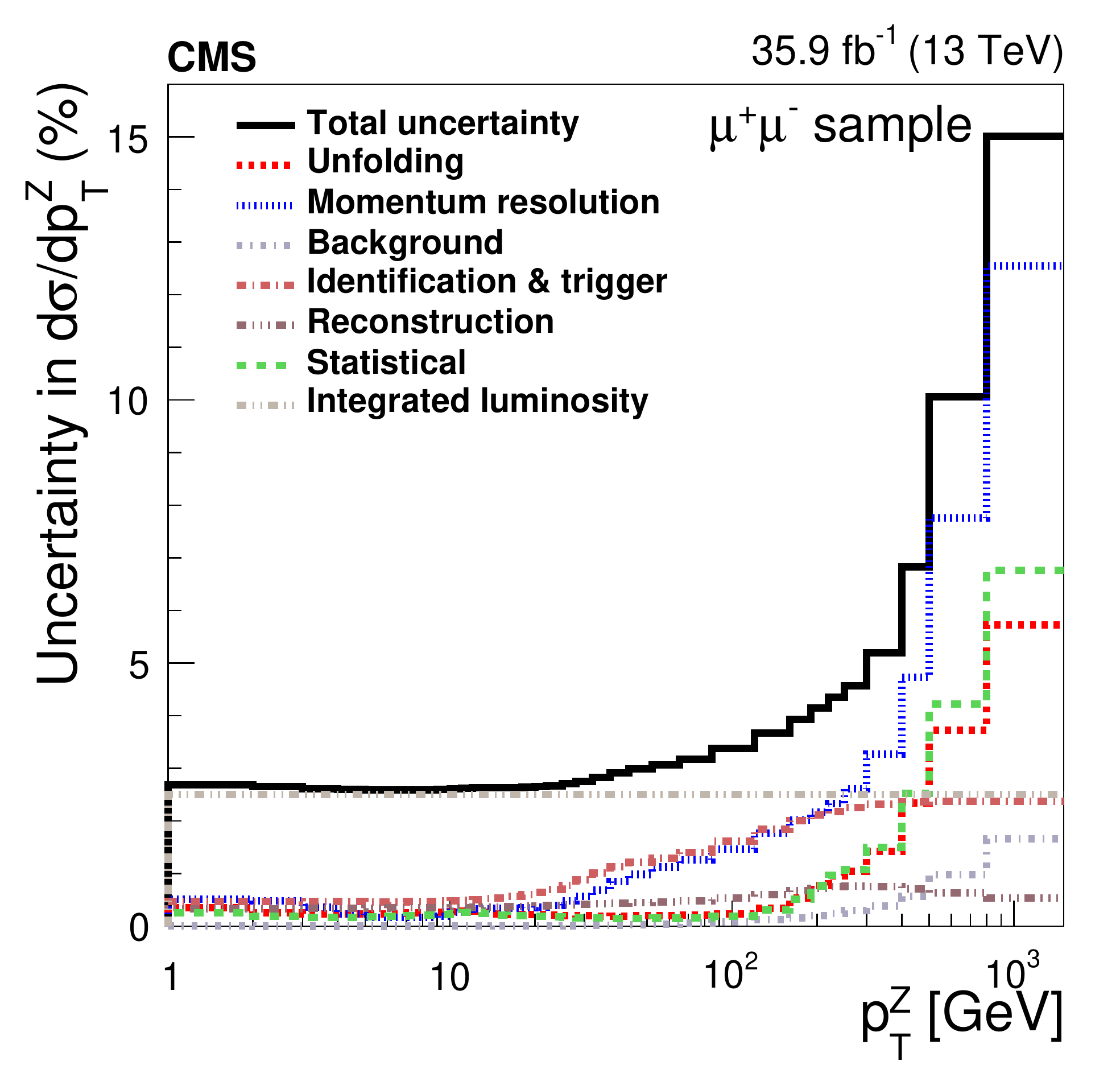}
        \includegraphics[width=0.45\textwidth]{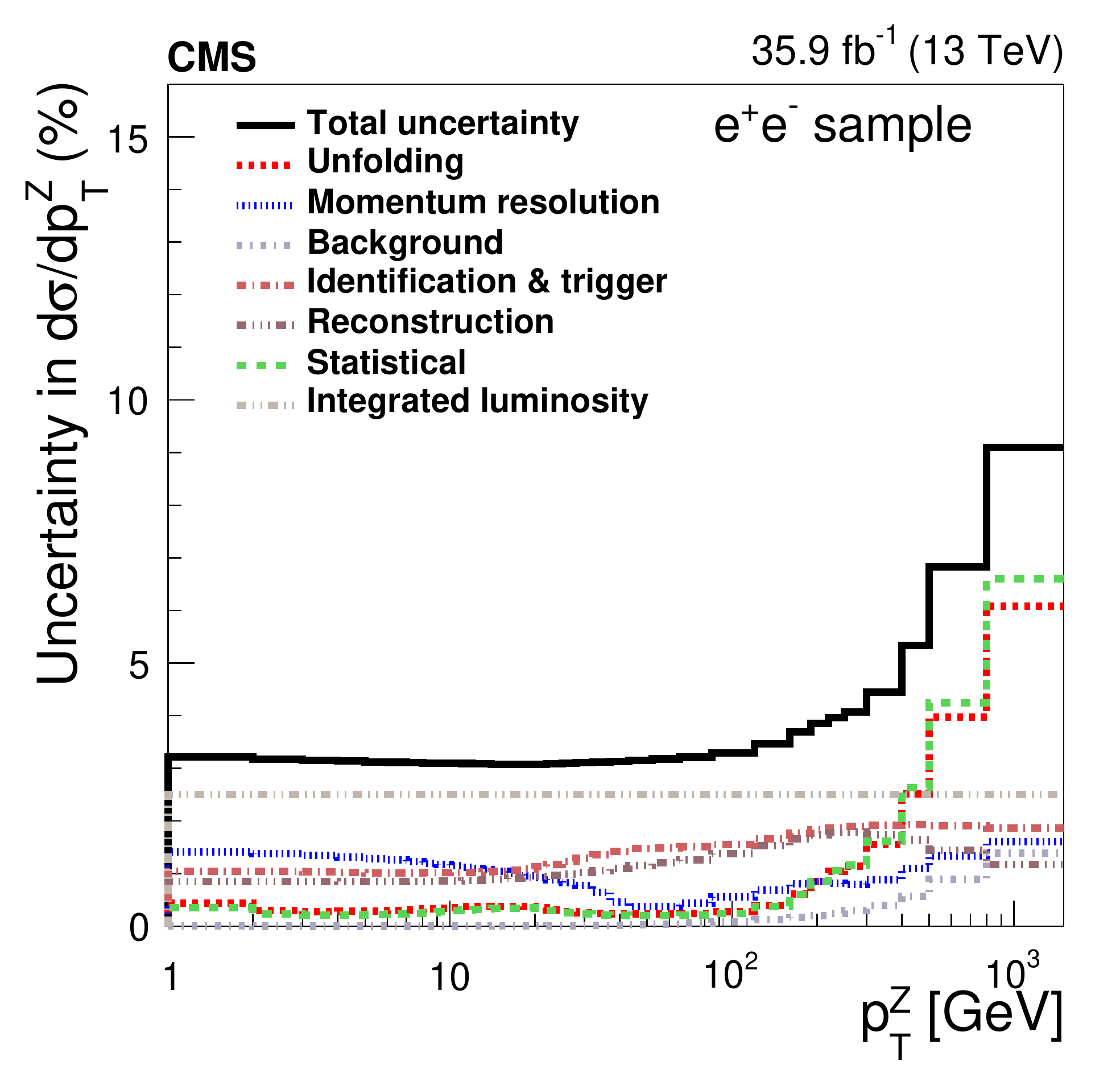}
	\includegraphics[width=0.45\textwidth]{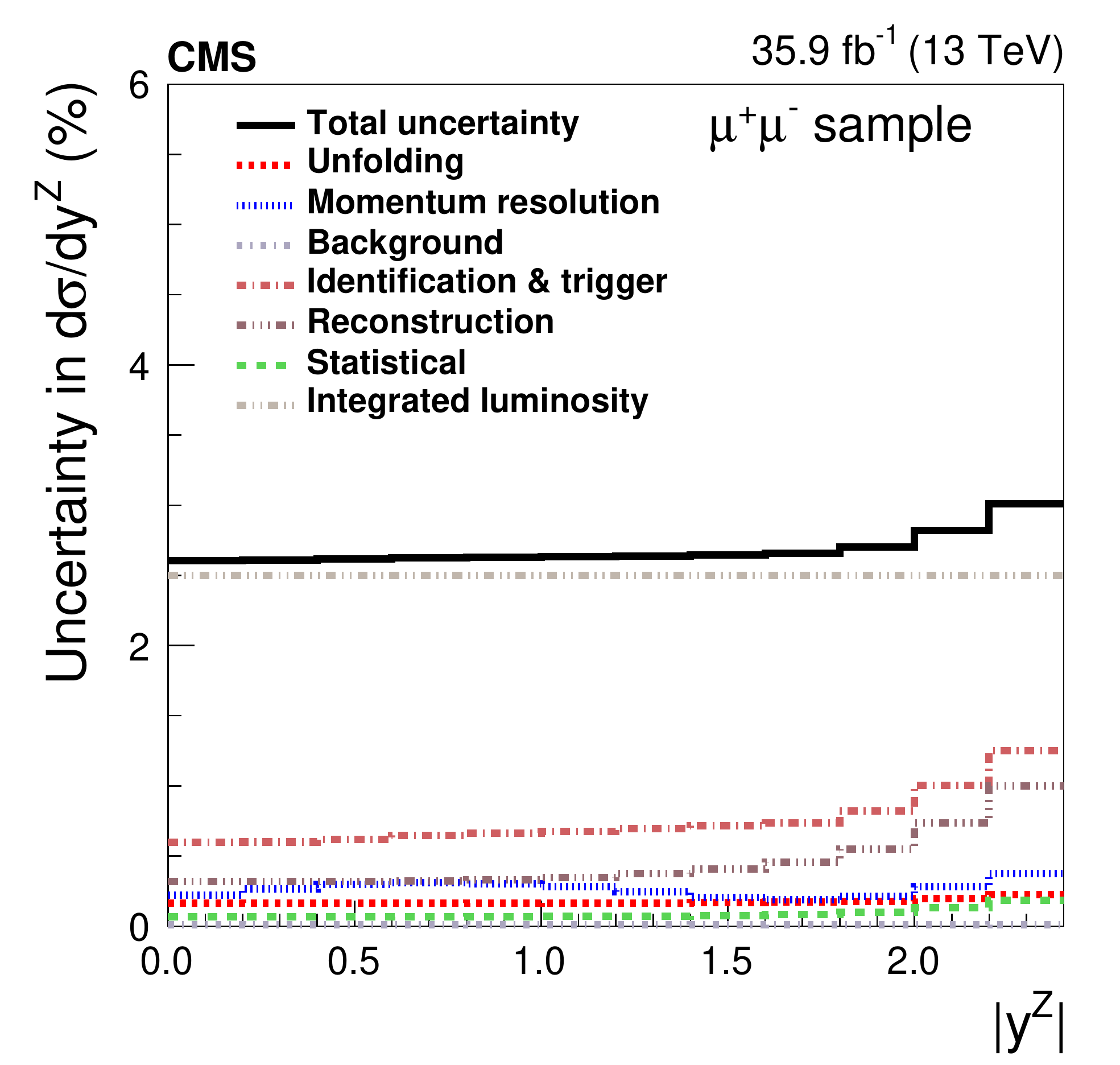}
        \includegraphics[width=0.45\textwidth]{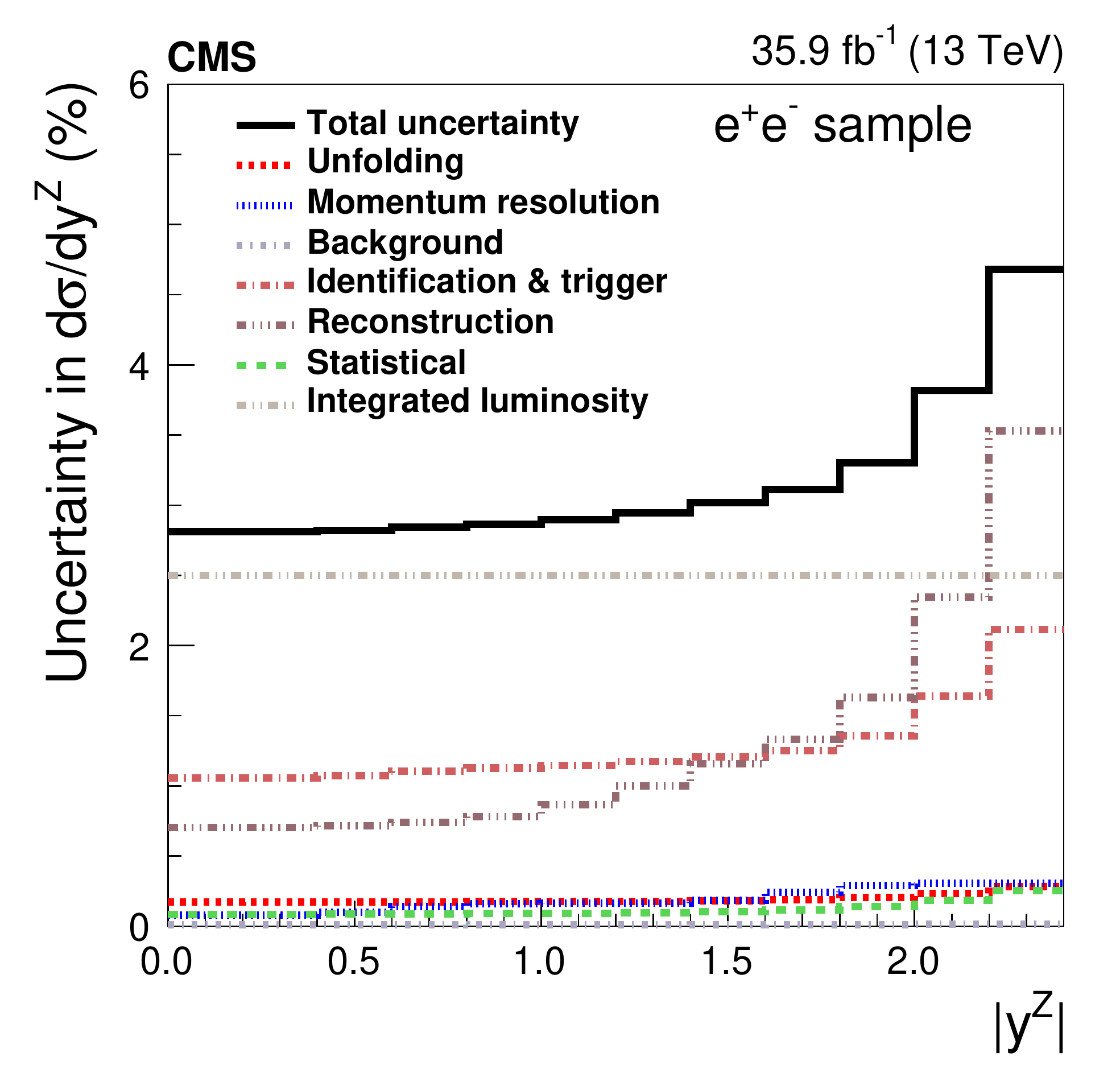}
	\includegraphics[width=0.45\textwidth]{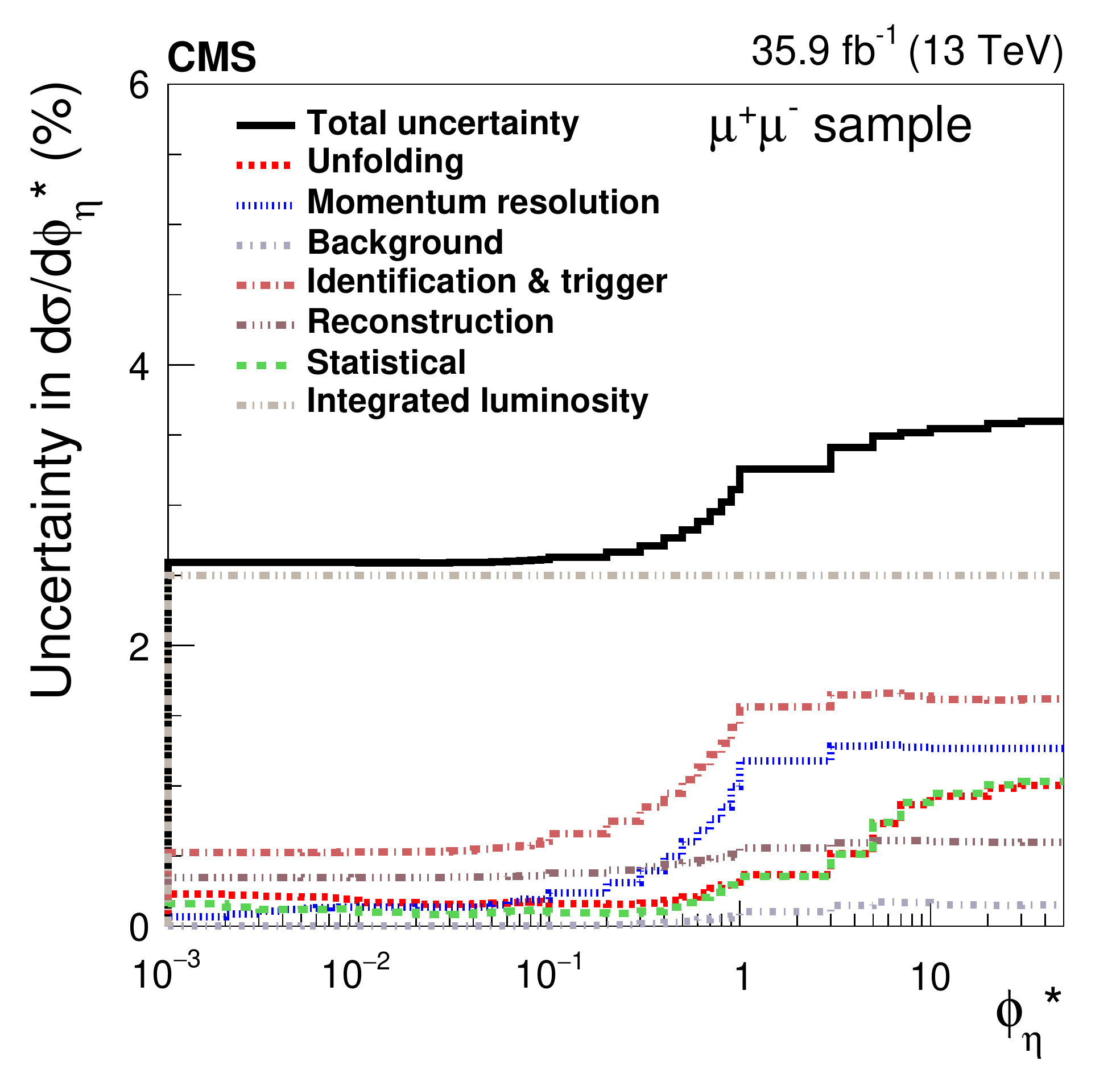}
        \includegraphics[width=0.45\textwidth]{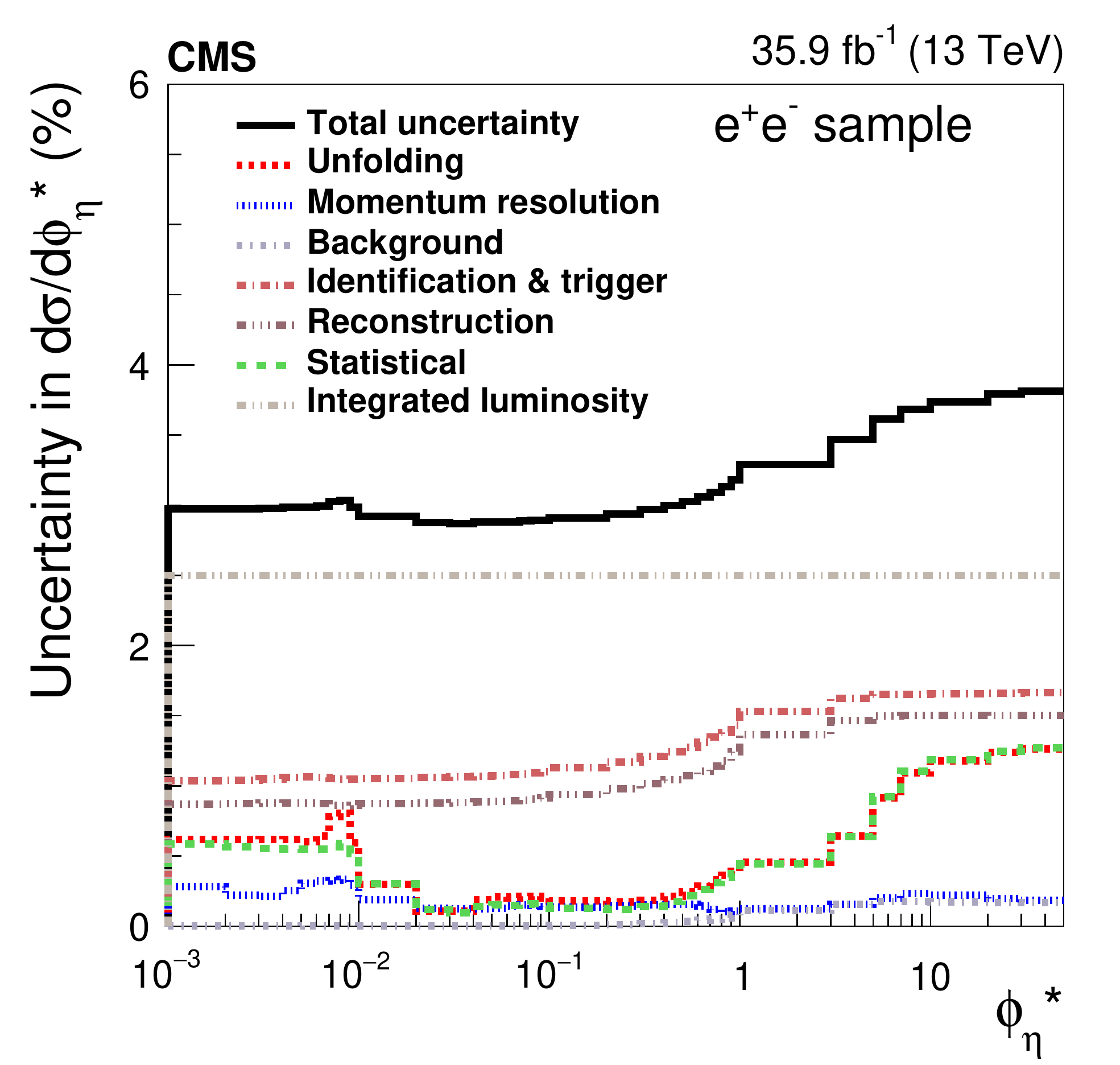}
	\caption{The relative statistical and systematic uncertainties from various
	sources for the absolute cross section measurements in bins of $\pt^{\cPZ}$ (upper),
	$\abs{\rapidity^{\cPZ}}$ (middle), and $\phiStar$ (lower). The left plots correspond to the dimuon final state and the right plots correspond to the
	dielectron final state. The uncertainty in the trigger efficiency is
	included as part of the lepton identification uncertainty.}
	\label{fig:syst0}
\end{figure}

The sources of systematic uncertainty in the measurement include the
uncertainties in the integrated luminosity, lepton efficiencies
(reconstruction, identification, and trigger), unfolding, lepton
momentum scale and resolution, and background
estimation. A summary of the total uncertainties for the absolute cross
section measurements in bins of $\pt^{\cPZ}$, $\abs{\rapidity^{\cPZ}}$, and $\phiStar$ is shown in
Fig.~\ref{fig:syst0}. The uncertainty in the trigger efficiency is
included as part of the lepton identification efficiency uncertainty.

Most of the sources of systematic uncertainty are considered fully correlated
between bins in all variables. The statistical
uncertainties due to the limited size of the data and simulated samples are considered uncorrelated between bins.
Some sources of systematic uncertainty have a significant statistical component,
such as the statistical uncertainties in the lepton efficiency measurement.
This statistical component is considered as uncorrelated between the lepton \pt
and $\eta$ bins used for the determination of the lepton efficiencies.

\begin{figure}
	\centering
	\includegraphics[width=0.45\textwidth]{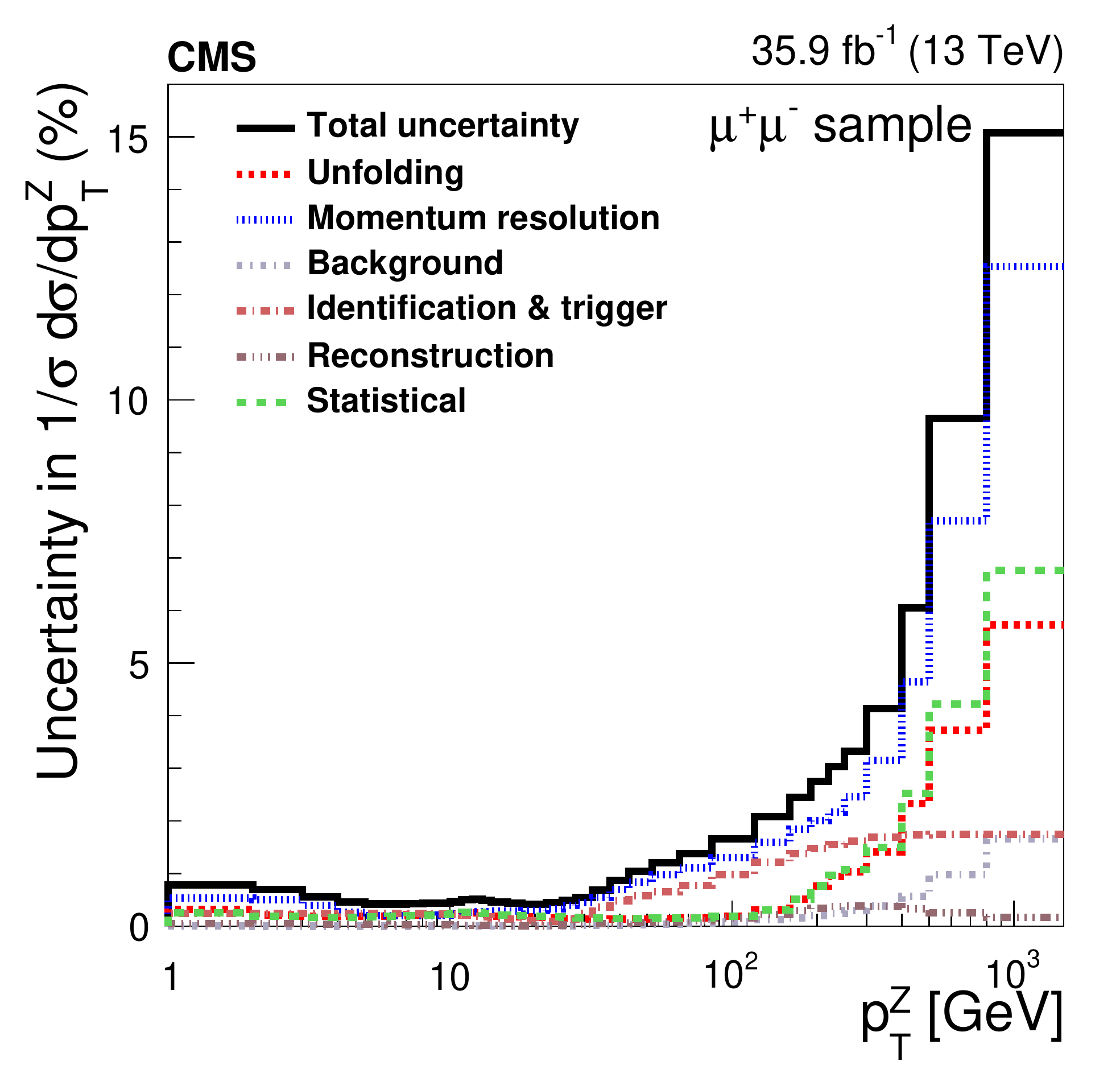}
        \includegraphics[width=0.45\textwidth]{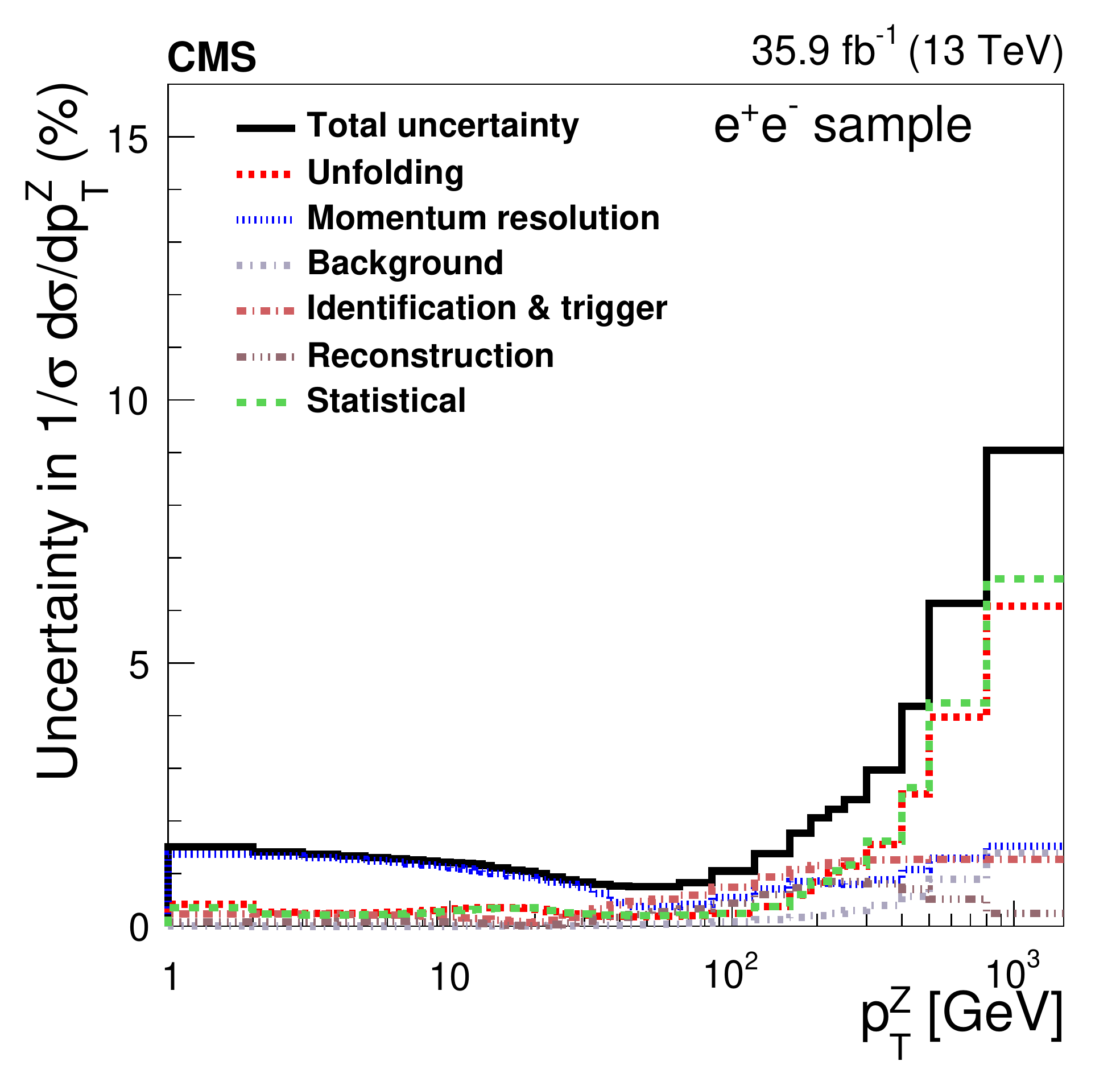}
	\includegraphics[width=0.45\textwidth]{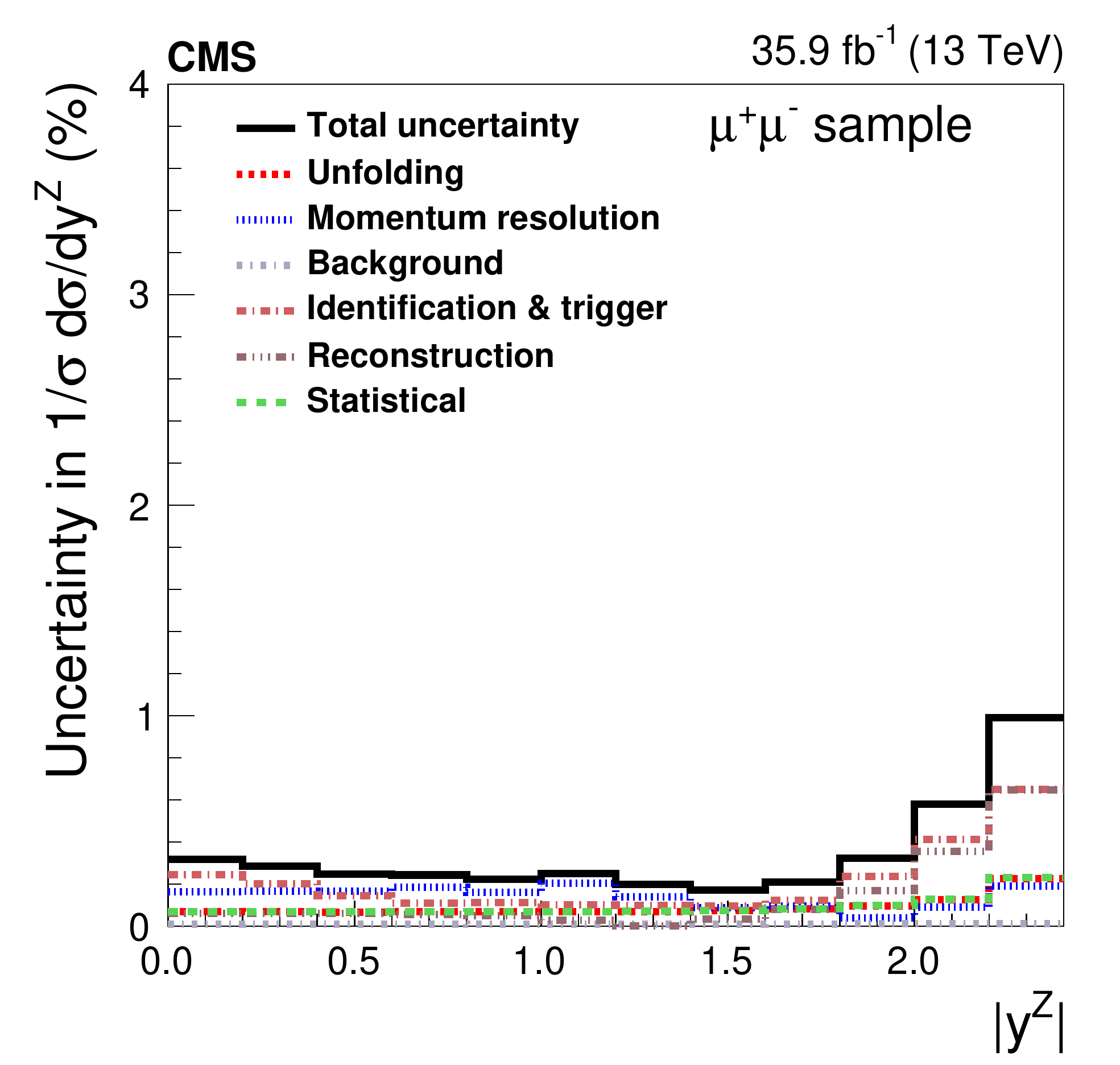}
        \includegraphics[width=0.45\textwidth]{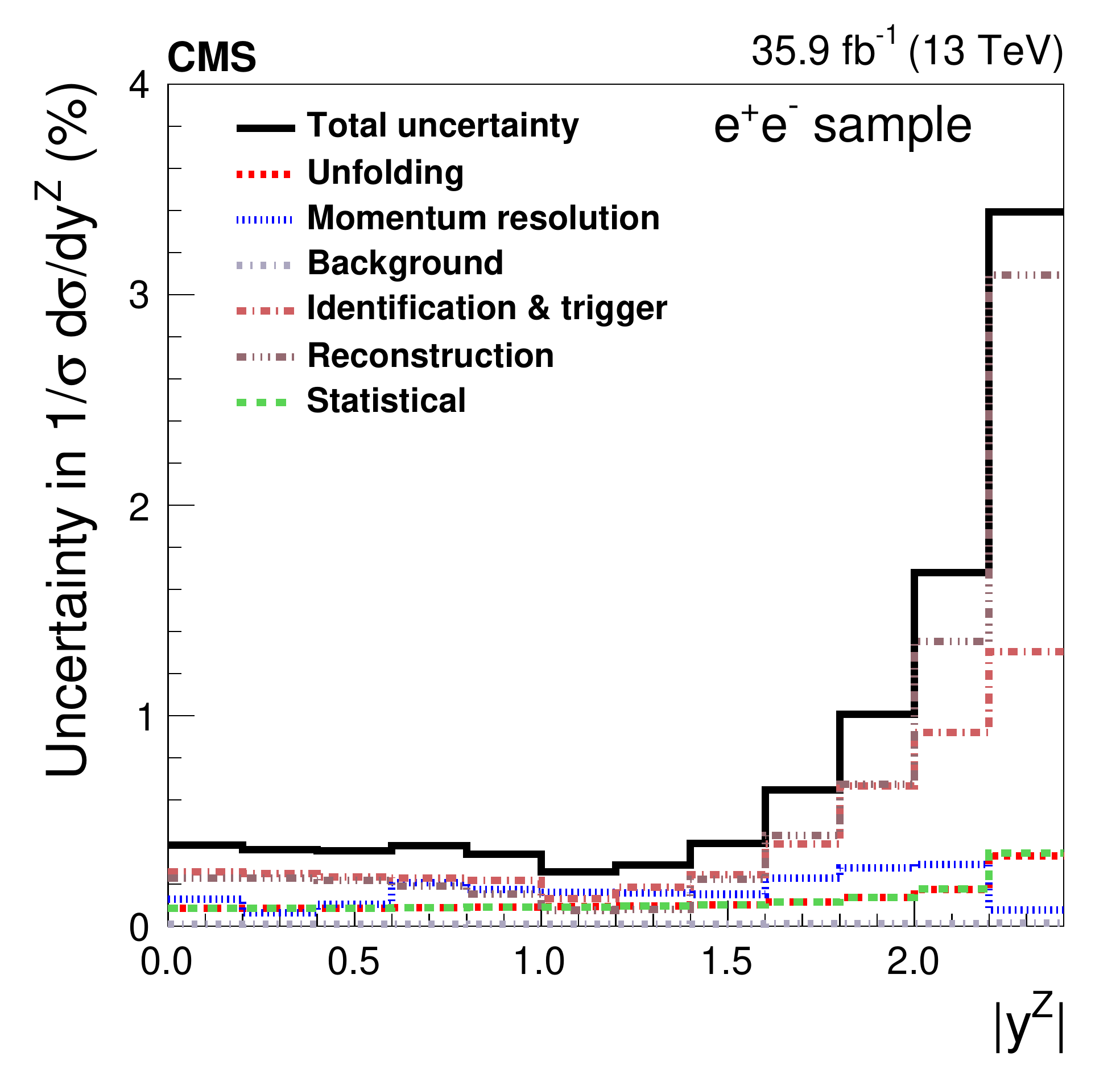}
	\includegraphics[width=0.45\textwidth]{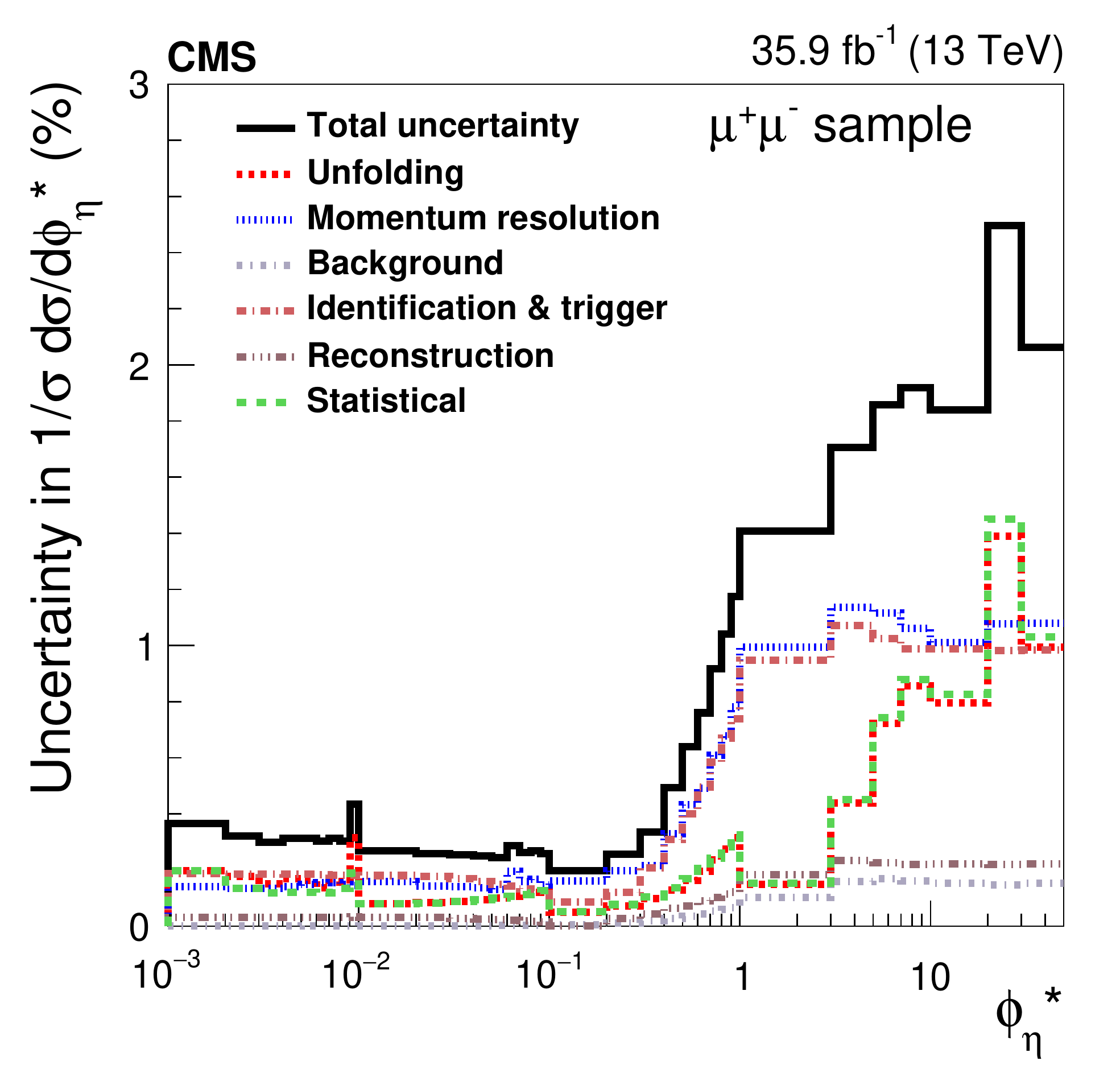}
        \includegraphics[width=0.45\textwidth]{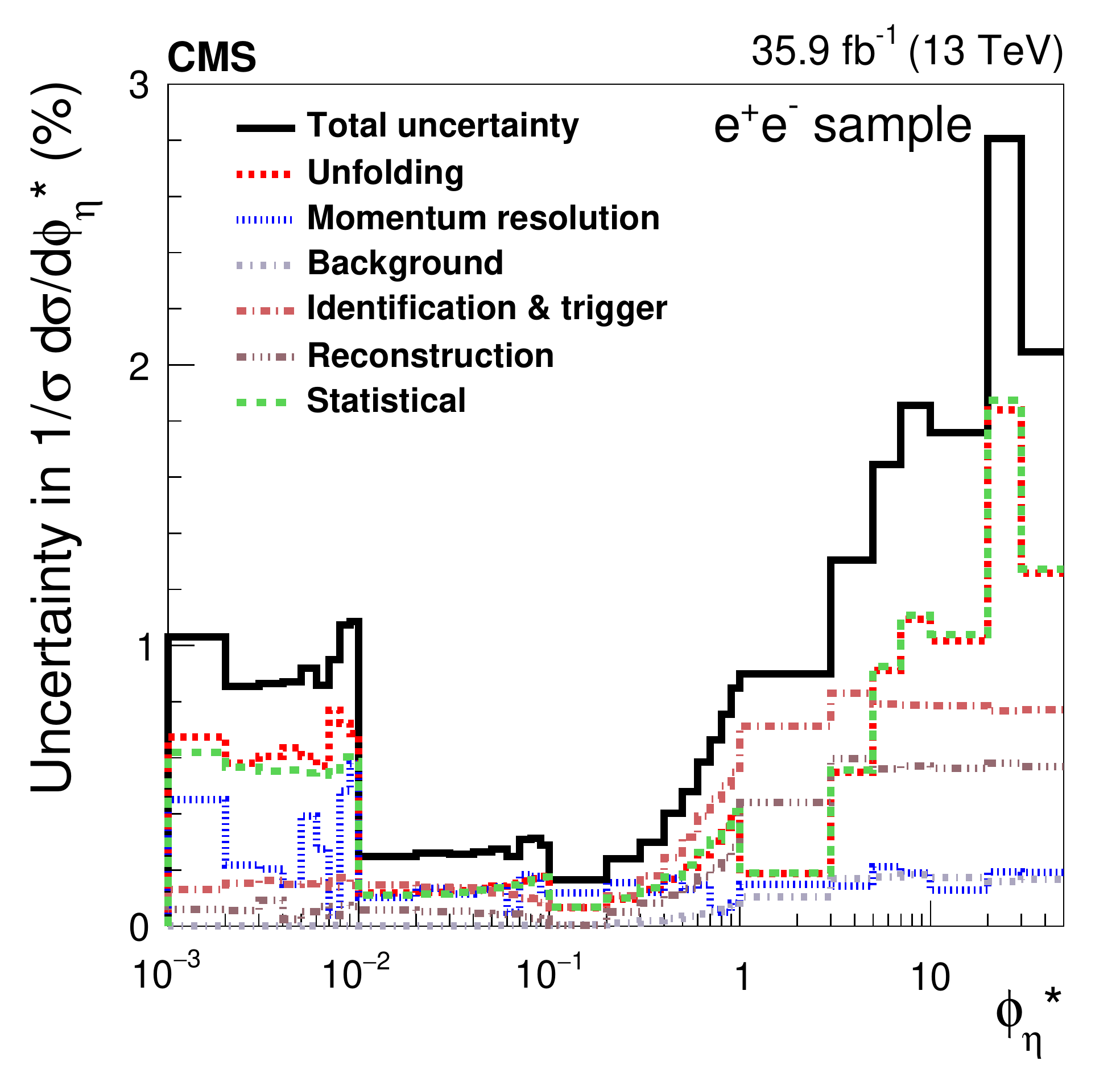}
	\caption{The relative statistical and systematic uncertainties from various
	sources for the normalized cross section measurements in bins of $\pt^{\cPZ}$ (upper),
	$\abs{\rapidity^{\cPZ}}$ (middle), and $\phiStar$ (lower). The left plots correspond to the dimuon final state and the right plots correspond to the dielectron final state.}
	\label{fig:syst_xratio}
\end{figure}

Measurements of the normalized differential cross sections $(1/\sigma)\rd\sigma/\rd\pt^{\cPZ}$, $(1/\sigma)\rd\sigma/\rd\abs{\rapidity^{\cPZ}}$, and $(1/\sigma)\rd\sigma/\rd\phiStar$ are also performed. Systematic uncertainties are largely reduced for the normalized cross section measurements. A summary of the total uncertainties for the normalized cross section measurements in bins of $\pt^{\cPZ}$, $\abs{\rapidity^{\cPZ}}$, and $\phiStar$ is
shown in Fig.~\ref{fig:syst_xratio}. 
Because of the binning in $\phiStar$, the uncertainty in this observable 
in the region around 1 is expected to follow a sharper behavior.

The largest source of uncertainty in the inclusive total cross section
measurement comes from the measurement of the integrated luminosity and
amounts to~2.5\%~\cite{LUM-17-001}. That uncertainty is relevant only for the
absolute cross section measurements. The leading uncertainties for the
normalized cross section measurements are related to the momentum
scale and the reconstruction efficiency.

A potential bias in the measurement of the reconstruction, identification, and
isolation efficiencies with the tag-and-probe technique is estimated by
studying the modeling of the background and signal parameterization in the
dilepton invariant mass fit. The uncertainty in the modeling of the electromagnetic 
FSR in the tag-and-probe fits is obtained by weighting the simulation to reflect 
the differences between $\PYTHIA$~\cite{Sjostrand:2014zea} and 
PHOTOS 3.56~\cite{Golonka:2005pn} modeling of the FSR. The exponentiation mode 
of PHOTOS is used. The tag selection in the tag-and-probe technique can also bias 
the efficiency measurement. An additional uncertainty is
considered by varying the tag selection requirements in the efficiency measurement. The uncertainty
in the trigger and lepton reconstruction and selection efficiency is about 0.8 (1.3)\% in dimuon
(dielectron) final states with a sizable dependence on $\pt^{\cPZ}$,
$\abs{\rapidity^{\cPZ}}$, and $\phiStar$.

The uncertainty in the dimuon (dielectron) reconstruction efficiency varies
between 0.1 (0.2)\% in the central part of the detector and 0.5 (2.5)\% at large
$\abs{\rapidity^{\cPZ}}$ values. The reconstruction efficiency uncertainty
also includes the effect of partial mistiming of signals in the forward region in the ECAL endcaps,
leading to a one percent reduction in the first-level trigger efficiency. The effect of statistical uncertainties
in the measured data-to-simulation scale factors is estimated by varying them within the uncertainties in a series of pseudo-experiments.

The systematic uncertainty due to the choice of the $\cPZ$ boson simulated
sample used to determine the response matrices is evaluated by repeating the
analysis using $\POWHEG$ as the signal sample. The dependence of the
measurements on the shapes of $\pt^{\cPZ}$, $\abs{\rapidity^{\cPZ}}$, and $\phiStar$
are about 0.3 and 0.5\% for the dimuon and dielectron final states, respectively.
The uncertainty due to the finite size of the simulated signal
sample used for the unfolding reaches about 5\% at large $\pt^{\cPZ}$, and the
variation with $\pt^{\cPZ}$, $\abs{\rapidity^{\cPZ}}$, and $\phiStar$ closely
resembles the statistical uncertainty in data.
The systematic uncertainties in the absolute cross section measurement arising from
the uncertainties in the lepton momentum scale and resolution are at a level of 0.1
(0.5)\% for the dimuon (dielectron) final state. 
These uncertainties also affect event selection and, because of the correlation 
between $\phiStar$ and $\pt^{\cPZ}$, follow a similar trend for both observables. 
The muon and electron momentum scales are corrected for the residual misalignment in the
detector and the uncertainty in the magnetic field measurements.

The uncertainty in the nonresonant background contribution is estimated
conservatively to be about 5\%, leading to an uncertainty in the total cross
section measurement below 0.1\%. The relative contribution of the
nonresonant background processes increases with $\abs{\rapidity^{\cPZ}}$ and \pt, resulting
in an uncertainty of 2\% at high \pt. The resonant background
processes are estimated from simulation and the uncertainties in the
background normalization are derived from variations of $\mu_{\mathrm{R}}$, $\mu_{\mathrm{F}}$,
$\alpS$, and PDFs~\cite{Ball:2014uwa,Butterworth:2015oua,Lai:2010vv,Martin:2009iq,Ball:2011mu,MCFM}
resulting in uncertainties below 0.1\% for the absolute cross section
measurement.

When combining the muon and electron channels, the luminosity, background estimation, and
modeling uncertainties are treated as correlated parameters, all others are considered
as uncorrelated.

\begin{figure}
	\centering
	\includegraphics[width=0.45\textwidth]{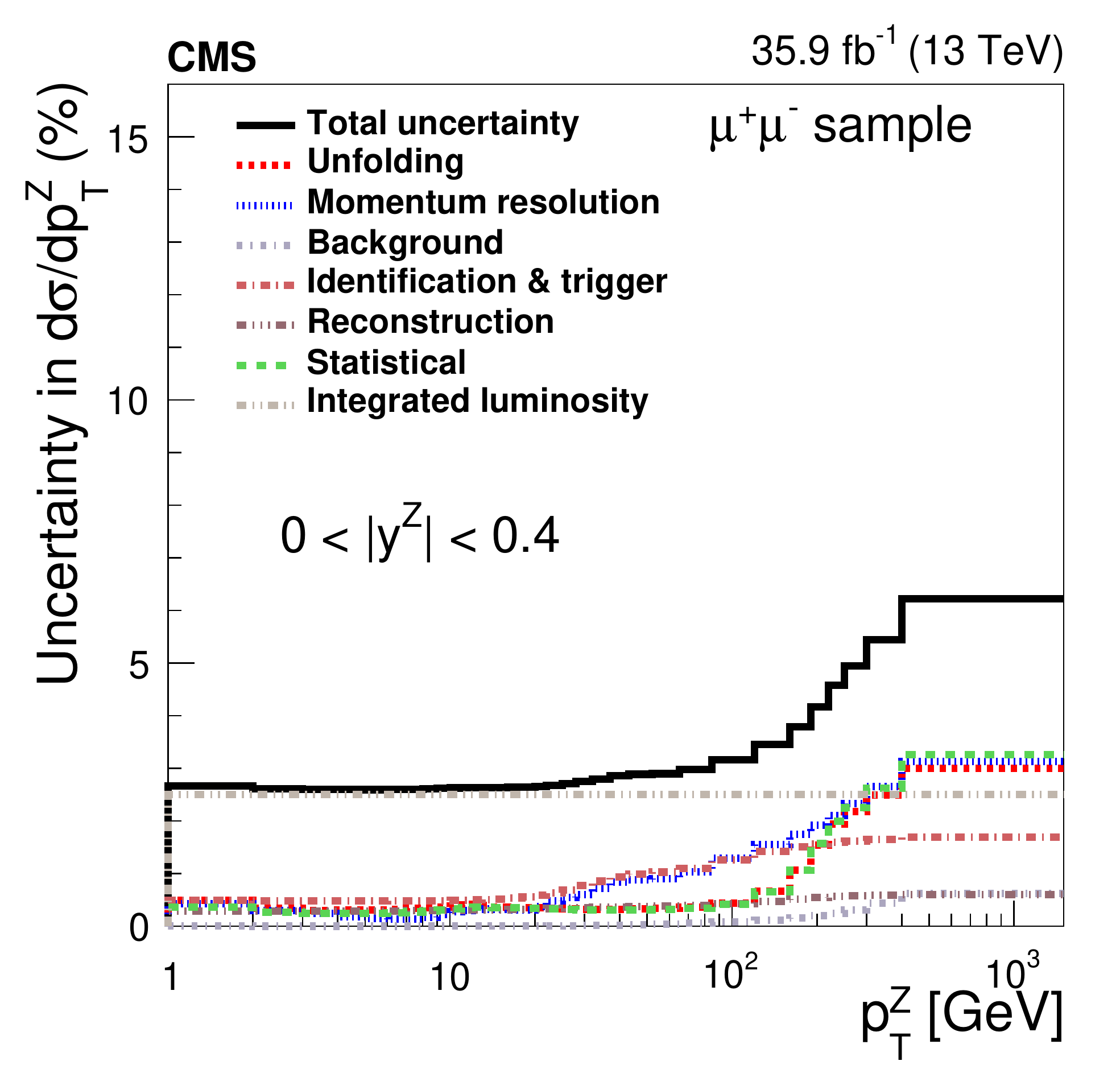}
	\includegraphics[width=0.45\textwidth]{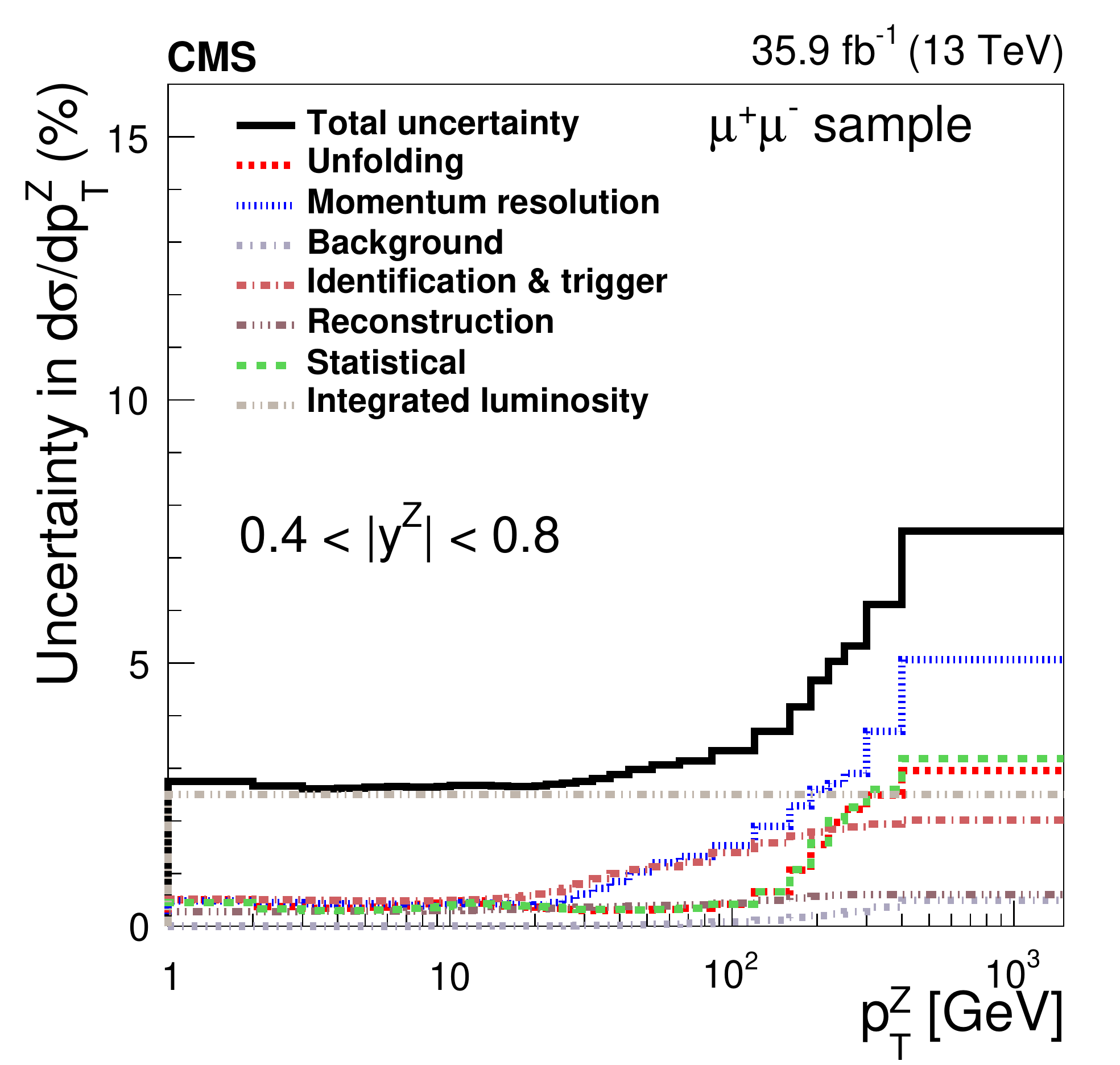}
	\includegraphics[width=0.45\textwidth]{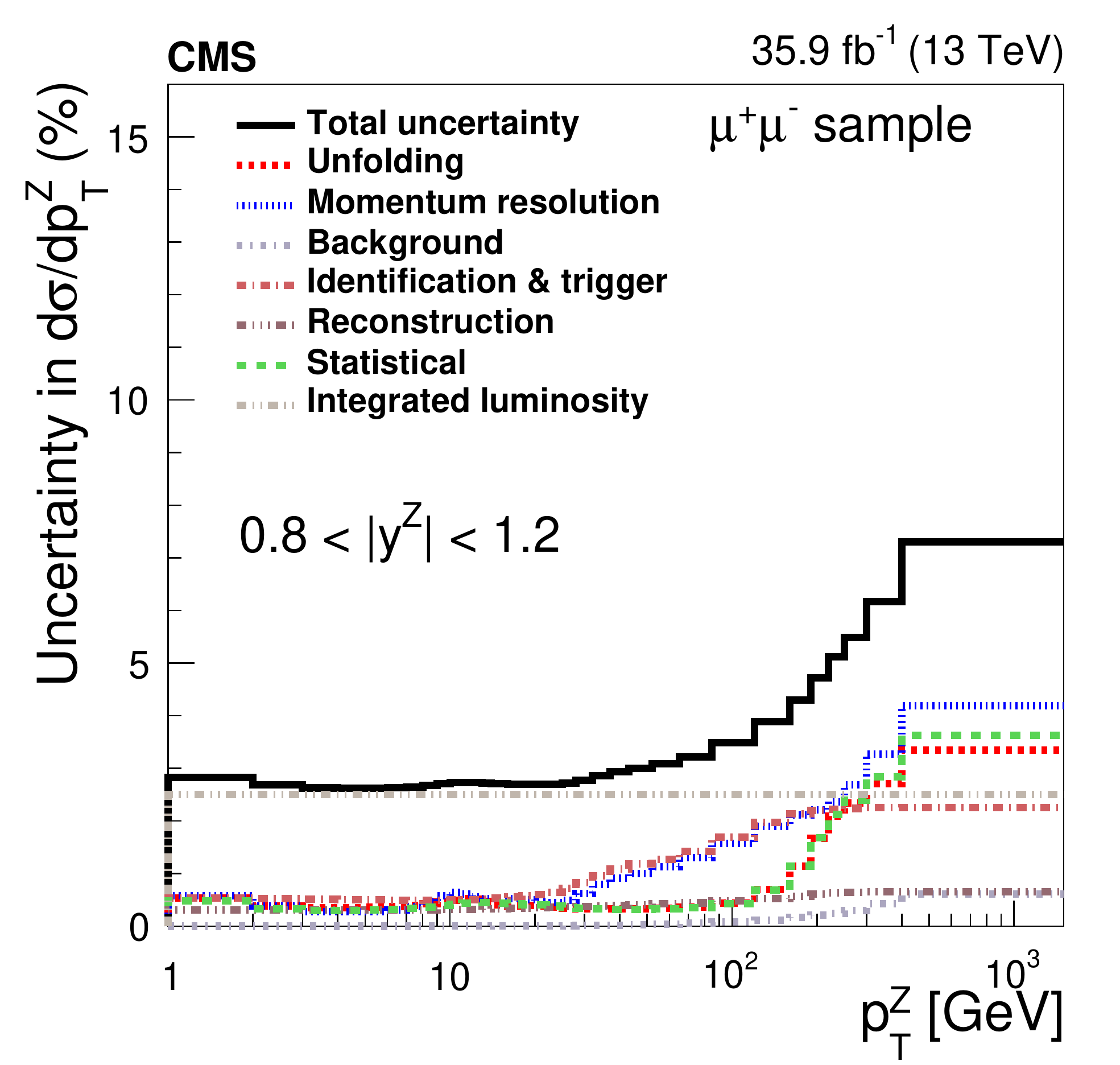}
	\includegraphics[width=0.45\textwidth]{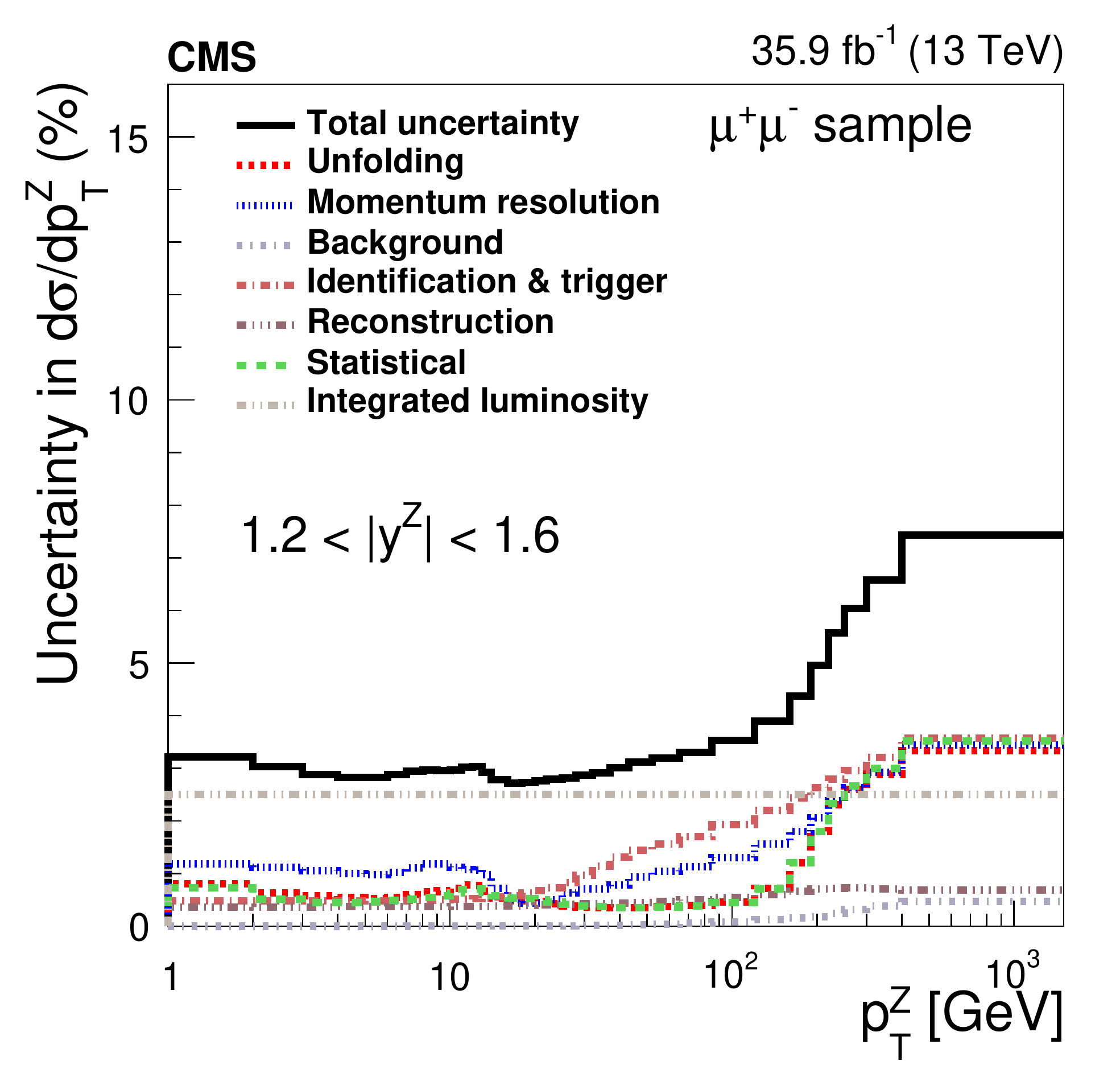}
	\includegraphics[width=0.45\textwidth]{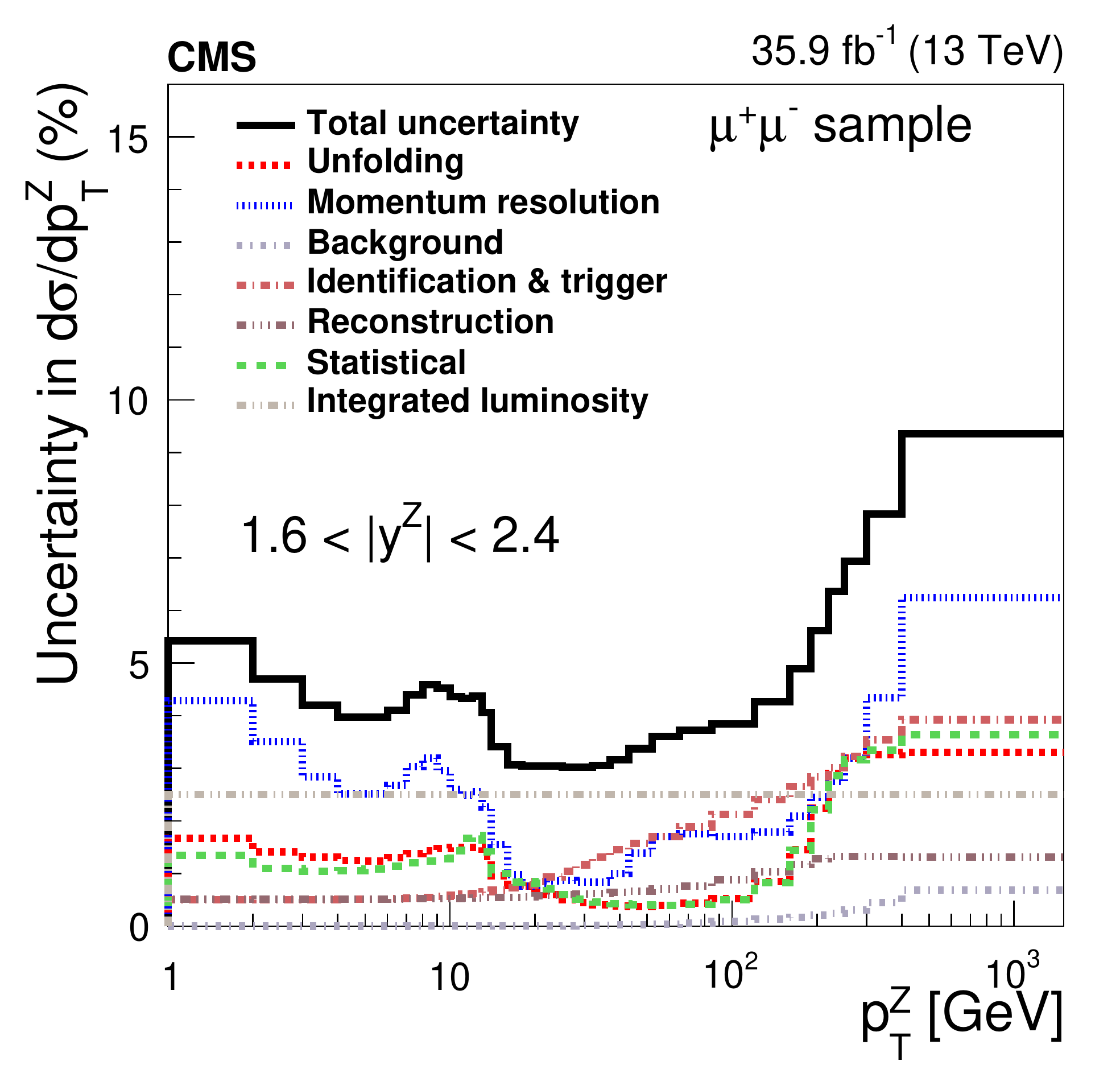}
	\caption{The relative statistical and systematic uncertainties from various sources for the absolute double-differential cross section measurements in bins of $\pt^{\cPZ}$ for the
	$0.0 < \abs{\rapidity^{\cPZ}} < 0.4$ bin (upper left), $0.4 < \abs{\rapidity^{\cPZ}} < 0.8$ bin (upper right),
	$0.8 < \abs{\rapidity^{\cPZ}} < 1.2$ bin (middle left), $1.2 < \abs{\rapidity^{\cPZ}} < 1.6$ bin (middle right), and $1.6 < \abs{\rapidity^{\cPZ}} < 2.4$ bin (lower) in the dimuon final state.
	}
	\label{fig:syst1}
\end{figure}

\begin{figure}
	\centering
	\includegraphics[width=0.45\textwidth]{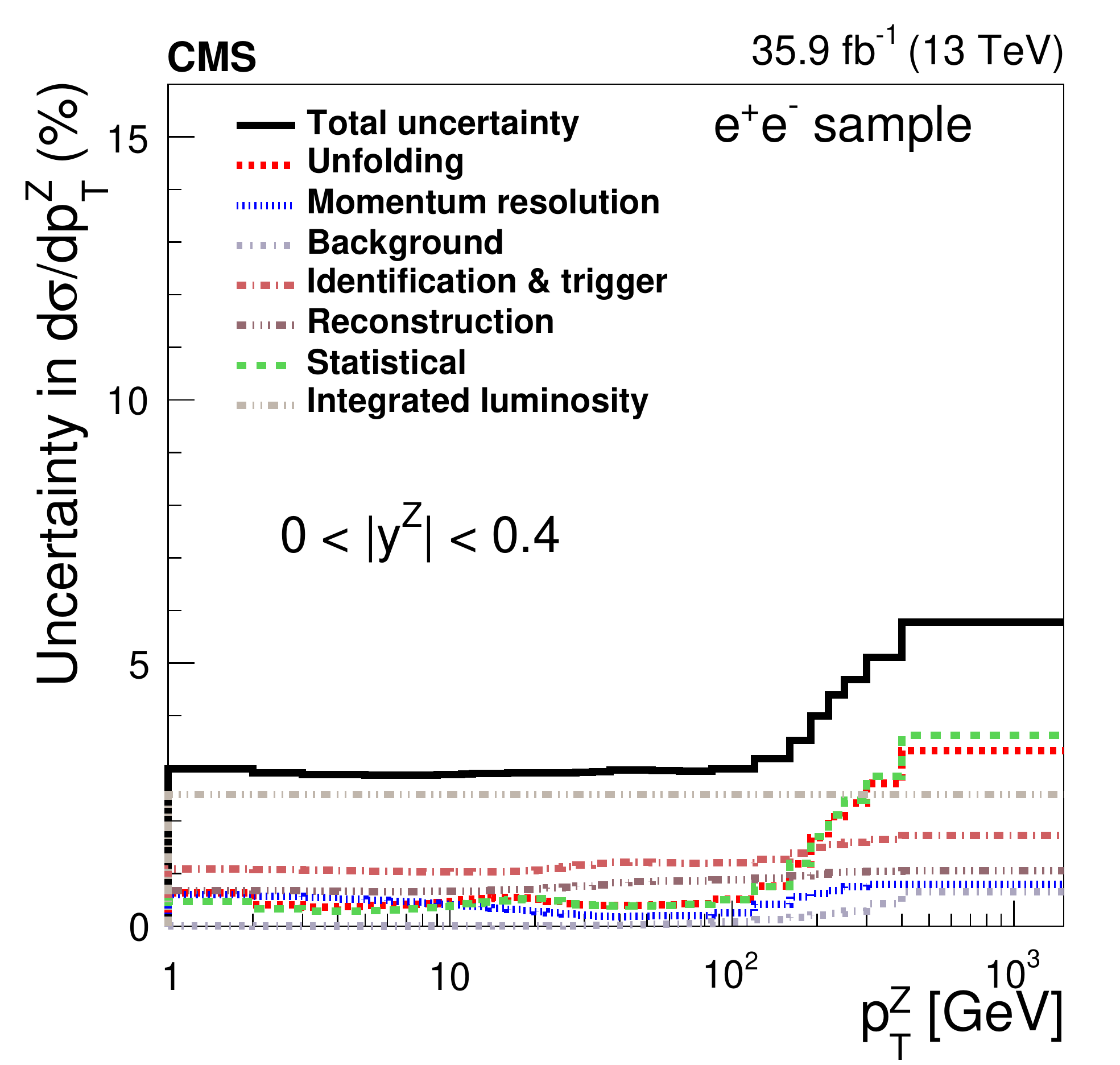}
	\includegraphics[width=0.45\textwidth]{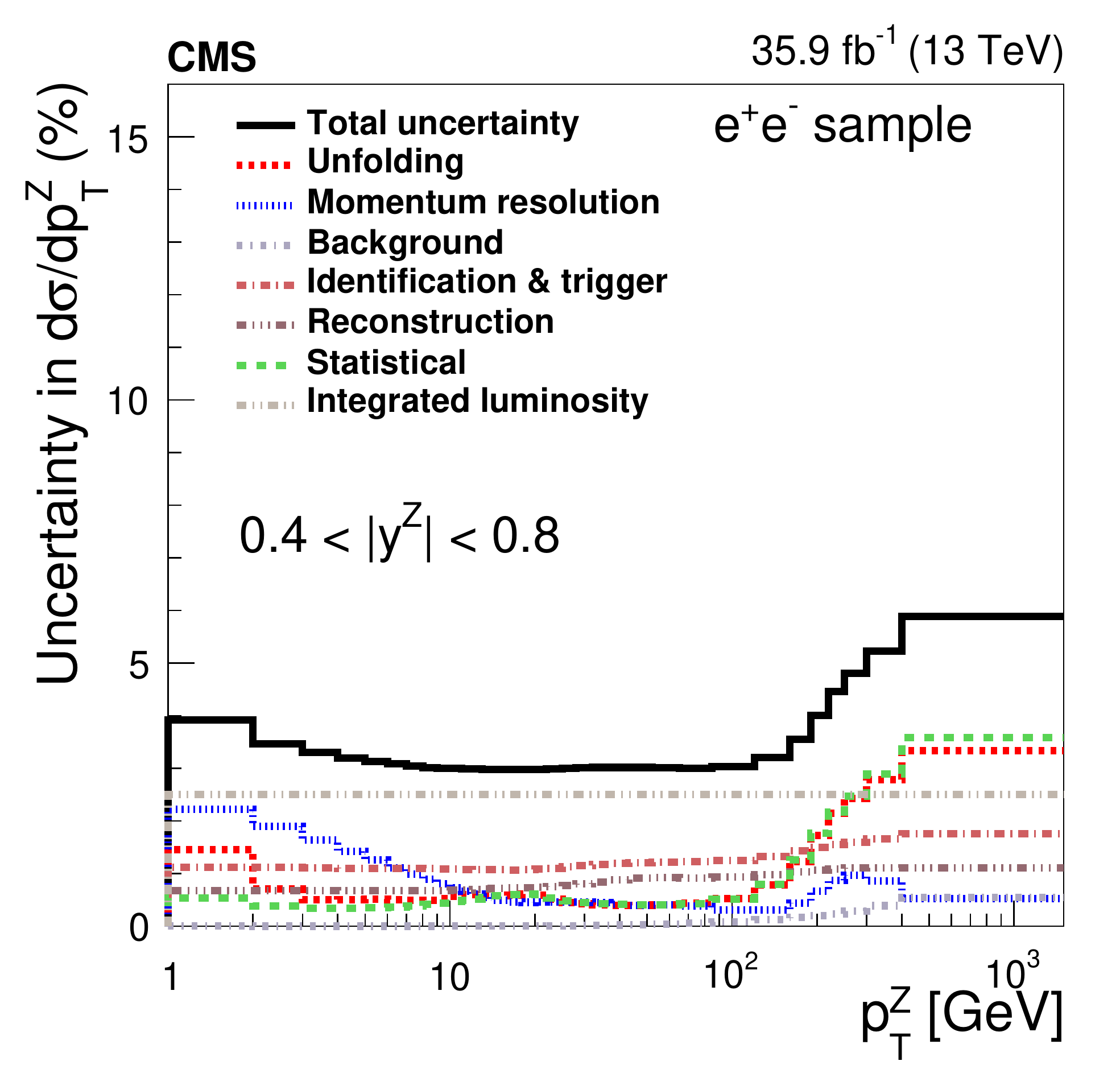}
	\includegraphics[width=0.45\textwidth]{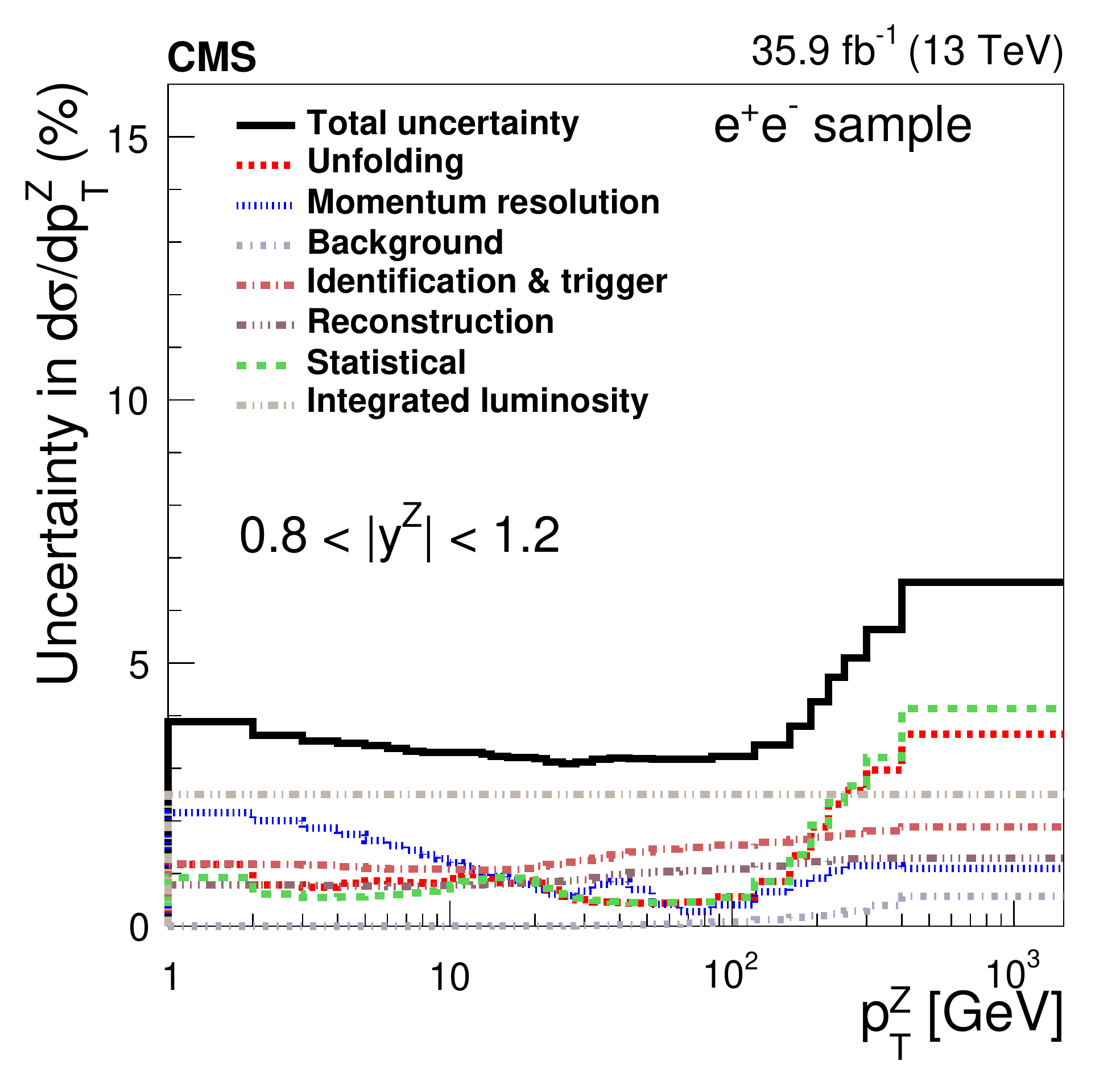}
	\includegraphics[width=0.45\textwidth]{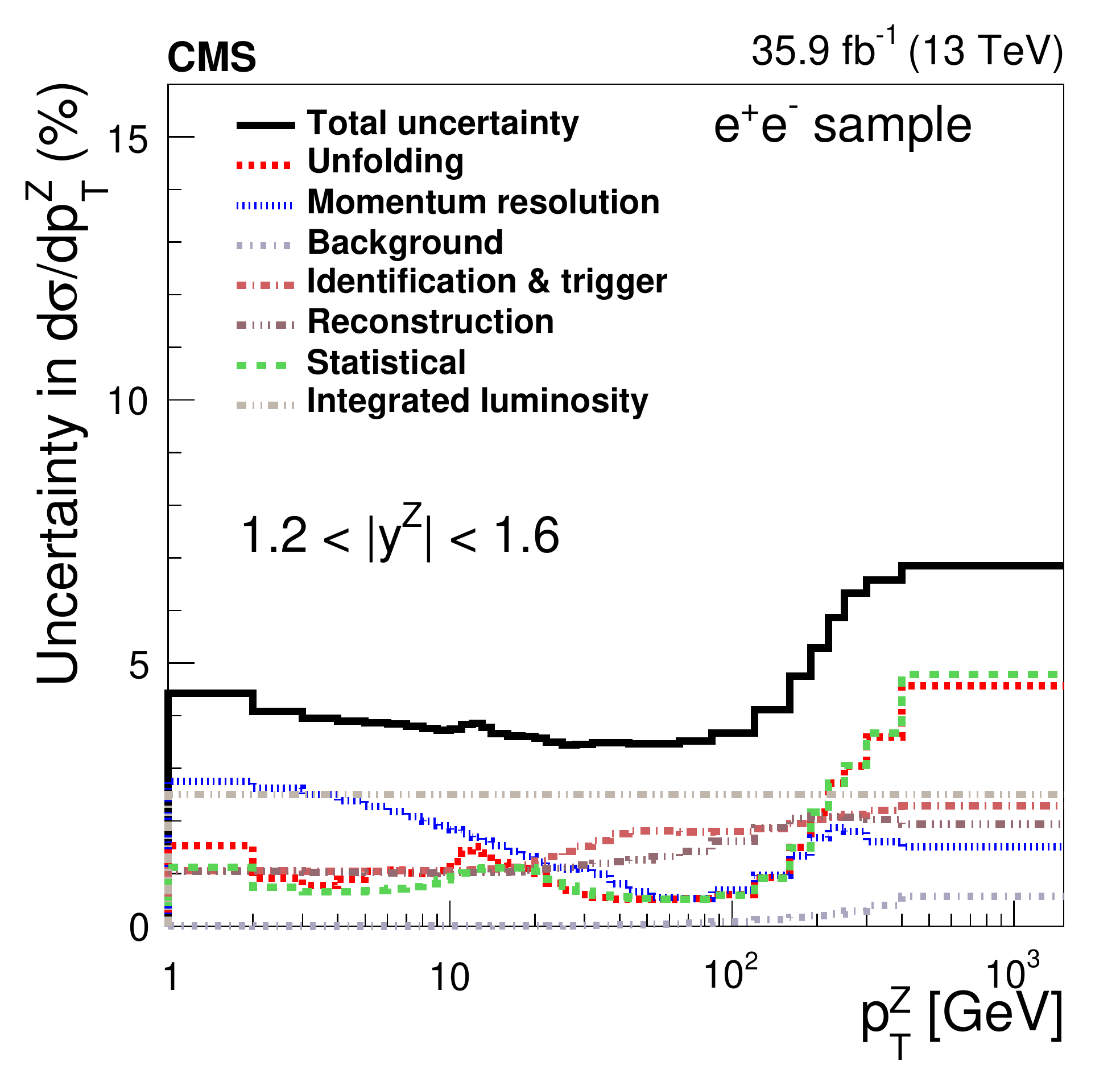}
	\includegraphics[width=0.45\textwidth]{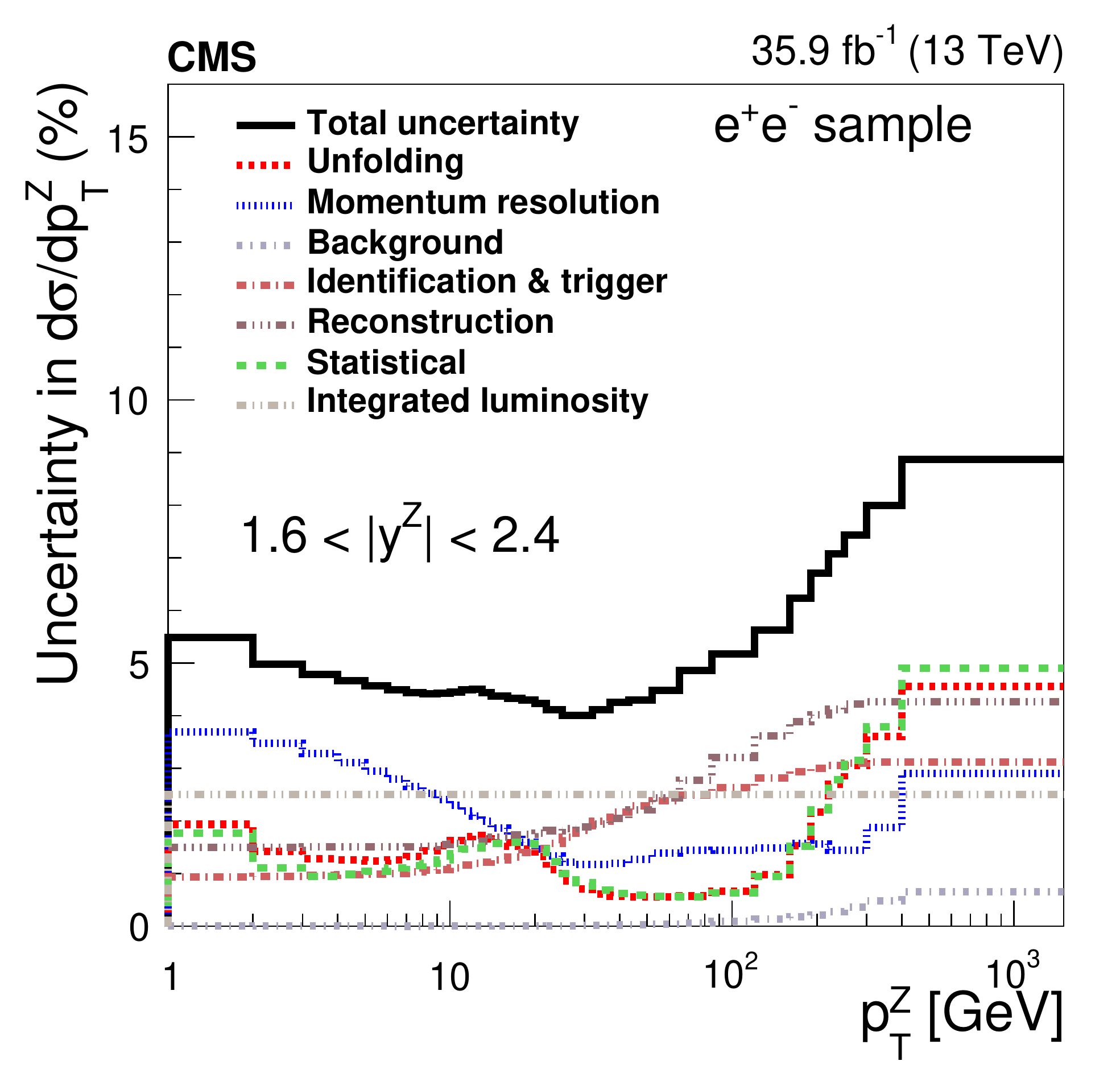}
	\caption{The relative statistical and systematic uncertainties from various sources for the absolute double-differential cross section measurements in bins of $\pt^{\cPZ}$ for the
	$0.0 < \abs{\rapidity^{\cPZ}} < 0.4$ bin (upper left), $0.4 < \abs{\rapidity^{\cPZ}} < 0.8$ bin (upper right),
	$0.8 < \abs{\rapidity^{\cPZ}} < 1.2$ bin (middle left), $1.2 < \abs{\rapidity^{\cPZ}} < 1.6$ bin (middle right), and $1.6 < \abs{\rapidity^{\cPZ}} < 2.4$ bin (lower) in the dielectron final state.
	}
	\label{fig:syst2}
\end{figure}

Summaries of the uncertainties of the absolute double-differential cross
section measurements in $\pt^{\cPZ}$ and $\abs{\rapidity^{\cPZ}}$ are shown in
Figs.~\ref{fig:syst1} and~\ref{fig:syst2}. The statistical uncertainties in
the data and the systematic uncertainties with a statistical component are
large compared to the single-differential cross section measurements. The
statistical uncertainty starts to dominate the total uncertainty
in the high $\pt^{\cPZ}$ regions.

\section{Results}
The inclusive fiducial cross section is measured in the dimuon and dielectron
final states, using the definition described in Section~\ref{sec:fiducial}. The combined cross section is obtained by treating
the systematic uncertainties, except the uncertainties due to the integrated
luminosity and background estimation, as uncorrelated between the two final states.
The integrated luminosity and background estimation uncertainties are treated as
fully correlated in the combined measurement. The combined cross section is obtained by unfolding simultaneously the dimuon and dielectron final states.
The uncertainties are dominated by the uncertainty in the integrated luminosity and the lepton  efficiency.
A summary of the systematic uncertainties is shown in Table~\ref{tab:syst_xs}.
The measured cross sections are shown in Table~\ref{tab:totCross}.

\begin{table}[hbtp]
  \begin{center}
\caption{Summary of the systematic uncertainties for the
inclusive fiducial cross section measurements.\label{tab:syst_xs}}
\begin{tabular}{lcc}
\hline
Source      & $\cPZ \to \mu\mu$ (\%) & $\cPZ \to \Pe\Pe$ (\%) \\
\hline
Luminosity                         & 2.5     & 2.5     \\[\cmsTabSkip]

Muon reconstruction efficiency     & 0.4     &  \NA      \\
Muon selection efficiency          & 0.7     &  \NA      \\
Muon momentum scale                & 0.1     &  \NA      \\
Electron reconstruction efficiency &  \NA      & 0.9     \\
Electron selection efficiency      &  \NA      & 1.0     \\
Electron momentum scale            &  \NA      & 0.2     \\
Background estimation              & 0.1     & 0.1     \\[\cmsTabSkip]

Total (excluding luminosity)       & 0.8     & 1.4     \\
\hline
  \end{tabular}
  \end{center}
\end{table}

\begin{table}[hbtp]
\topcaption{
The measured inclusive fiducial cross sections in the dimuon and dielectron final states.
The combined measurement is also shown. $\mathcal{B}$ is the $\cPZ \to \ell\ell$
branching fraction.
}
\begin{center}
\begin{tabular}{lrrrrrrr}
\hline
Cross section                             & \multicolumn{7}{c}{$\sigma \, \mathcal{B}$ [pb]} \\[\cmsTabSkip]

$\sigma_{\cPZ \to \mu\mu}$         & 694 & $\pm$ &  6 & $\syst$ & $\pm$ & 17 & $\lum$ \\
$\sigma_{\cPZ \to \Pe\Pe}$         & 712 & $\pm$ & 10 & $\syst$ & $\pm$ & 18 & $\lum$ \\
$\sigma_{\cPZ \to \ell\ell}$       & 699 & $\pm$ &  5 & $\syst$ & $\pm$ & 17 & $\lum$ \\
\hline
\end{tabular}
\label{tab:totCross}
\end{center}
\end{table}

The measured cross section values agree with the theoretical predictions within uncertainties. The
predicted values are $\sigma_{\cPZ \to \ell\ell} = 682 \pm 55\unit{pb}$ with \MGvATNLO using the NNPDF 3.0~\cite{Ball:2014uwa} NLO PDF set,
and $\sigma_{\cPZ \to \ell\ell} = 719 \pm 8\unit{pb}$ with fixed order
$\FEWZ$~\cite{FEWZ, Gavin:2010az, Gavin:2012sy, Li:2012wna} at NNLO accuracy in
QCD using the NNPDF 3.1~\cite{Ball:2017nwa} NNLO PDF set. The theoretical
uncertainties for \MGvATNLO and $\FEWZ$ include statistical,
PDF, and scale uncertainties. The scale uncertainties are estimated by varying
$\mu_{\mathrm{R}}$ and $\mu_{\mathrm{F}}$ independently
up and down by a factor of two from their nominal values (excluding the two extreme variations)
and taking the largest cross section variations as the uncertainty.

\begin{figure}
	\centering
        \includegraphics[width=0.45\textwidth]{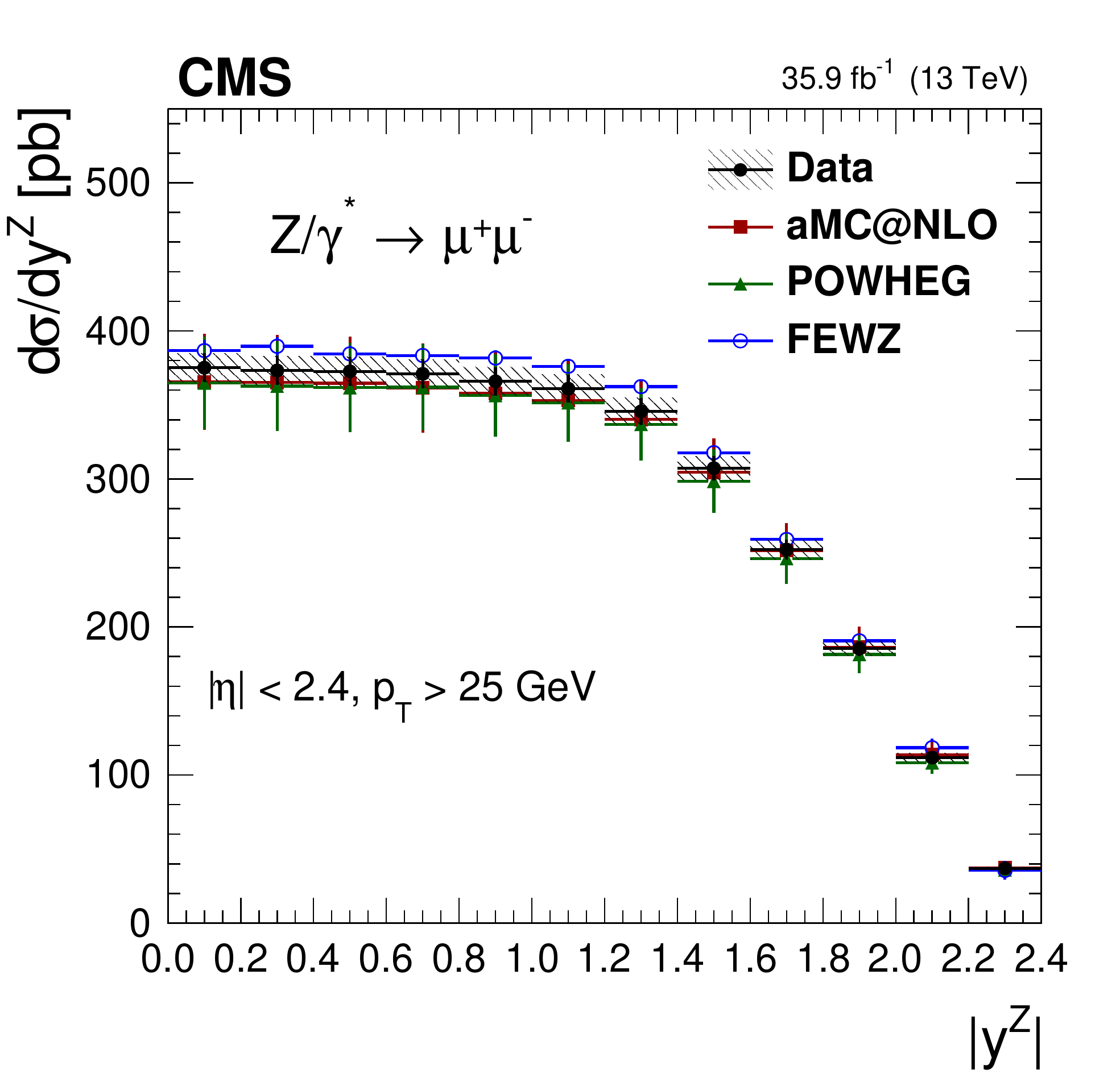}
	\includegraphics[width=0.45\textwidth]{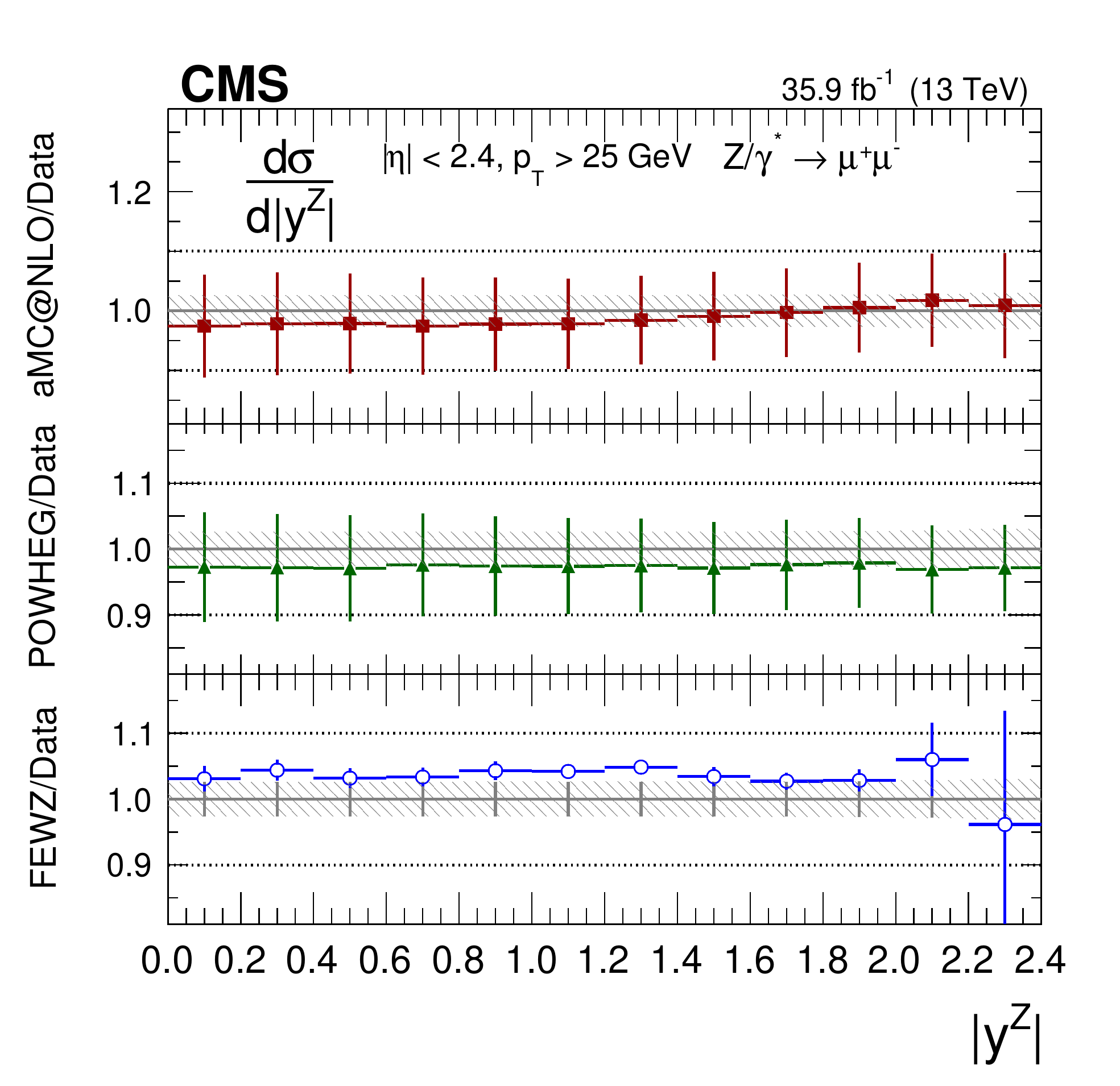}
	\includegraphics[width=0.45\textwidth]{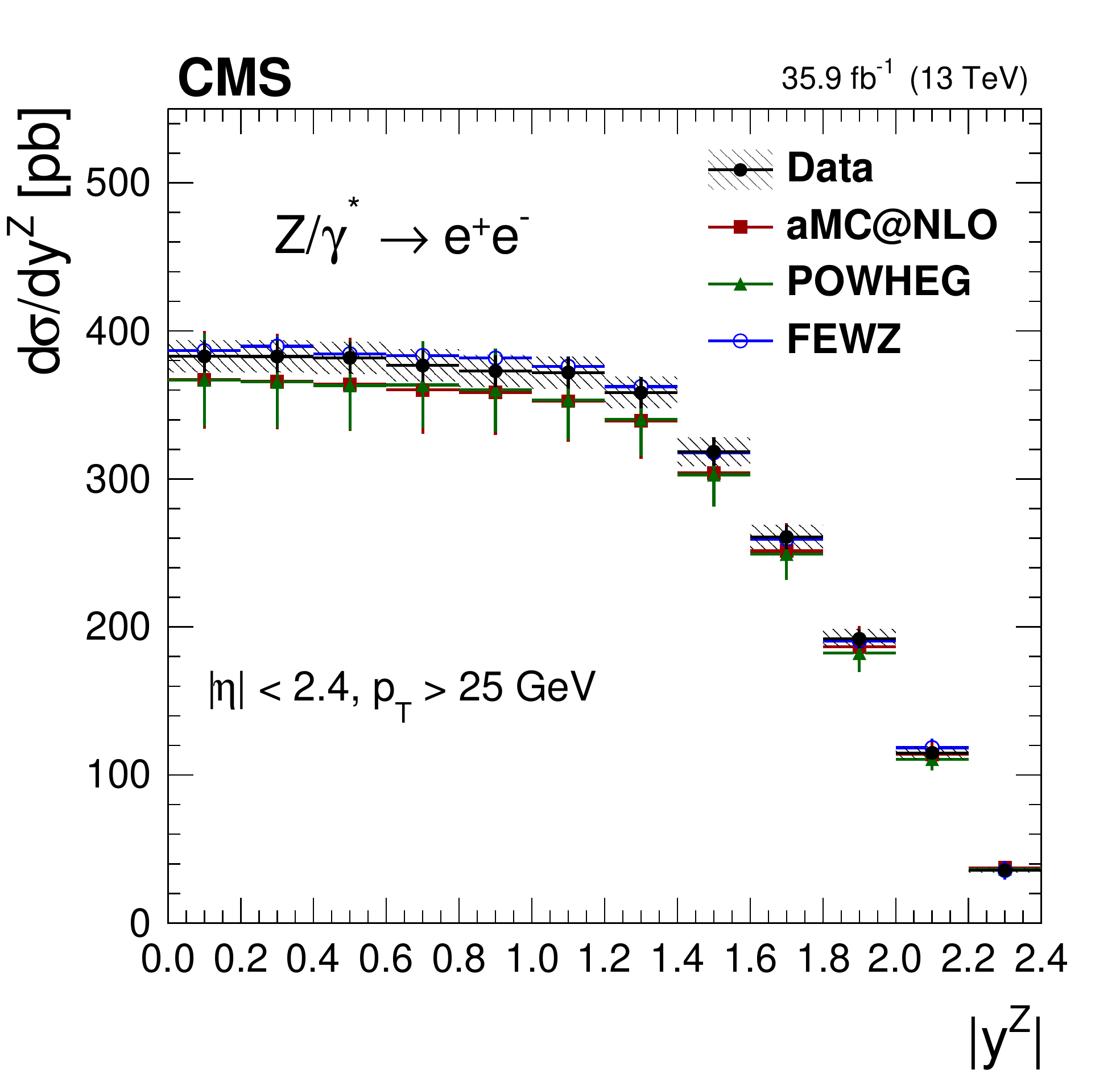}
	\includegraphics[width=0.45\textwidth]{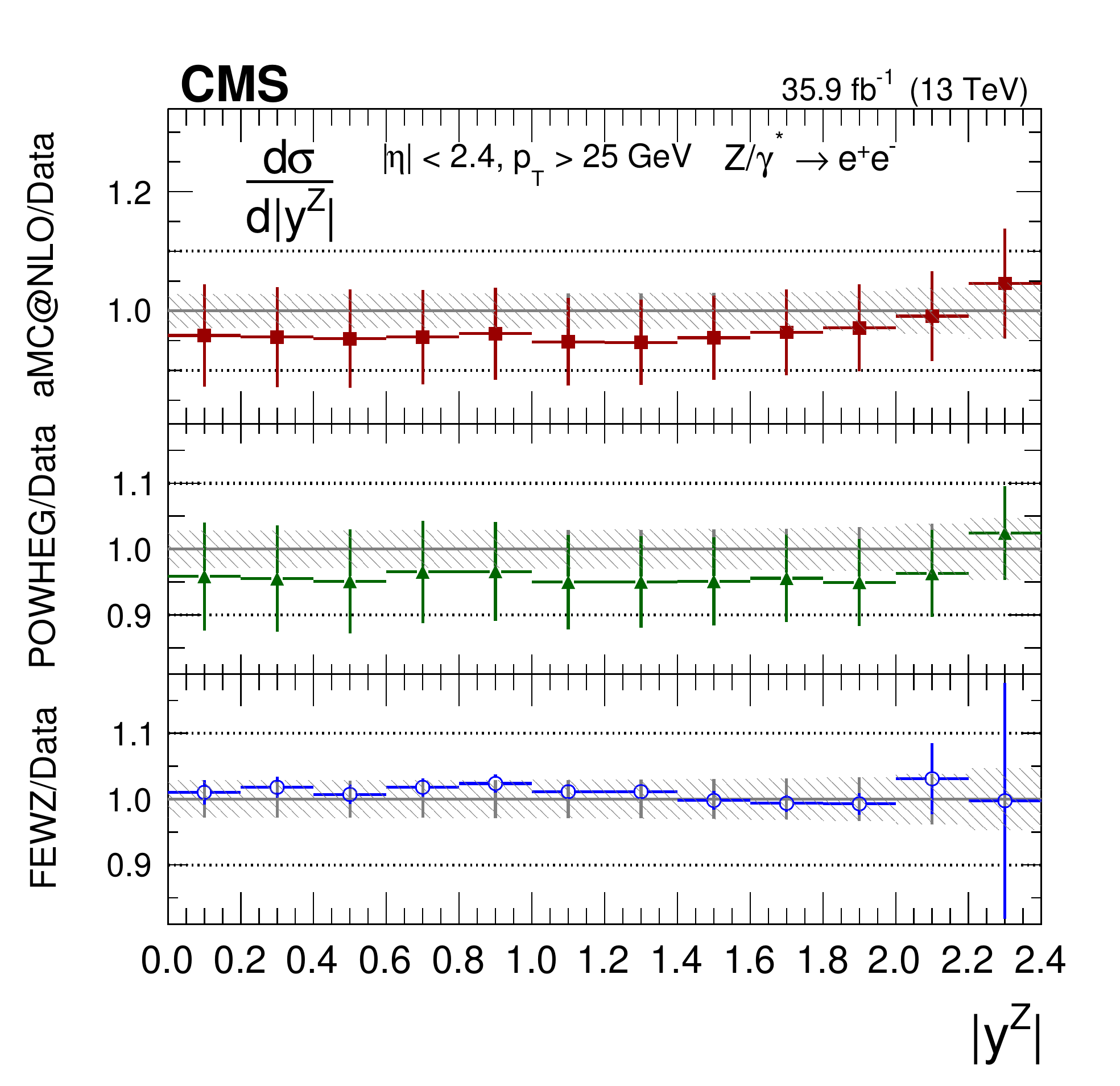}
	\includegraphics[width=0.45\textwidth]{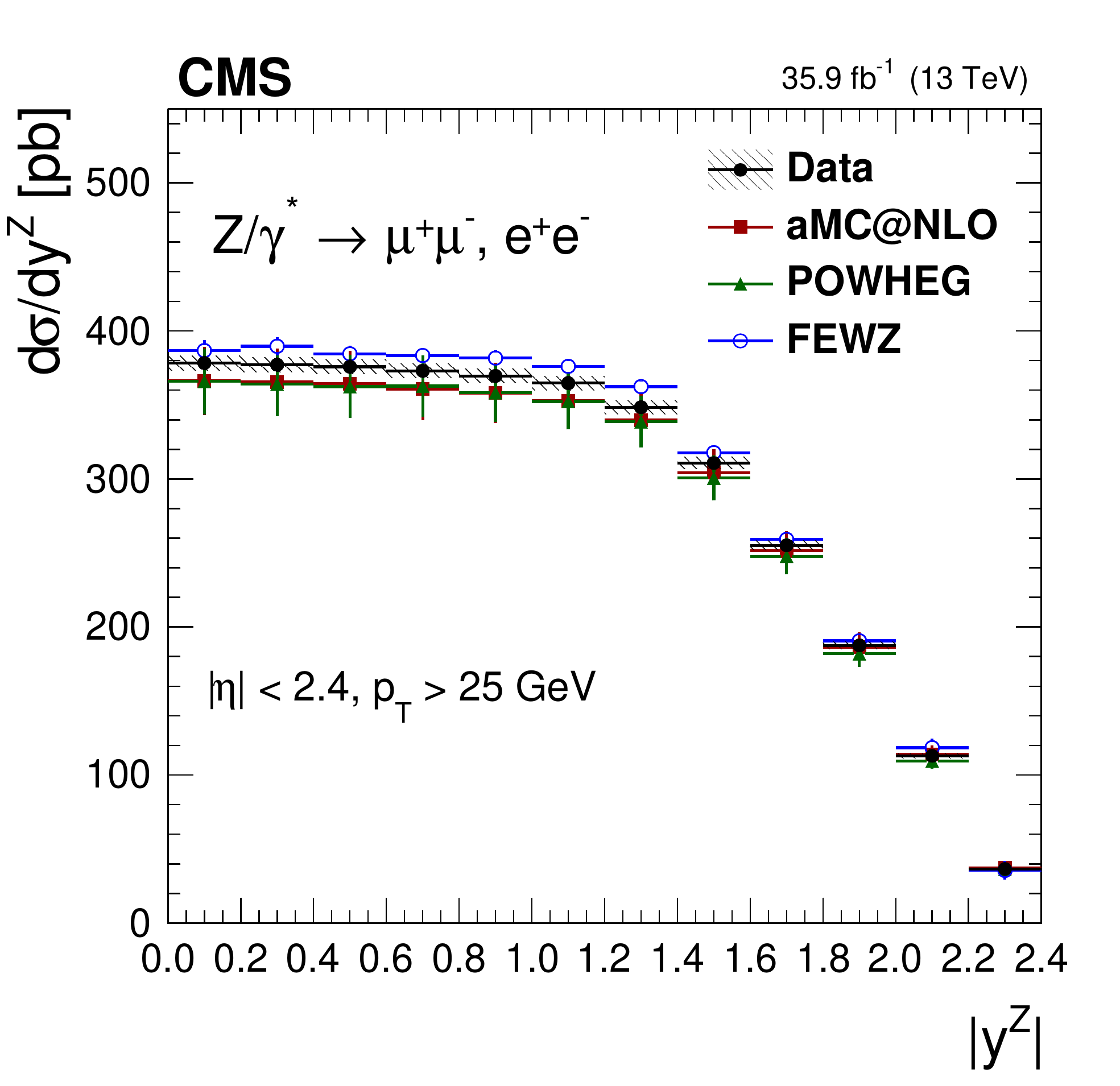}
	\includegraphics[width=0.45\textwidth]{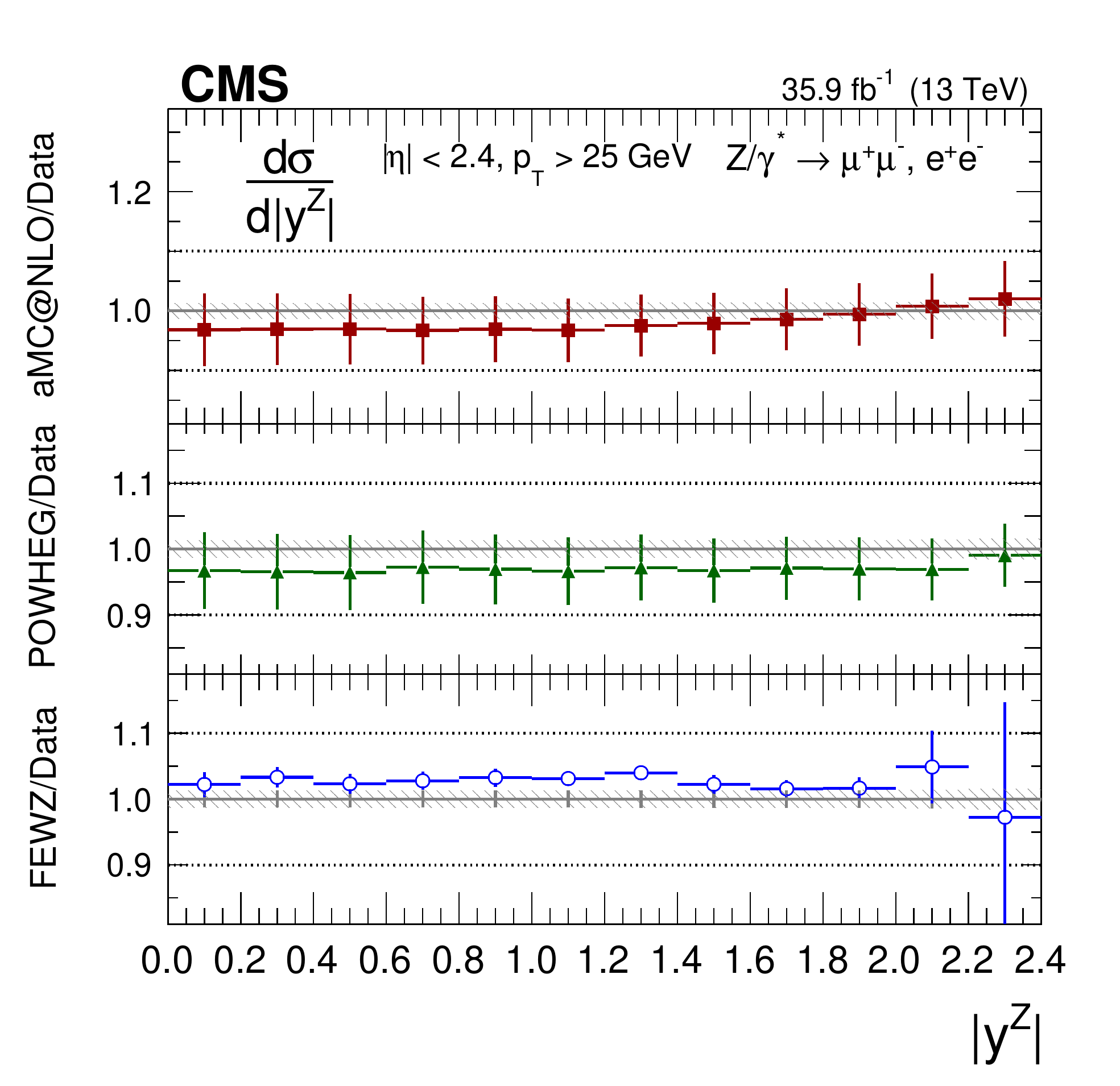}
	\caption{The measured absolute cross sections (left) in bins of $\abs{\rapidity^{\cPZ}}$ for
	the dimuon (upper) and dielectron (middle) final states, and for the combination (lower).
	The ratios of the predictions to the data are also shown (right). The shaded bands around the
	data points (black) correspond to the total experimental uncertainty. The measurement is
	compared to the predictions with \MGvATNLO (square red markers),
	$\POWHEG$ (green triangles), and FEWZ (blue circles). The error bars around
	the predictions correspond to the combined statistical, PDF, and scale uncertainties.}
	\label{fig:unf_rap}
\end{figure}

{\tolerance=8000 The measured differential cross sections corrected for detector effects are
compared to various theoretical predictions. The measured absolute
cross sections in bins of $\abs{\rapidity^{\cPZ}}$ are shown in
Fig.~\ref{fig:unf_rap} for dimuon and dielectron final states, and their combination.
The measurement is compared to the predictions using
parton shower modeling with both \MGvATNLO and $\POWHEG$ at
NLO accuracy in QCD using the NNPDF 3.0 PDF set. The \MGvATNLO prediction includes up to two additional partons at Born level in the matrix element
calculations, merged with the parton shower description using the $\FxFx$ scheme~\cite{Frederix:2012ps}.s
A comparison with a fixed order prediction at NNLO accuracy with $\FEWZ$ using the NNPDF 3.1 NNLO PDF set is also
shown. The \MGvATNLO and $\POWHEG$ predictions are
consistent with the data within the theoretical uncertainties. The $\FEWZ$
prediction with the NNPDF 3.1 PDF set is within 5\% of the measurement over
the entire $\abs{\rapidity^{\cPZ}}$ range, which is roughly within the uncertainties.\par}

\begin{figure}
	\centering
	\includegraphics[width=0.45\textwidth]{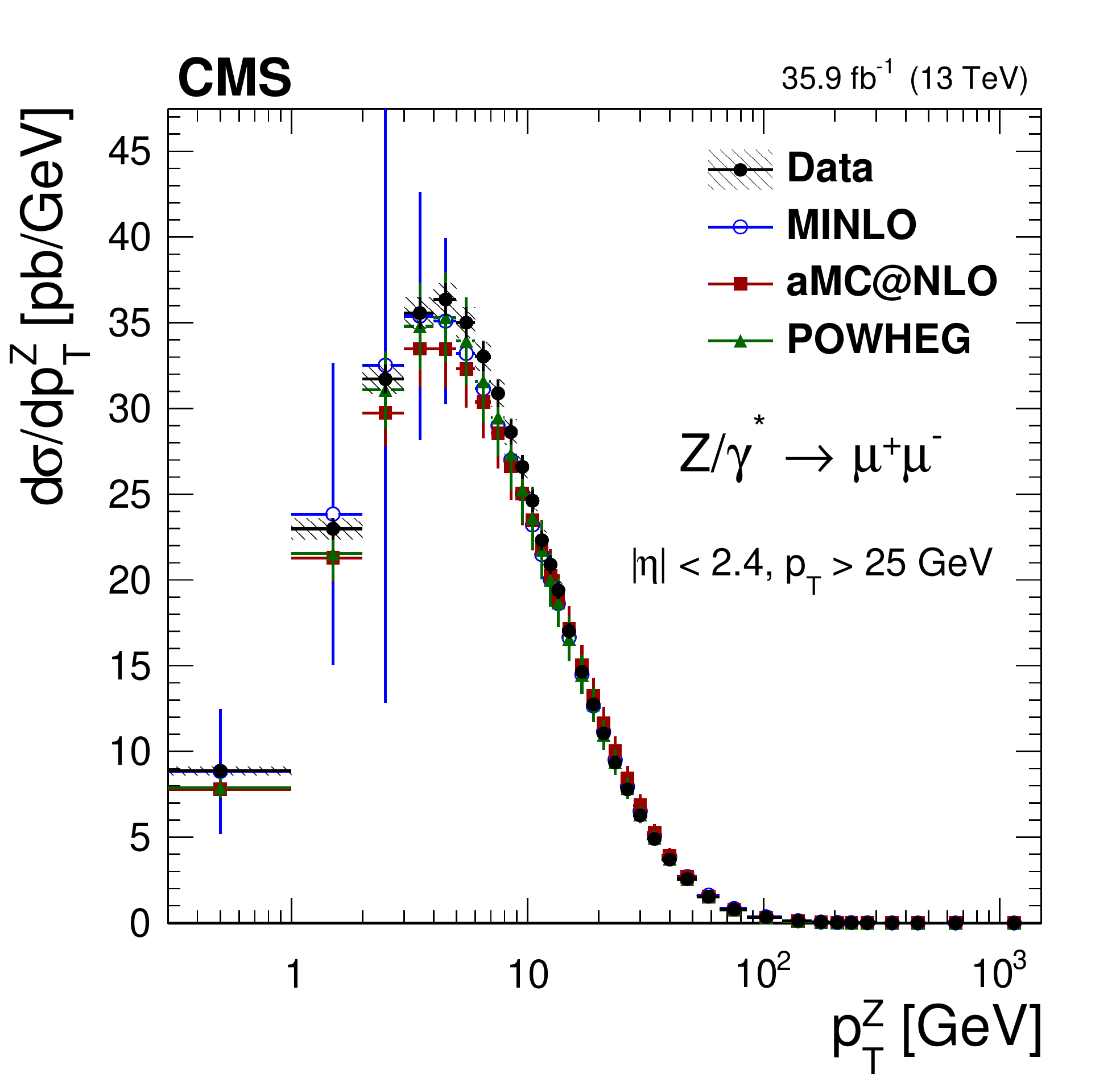}
	\includegraphics[width=0.45\textwidth]{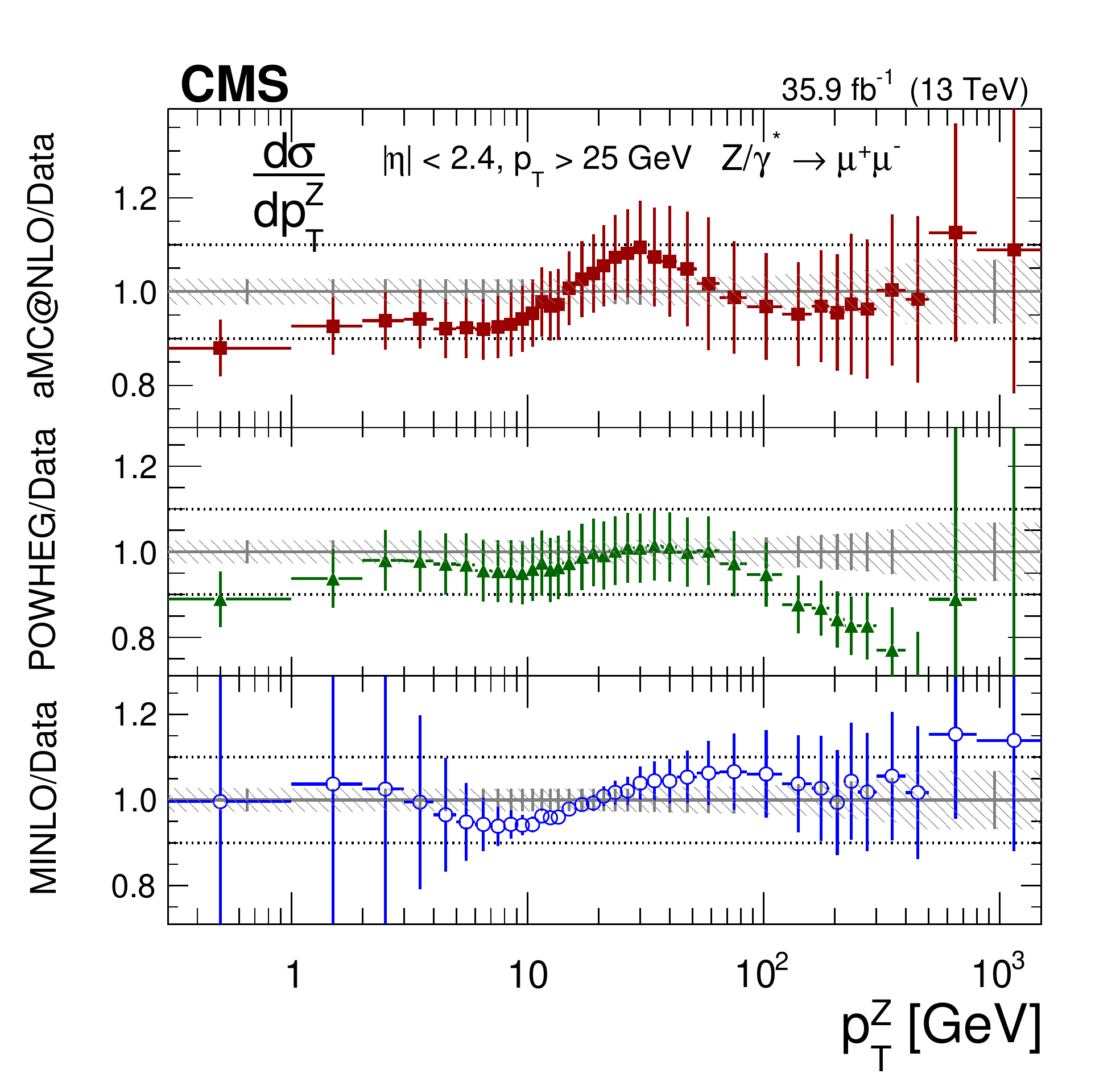}
	\includegraphics[width=0.45\textwidth]{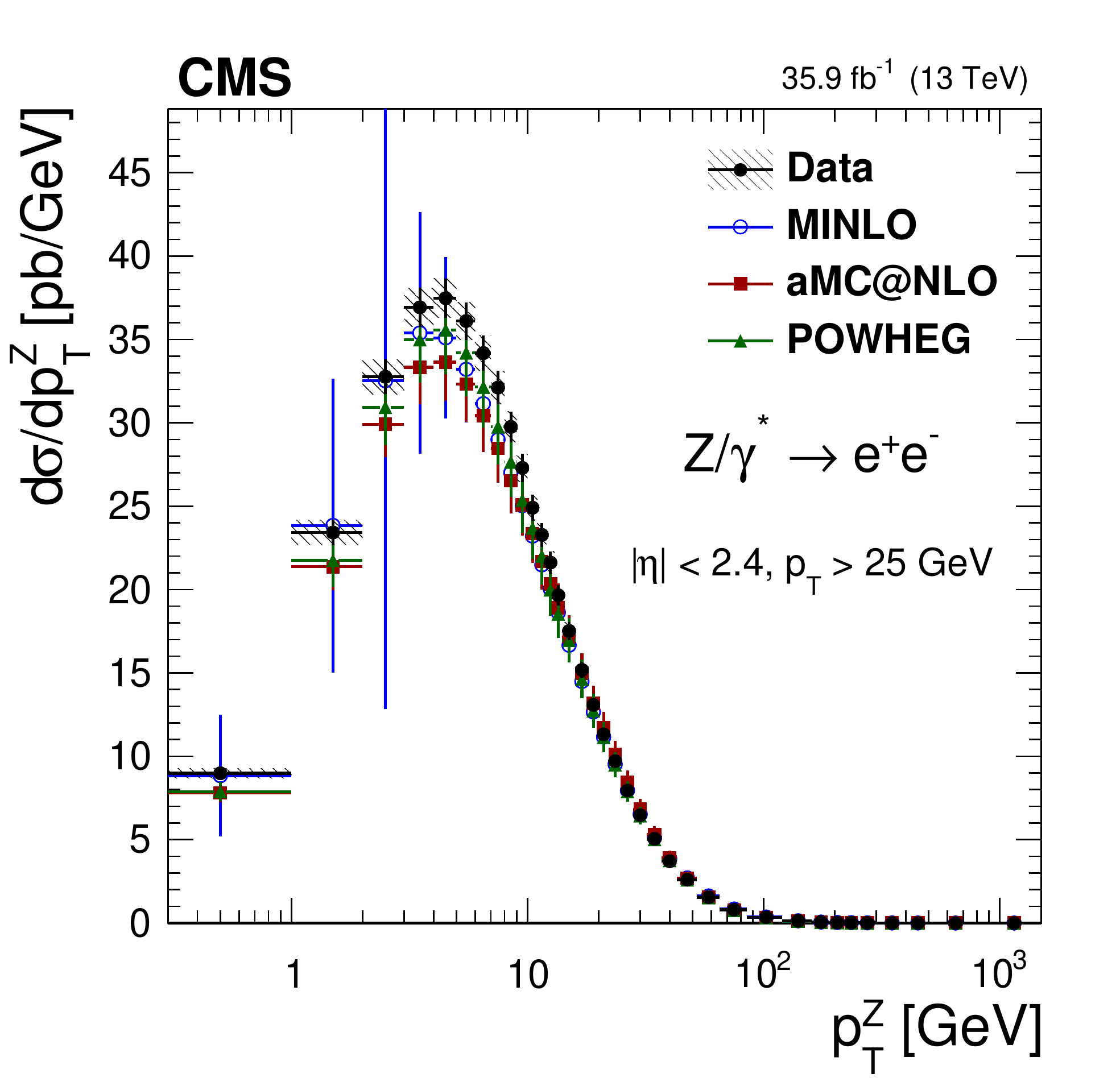}
	\includegraphics[width=0.45\textwidth]{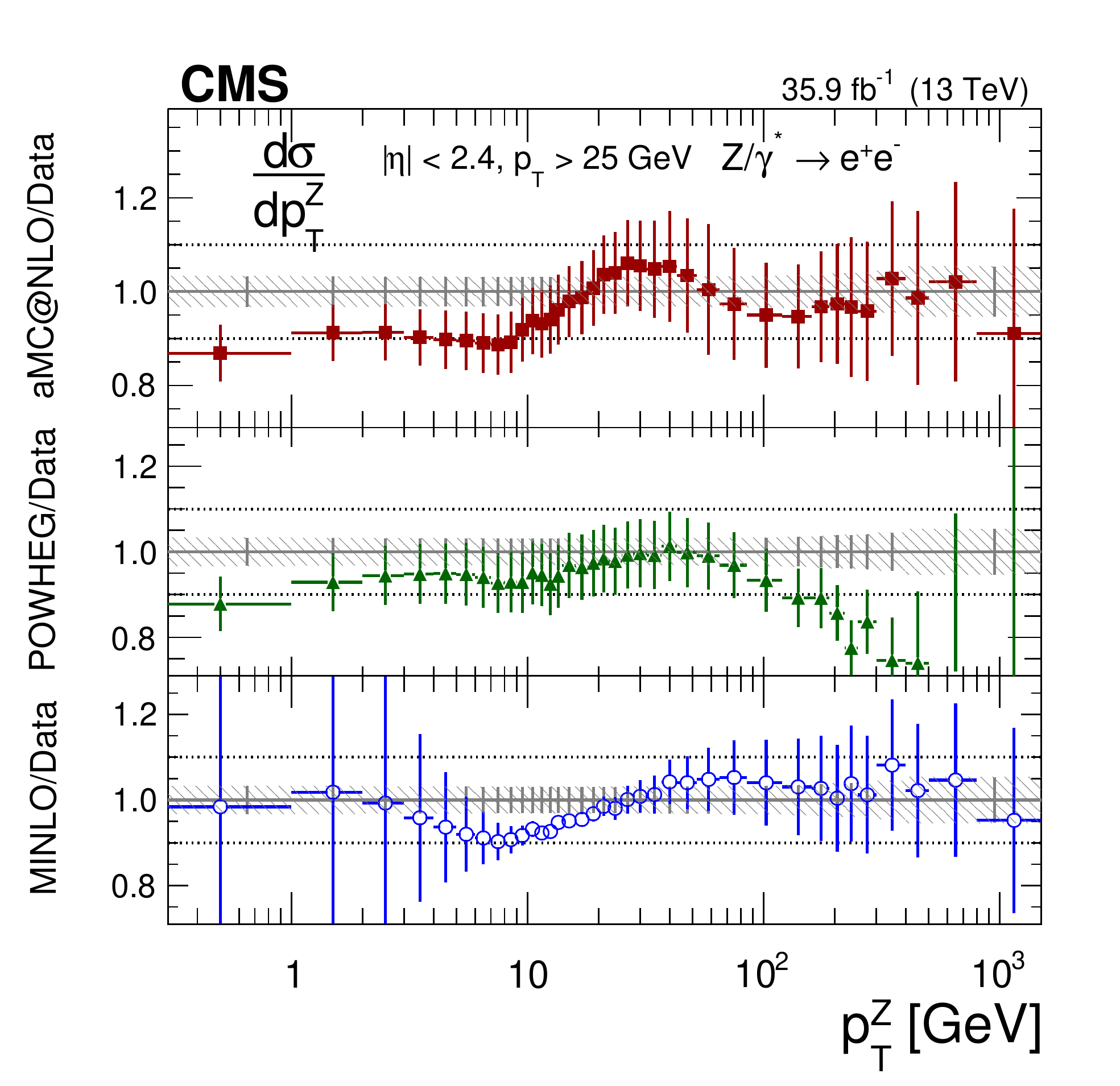}
	\includegraphics[width=0.45\textwidth]{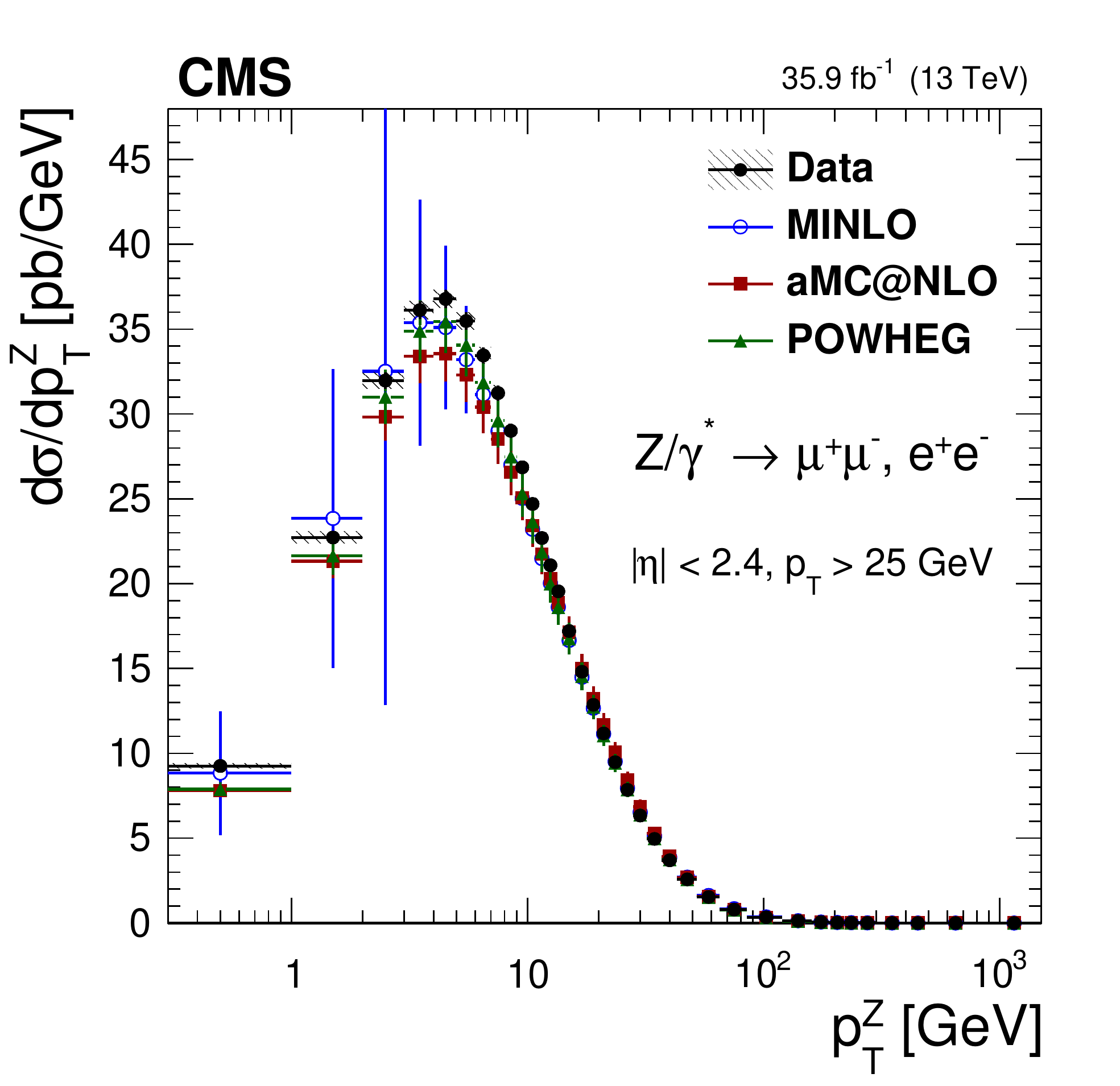}
	\includegraphics[width=0.45\textwidth]{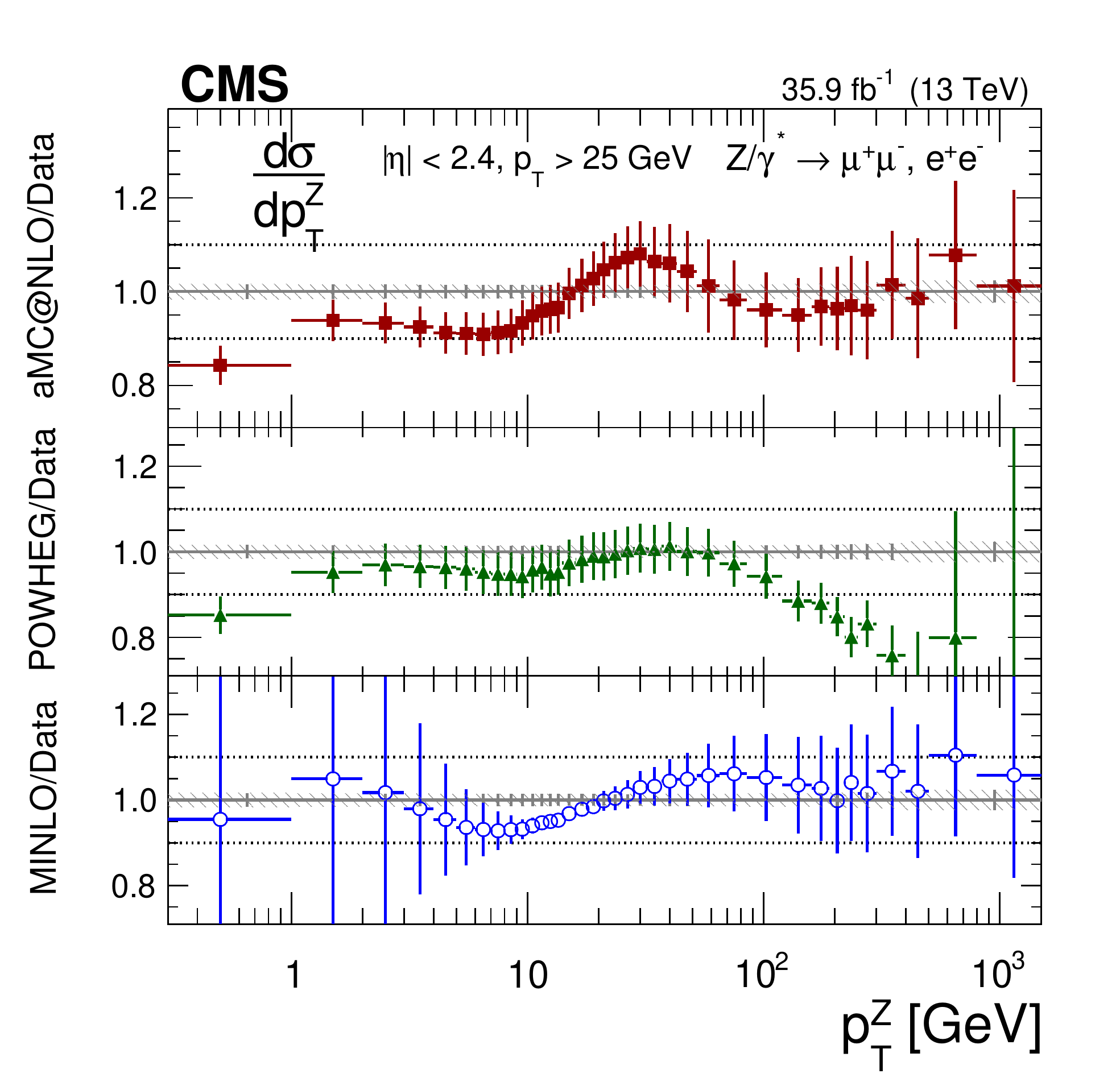}
	\caption{The measured absolute cross sections (left) in bins of $\pt^{\cPZ}$ for
	the dimuon (upper) and dielectron (middle) final states, and for the combination (lower).
	The ratios of the predictions to the data are also shown (right). The shaded bands around
	the data points (black) correspond to the total experimental uncertainty. The measurement
	is compared to the predictions with \MGvATNLO (square red markers), $\POWHEG$
	(green triangles), and $\POWHEG$-\textsc{MINLO} (blue circles). The error bars around the
	predictions correspond to the combined statistical, PDF, and scale uncertainties.}
	\label{fig:unf_pt_shower}
\end{figure}

Figure~\ref{fig:unf_pt_shower} shows the measured absolute cross sections in
bins of $ \pt^{\cPZ}$ for dimuon and dielectron final states, and their combination. The measurement is compared to the predictions using parton shower modeling with both \MGvATNLO and $\POWHEG$. A comparison
with $\POWHEG$ using the \textsc{MINLO} procedure~\cite{Hamilton:2012rf} and using the
NNPDF 3.1 NLO PDF set is also shown. The predictions are consistent with the
measurements within the theoretical uncertainties. The scale uncertainties for the $\POWHEG$-\textsc{MINLO}
predictions are evaluated by simultaneously varying $\mu_{\mathrm{R}}$ and $\mu_{\mathrm{F}}$
up and down by a factor of two~\cite{Hamilton:2012rf}. The $\POWHEG$ predictions at
high \pt, above  100\GeV, disagree with data. The better
accuracy of the \MGvATNLO and $\POWHEG$-\textsc{MINLO}
predictions at high \pt lead to an improved agreement with data.

\begin{figure}
	\centering
	\includegraphics[width=0.45\textwidth]{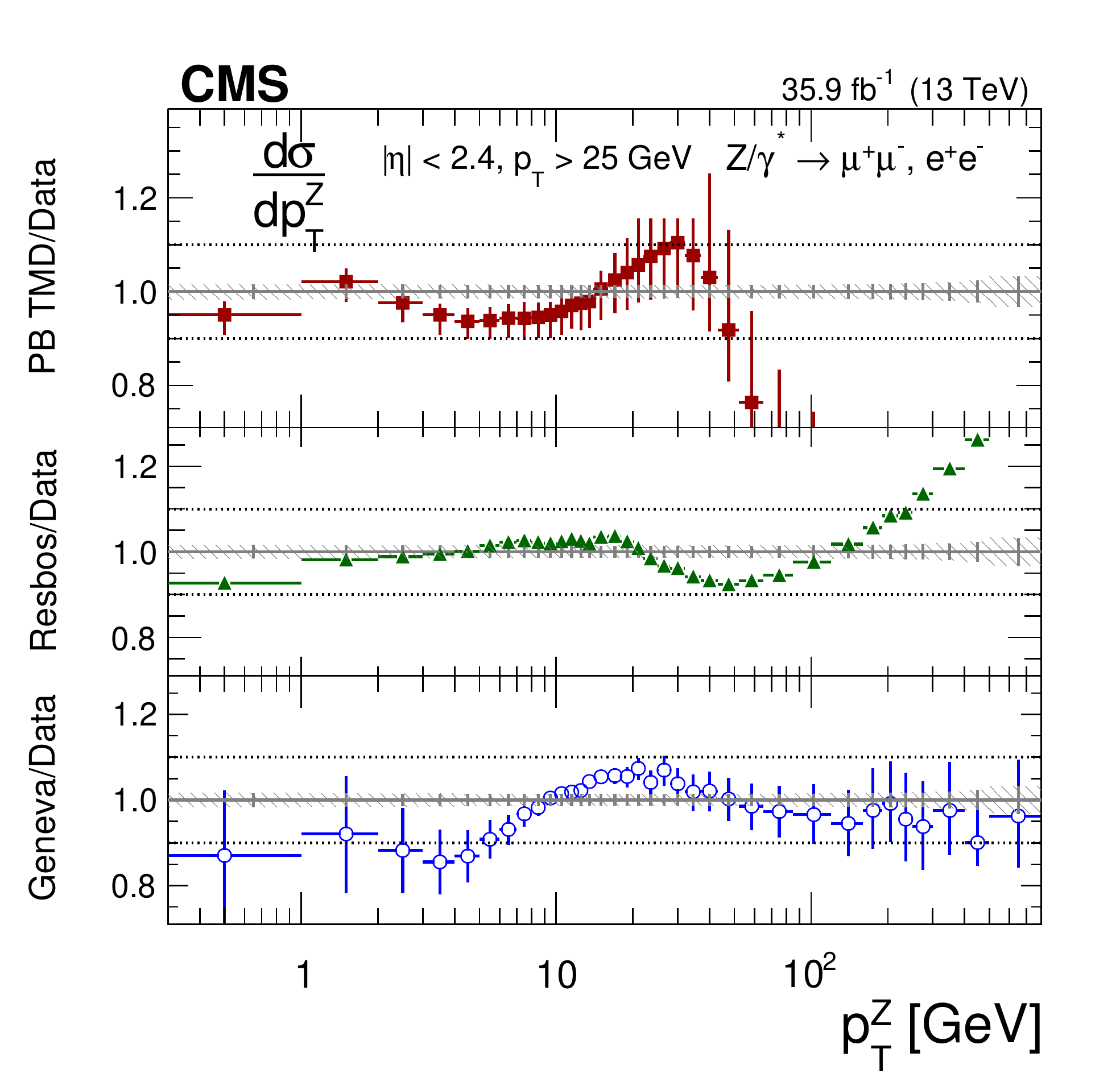}
	\includegraphics[width=0.45\textwidth]{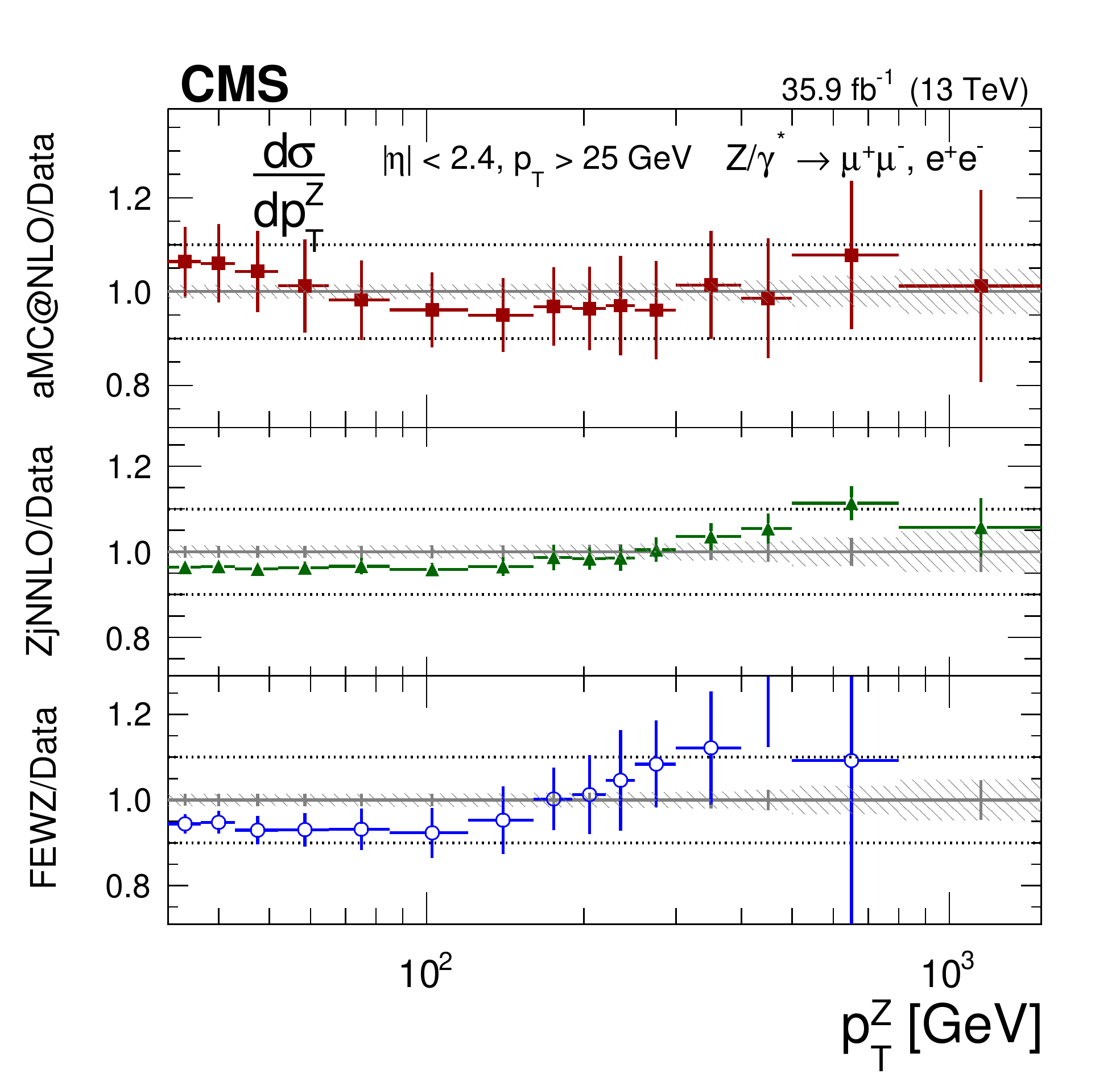}
	\caption{The ratios of the predictions to the data in bins of $\pt^{\cPZ}$
	for the combination of the dimuon and dielectron final states. The shaded
	bands around the data points (black) correspond to the total experimental
	uncertainty. The left plot shows comparisons to the predictions with PB TMD (square red markers),
	\textsc{RESBOS} (green triangles), and \textsc{GENEVA} (blue circles). The right plot shows the
	$\pt^{\cPZ}$ distribution for $\pt > 32\GeV$ compared to the predictions with \MGvATNLO (square red markers),  $\cPZ$ + 1 jet at NNLO (green triangles),
	and FEWZ (blue circles). The error bars around the predictions correspond to the combined statistical,
	PDF, and scale uncertainties. Only the statistical uncertainties are shown for the predictions with \textsc{RESBOS}.}
	\label{fig:unf_pt_fixed}
\end{figure}

Figure~\ref{fig:unf_pt_fixed} (left) shows comparisons to the resummed calculations
with both \textsc{RESBOS}~\cite{Ladinsky:1993zn, Balazs:1997xd, Landry:2002ix}
and \textsc{GENEVA}~\cite{Alioli:2015toa}. A comparison to the predictions with TMD PDFs
obtained~\cite{Martinez:2018jxt} from the parton branching method (PB TMD)~\cite{Hautmann:2017fcj,Hautmann:2017xtx}
and combined with \MGvATNLO at NLO is also
shown~\cite{Martinez:2019mwt}. The \textsc{RESBOS} predictions are obtained
at NNLL accuracy with the CT14 NNLO PDF set and are consistent with the data within
the uncertainties at low \pt but disagree with the measurements
at high \pt. The \textsc{GENEVA} predictions include resummation to NNLL accuracy where the resulting parton-level events are further combined with parton
showering and hadronization provided by $\PYTHIA$. The \textsc{GENEVA}
predictions with the NNPDF 3.1 PDF set and $\alpS(m_{\cPZ})=0.114$ are generally consistent with data within the theoretical uncertainties, but disagree with data at \pt below  30\GeV.
The PB TMD predictions include resummation to NLL accuracy and fixed-order results at NLO,
and take into account nonperturbative contributions from TMD parton distributions through
fits~\cite{Martinez:2018jxt} to precision deep inelastic scattering data. The theoretical uncertainties come from
variation of scales and from TMD uncertainties. The PB TMD prediction describes
data well at low \pt, but deviates from the measurements at high \pt because
of missing contributions from $\cPZ$+jets matrix element calculations.

The $\pt^{\cPZ}$ distribution for $\pt > 32\GeV$ is compared to fixed order
predictions, as shown in Fig.~\ref{fig:unf_pt_fixed} (right). A comparison to
the \MGvATNLO prediction is included as a reference. The
data is compared to the $\FEWZ$ predictions at NNLO in QCD and to the complete
NNLO predictions of vector boson production in association with a
jet~\cite{Boughezal:2015ded,Boughezal:2015dva}. The comparison
is performed for $\pt > 32\GeV$ because the $\cPZ$ + 1 jet at NNLO prediction
does not exist below that value.

The central values of the $\mu_{\mathrm{F}}$ and $\mu_{\mathrm{R}}$ are chosen to be
$\mu_{\mathrm{F/R}}=\sqrt{\smash[b]{(\pt^{\cPZ})^2+\mll^{2}}}$ for the $\FEWZ$ and $\cPZ$+1 jet at NNLO
predictions. The scale uncertainties are estimated by simultaneously varying the $\mu_{\mathrm{F}}$ and $\mu_{\mathrm{R}}$ up and down together by a factor
of two. The CT14~\cite{Dulat:2015mca} NNLO PDF set is used for the $\cPZ$+1 jet
at NNLO predictions. The predictions are consistent with the measurements within the
theoretical uncertainties. As can be seen, the $\cPZ$+1 jet at NNLO calculations
significantly reduce the scale uncertainties.
The electroweak corrections are important at high \pt with expected correction factors of up to
$~0.9$ at $\pt = 500\GeV$ and $~0.8$ at
$\pt = 1000\GeV$~\cite{Dittmaier:2014qza,Lindert:2017olm}.
They are not included in the predictions shown in
Fig.~\ref{fig:unf_pt_fixed}.

\begin{figure}
	\centering
	\includegraphics[width=0.45\textwidth]{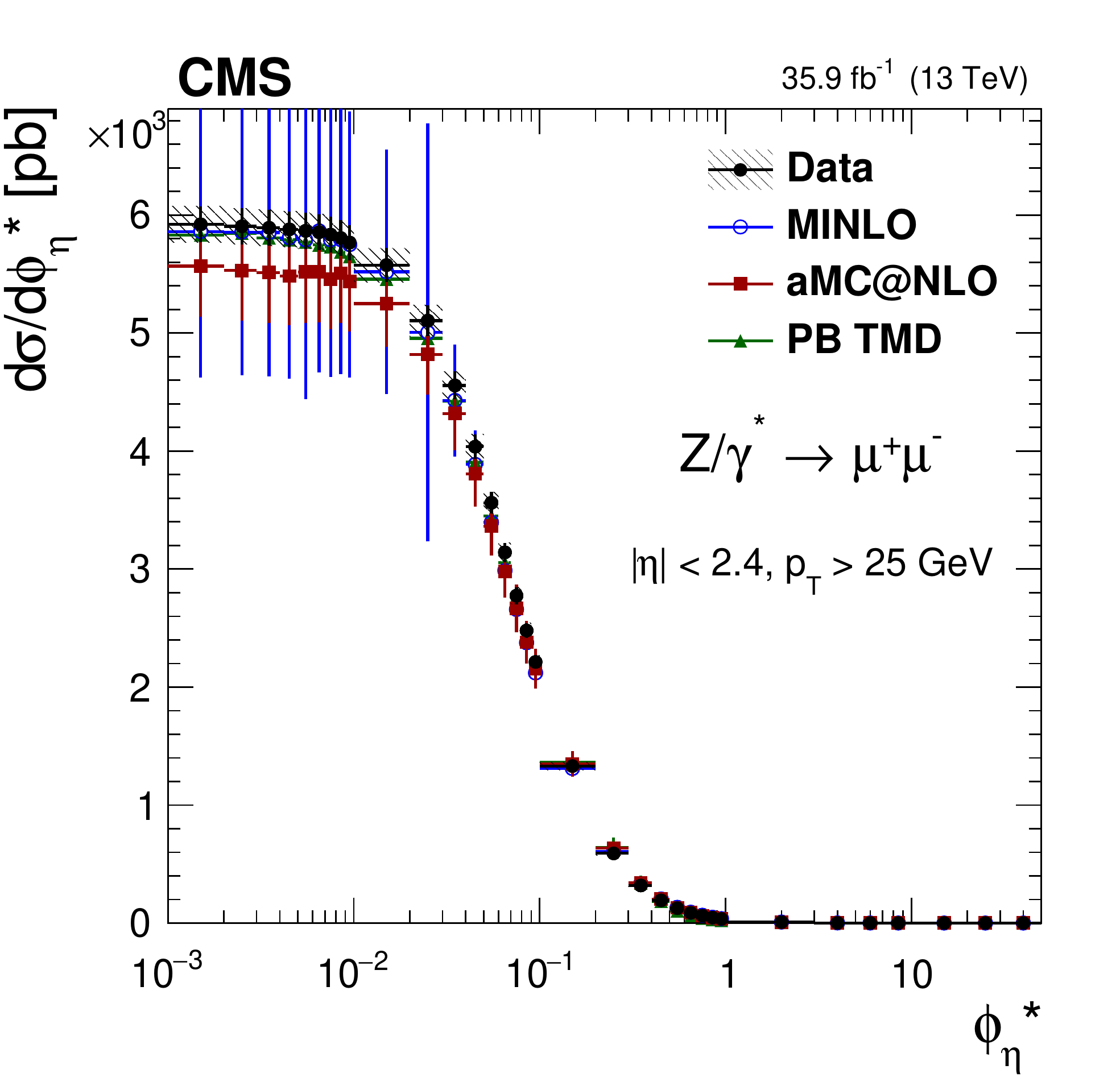}
	\includegraphics[width=0.45\textwidth]{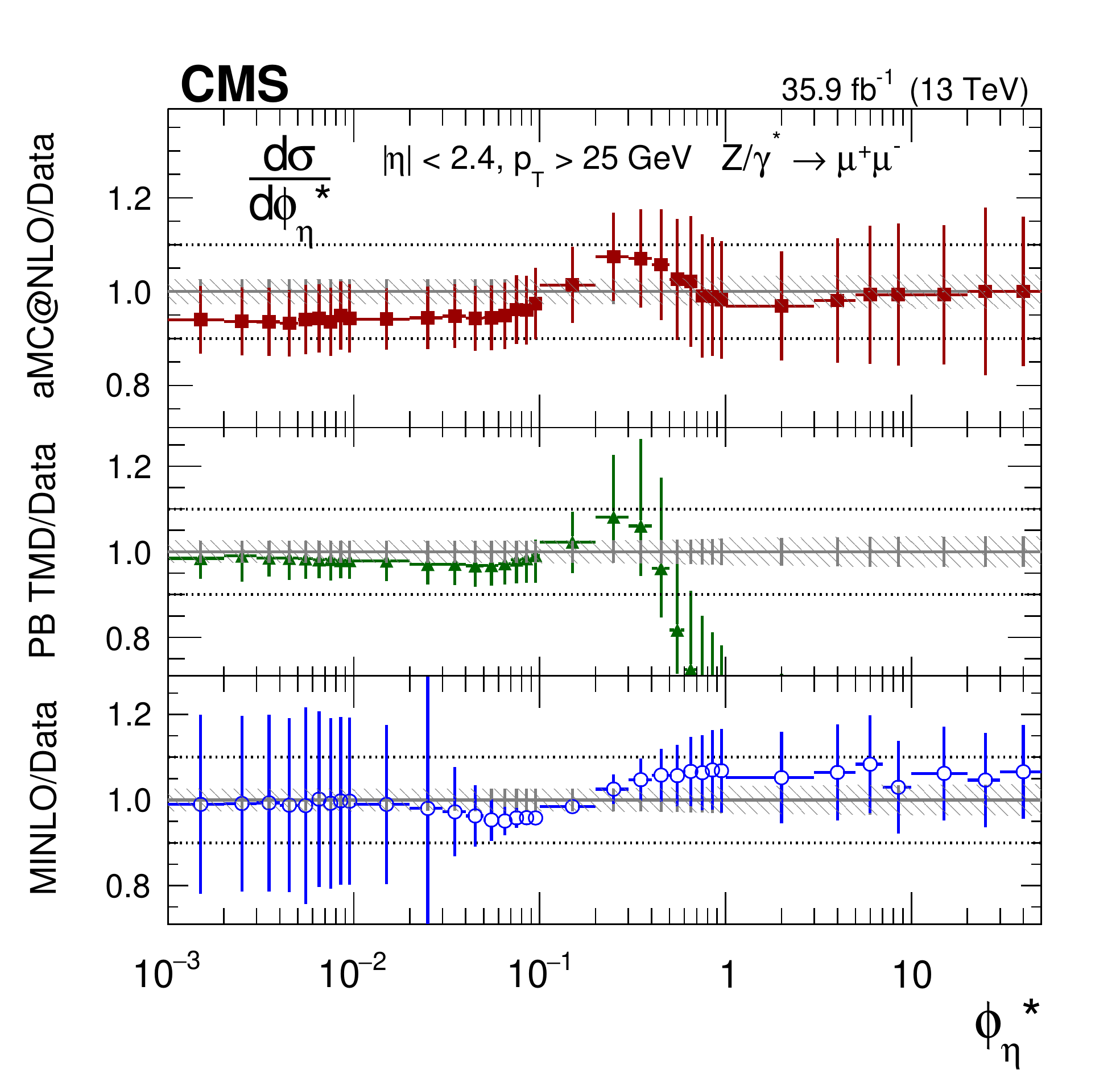}
	\includegraphics[width=0.45\textwidth]{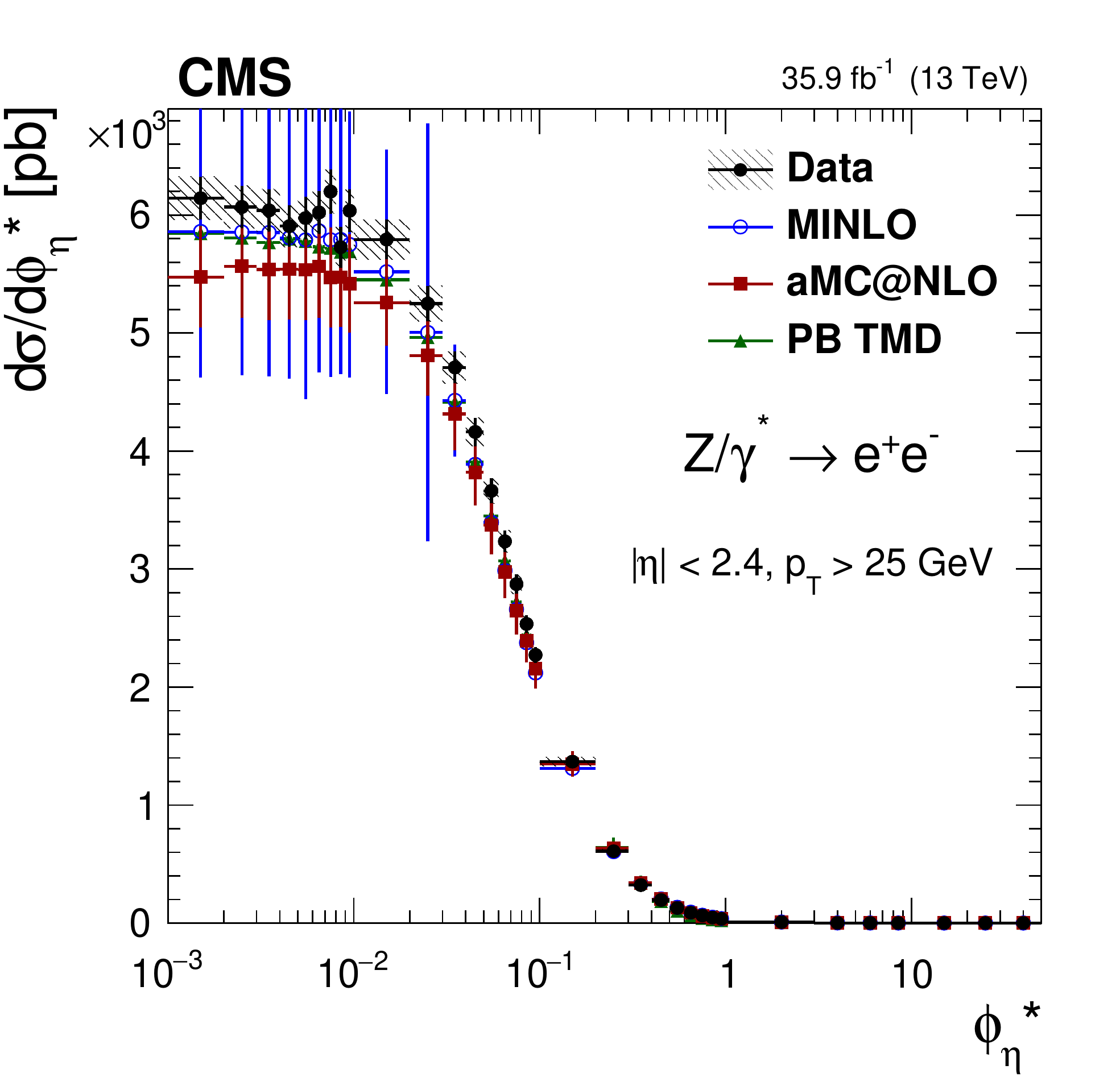}
	\includegraphics[width=0.45\textwidth]{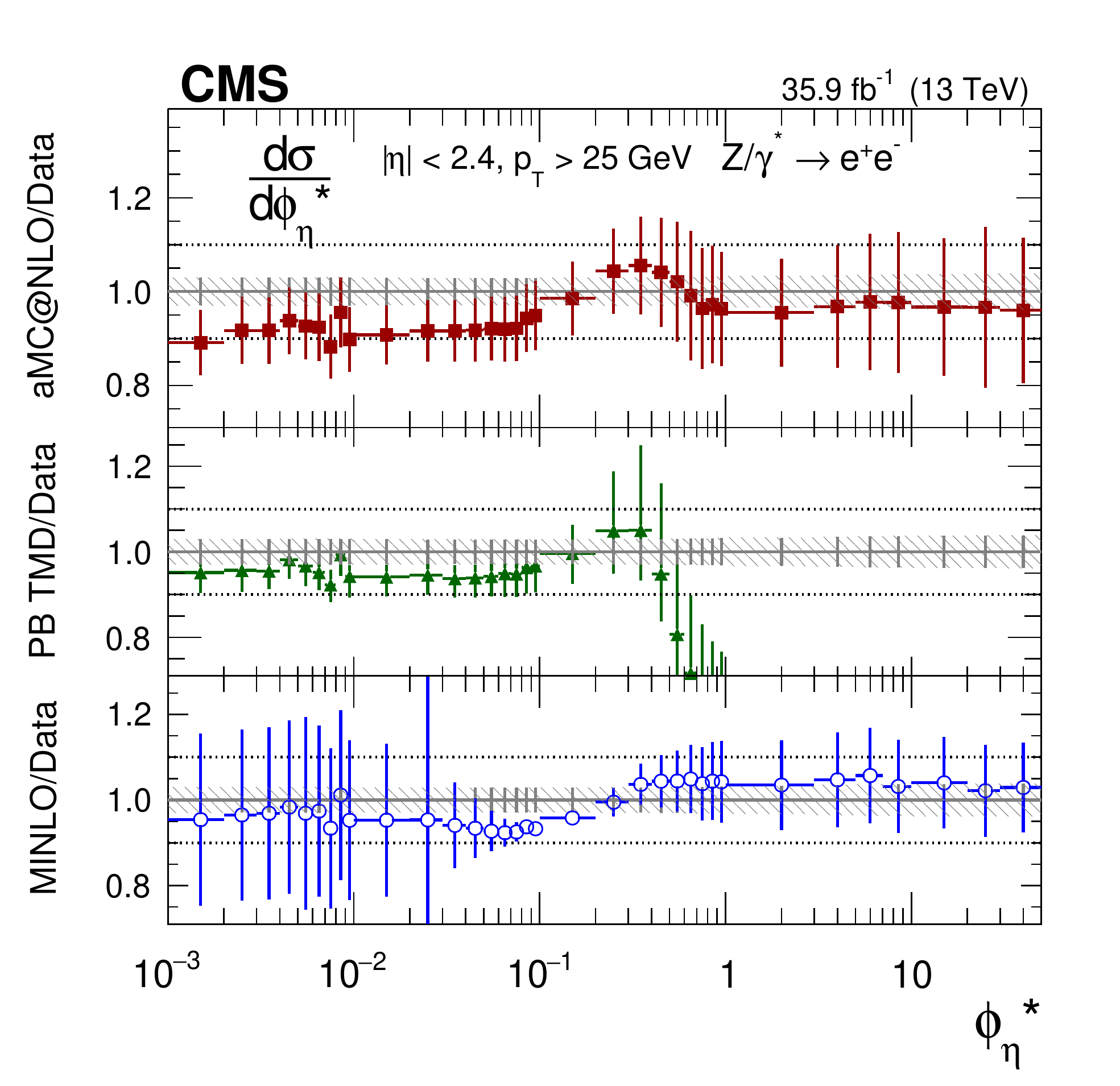}
	\includegraphics[width=0.45\textwidth]{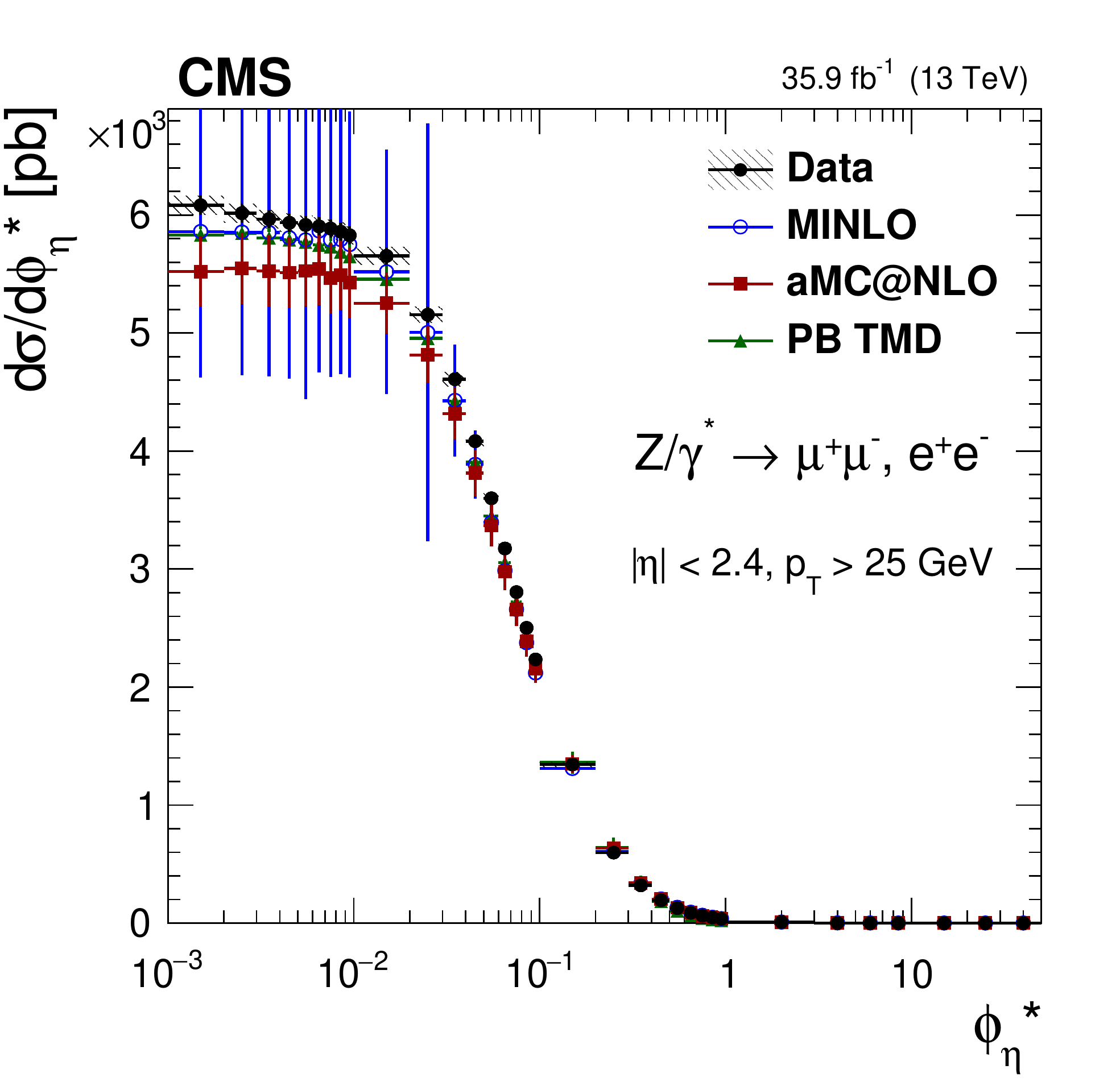}
	\includegraphics[width=0.45\textwidth]{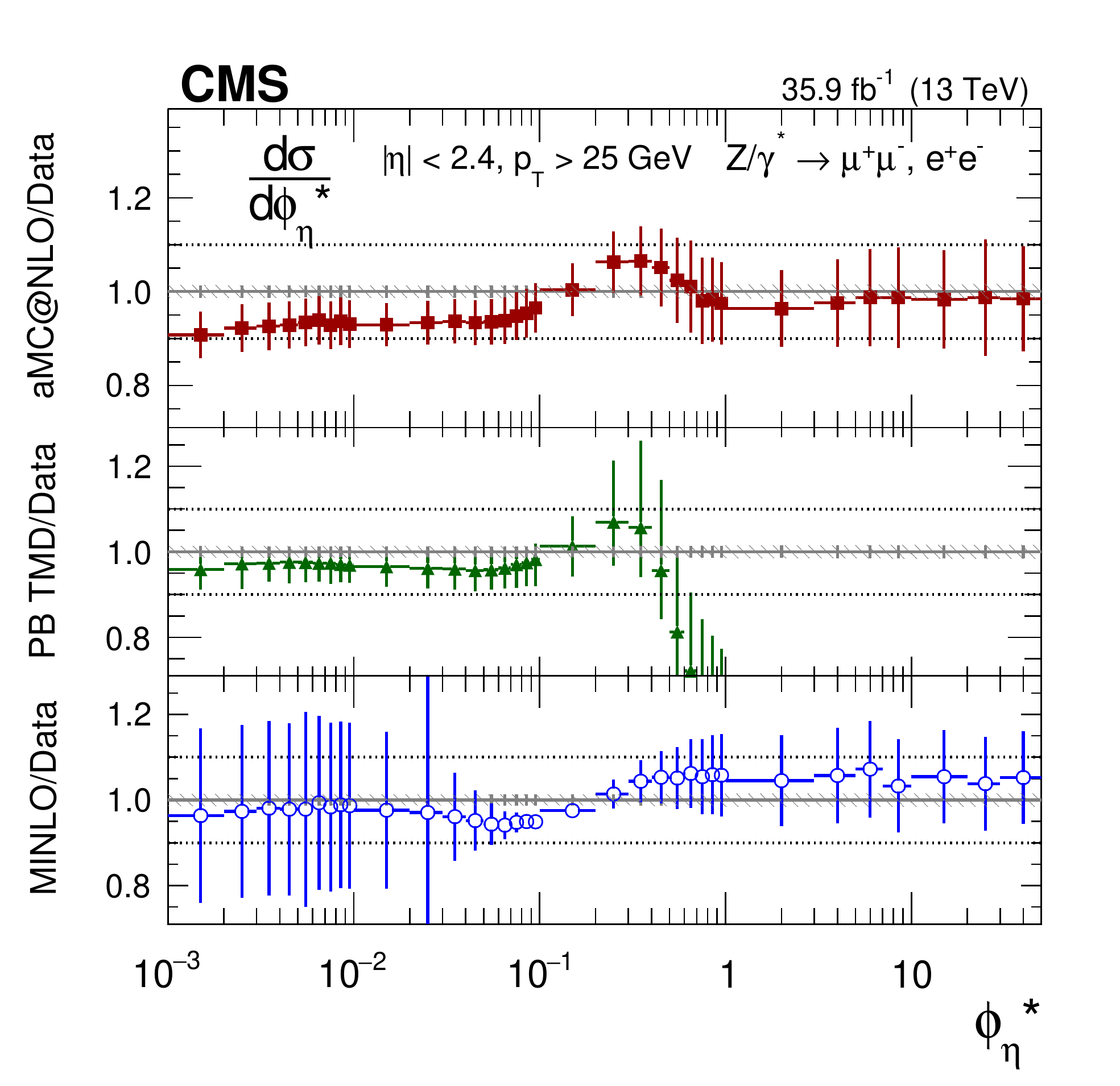}
	\caption{The measured absolute cross sections (left) in bins of $\phiStar$ for the dimuon (upper) and dielectron (middle) final states, and for the combination (lower). The ratios of the predictions to the data are also shown (right). The shaded bands around the data points (black) correspond to the total experimental uncertainty. The measurement is compared to the predictions with \MGvATNLO (square red markers), PB TMD (green triangles), and $\POWHEG$-\textsc{MINLO} (blue circles). The error bars around the predictions correspond to the combined statistical, PDF, and scale uncertainties.}
	\label{fig:unf_phi}
\end{figure}

Figure~\ref{fig:unf_phi} shows the measured absolute cross sections in bins
of $\phiStar$. The measurements are compared to the predictions from \MGvATNLO, PB TMD, and $\POWHEG$-\textsc{MINLO}.
The predictions are consistent with the measurements within the theoretical
uncertainties and describe data well at low \pt. As expected the PB
TMD predictions deviate from data at high \pt.

\begin{figure}
	\centering
	\includegraphics[width=0.45\textwidth]{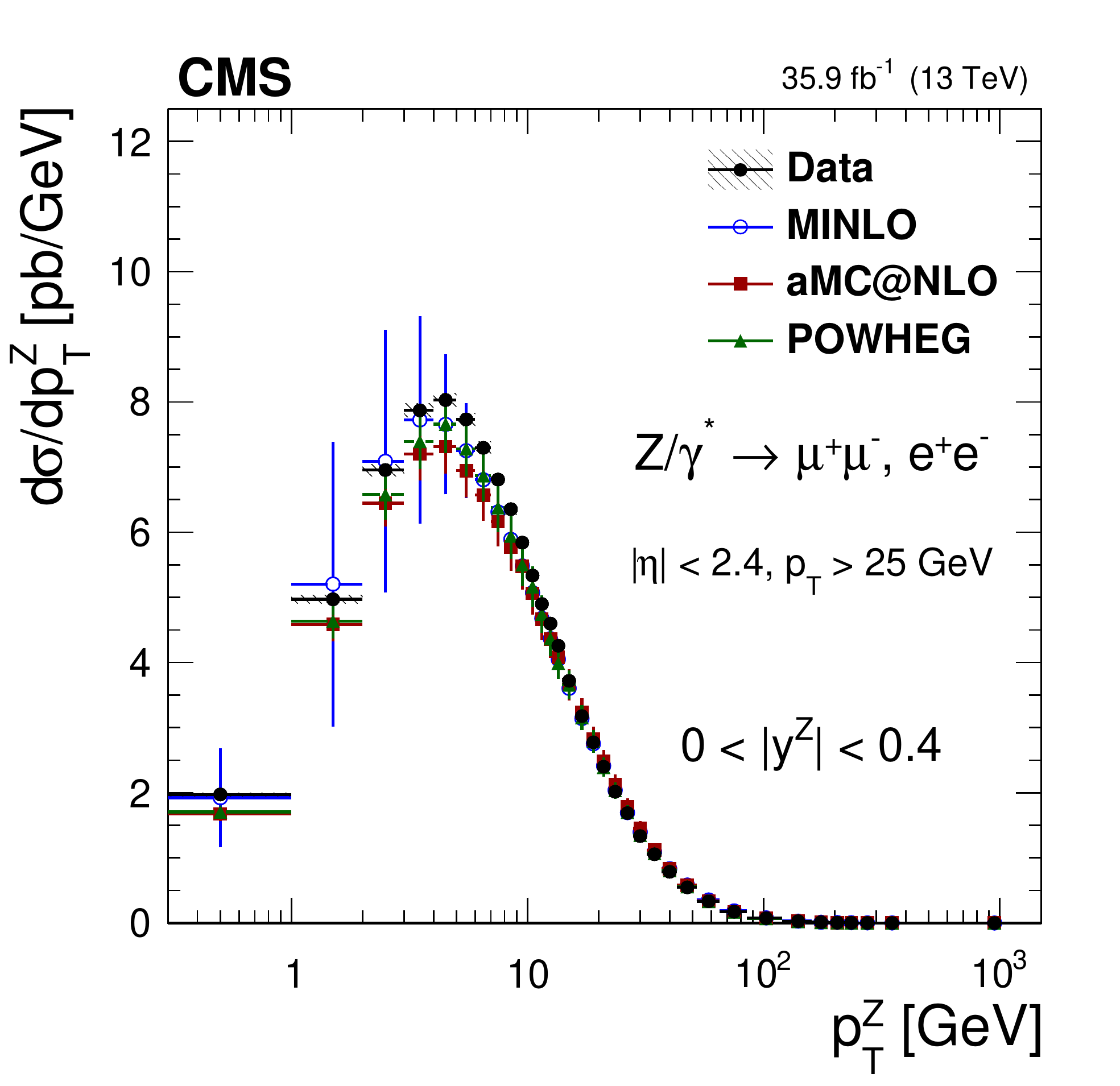}
        \includegraphics[width=0.45\textwidth]{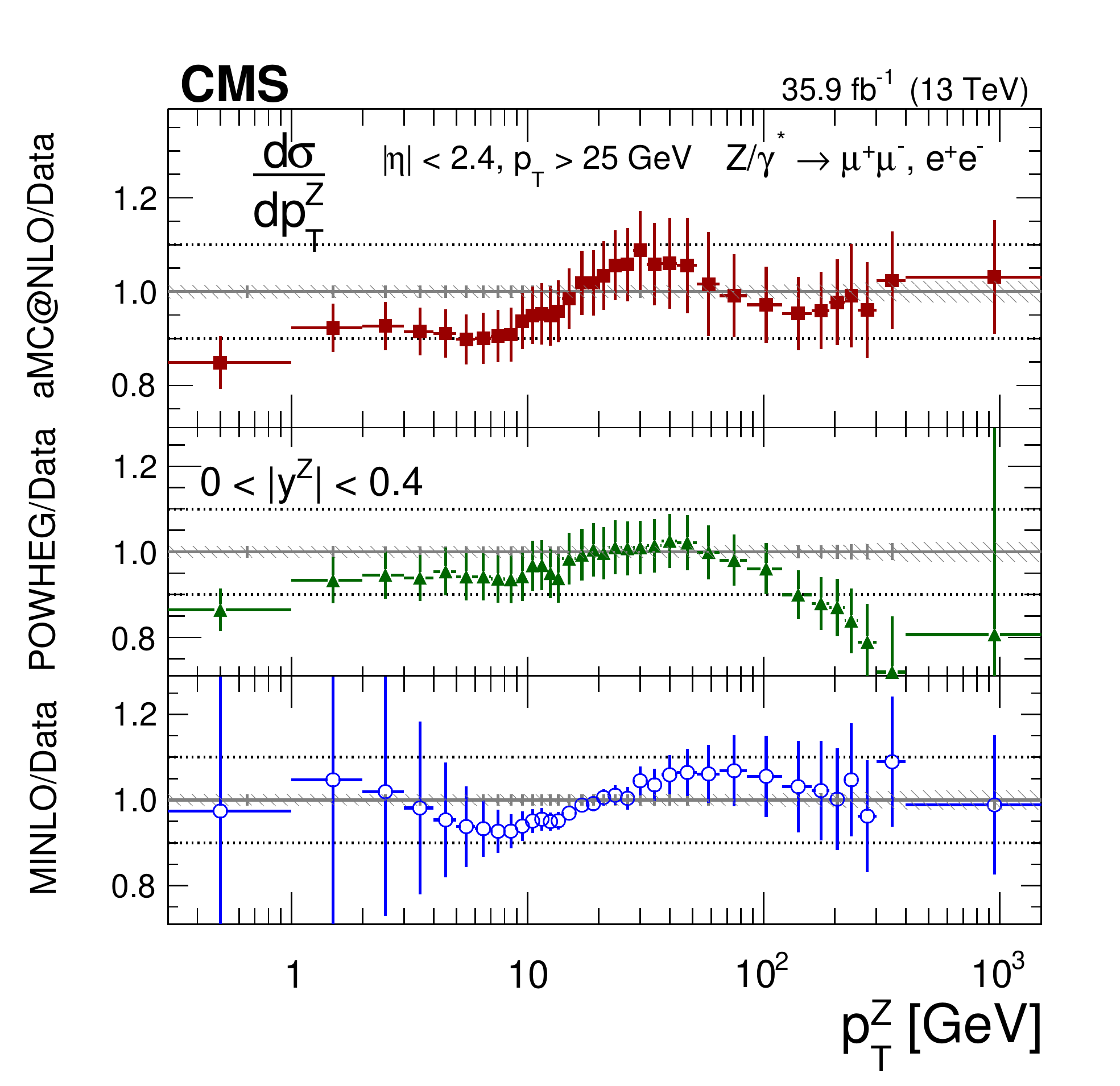}
	\caption{The measured absolute cross sections (left) in bins of $\pt^{\cPZ}$ for the $0.0 < \abs{\rapidity^{\cPZ}} < 0.4$ region. The ratios of the predictions to the data are also shown (right). The shaded bands around the data points (black) correspond to the total experimental uncertainty. The measurement is compared to the predictions with \MGvATNLO (square red markers),  $\POWHEG$ (green triangles), and $\POWHEG$-\textsc{MINLO} (blue circles). The error bands around the predictions correspond to the combined statistical, PDF, and scale uncertainties.}
	\label{fig:zll_double0}
\end{figure}

\begin{figure}
	\centering
	\includegraphics[width=0.45\textwidth]{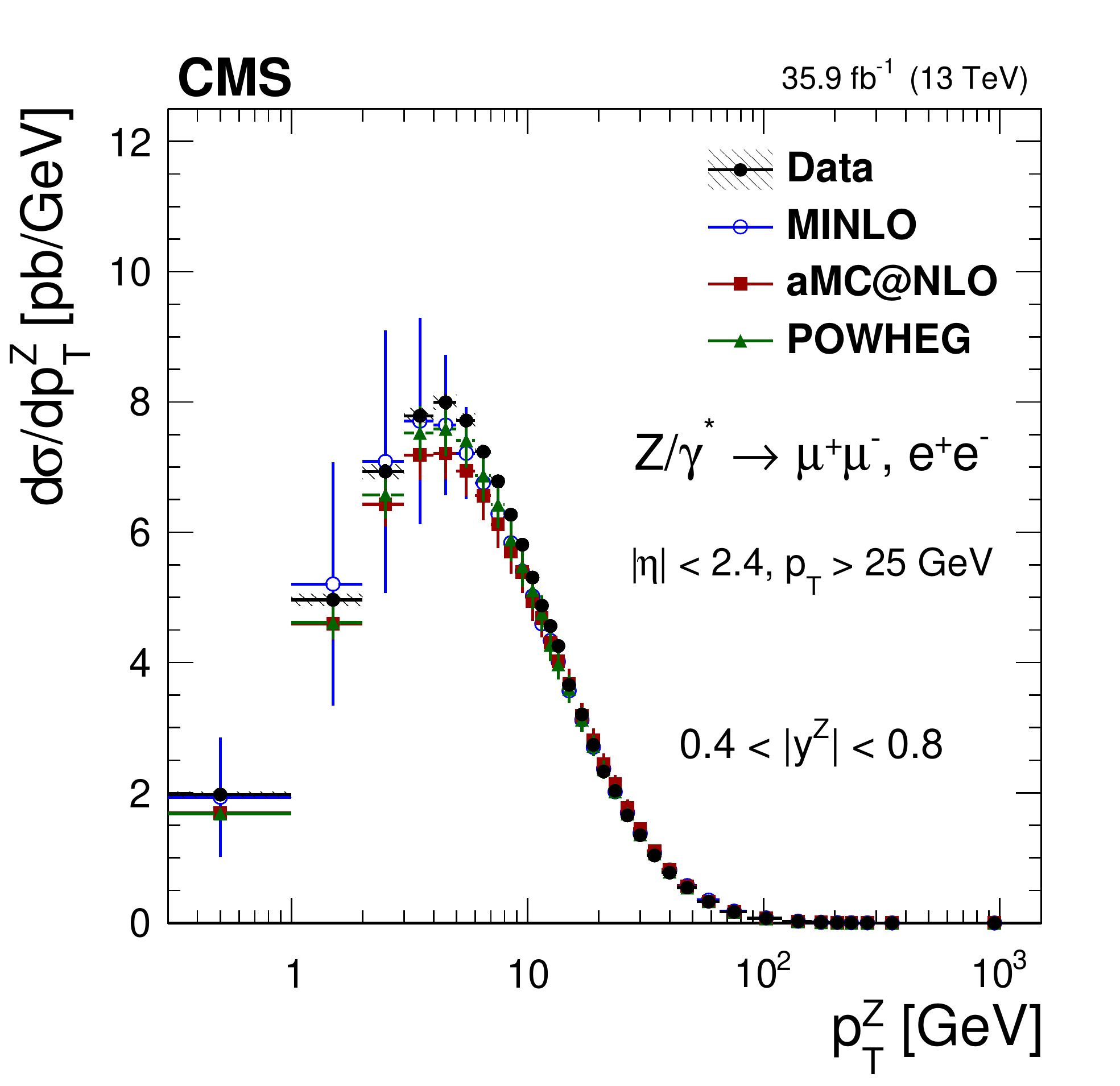}
        \includegraphics[width=0.45\textwidth]{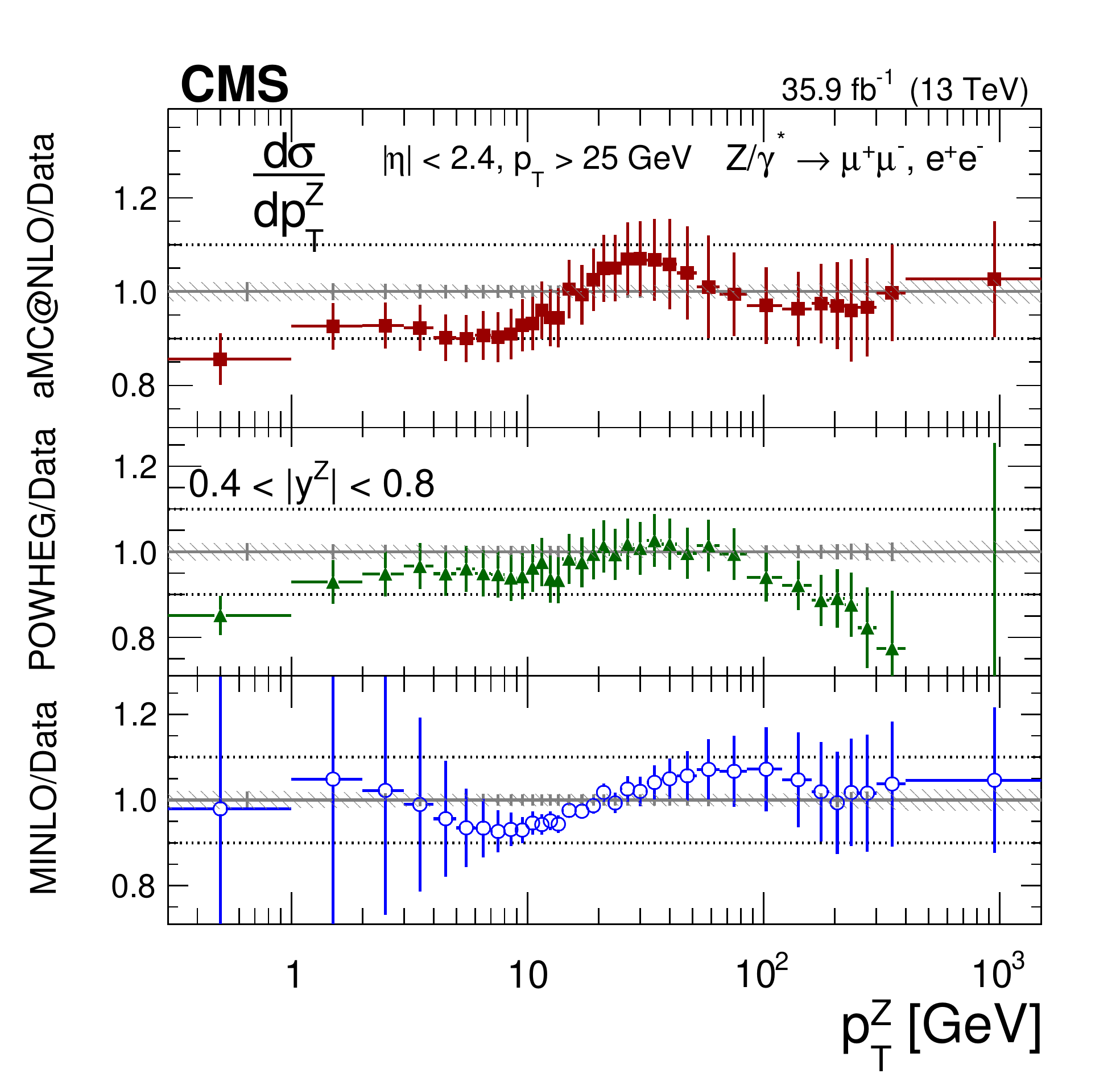}
	\caption{The measured absolute cross sections (left) in bins of $\pt^{\cPZ}$ for the $0.4 < \abs{\rapidity^{\cPZ}} < 0.8$ region. The ratios of the predictions to the data are also shown (right). The shaded bands around the data points (black) correspond to the total experimental uncertainty. The measurement is compared to the predictions with \MGvATNLO (square red markers),  $\POWHEG$ (green triangles), and $\POWHEG$-\textsc{MINLO} (blue circles). The error bands around the predictions correspond to the combined statistical, PDF, and scale uncertainties.}
	\label{fig:zll_double1}
\end{figure}

\begin{figure}
	\centering
	\includegraphics[width=0.45\textwidth]{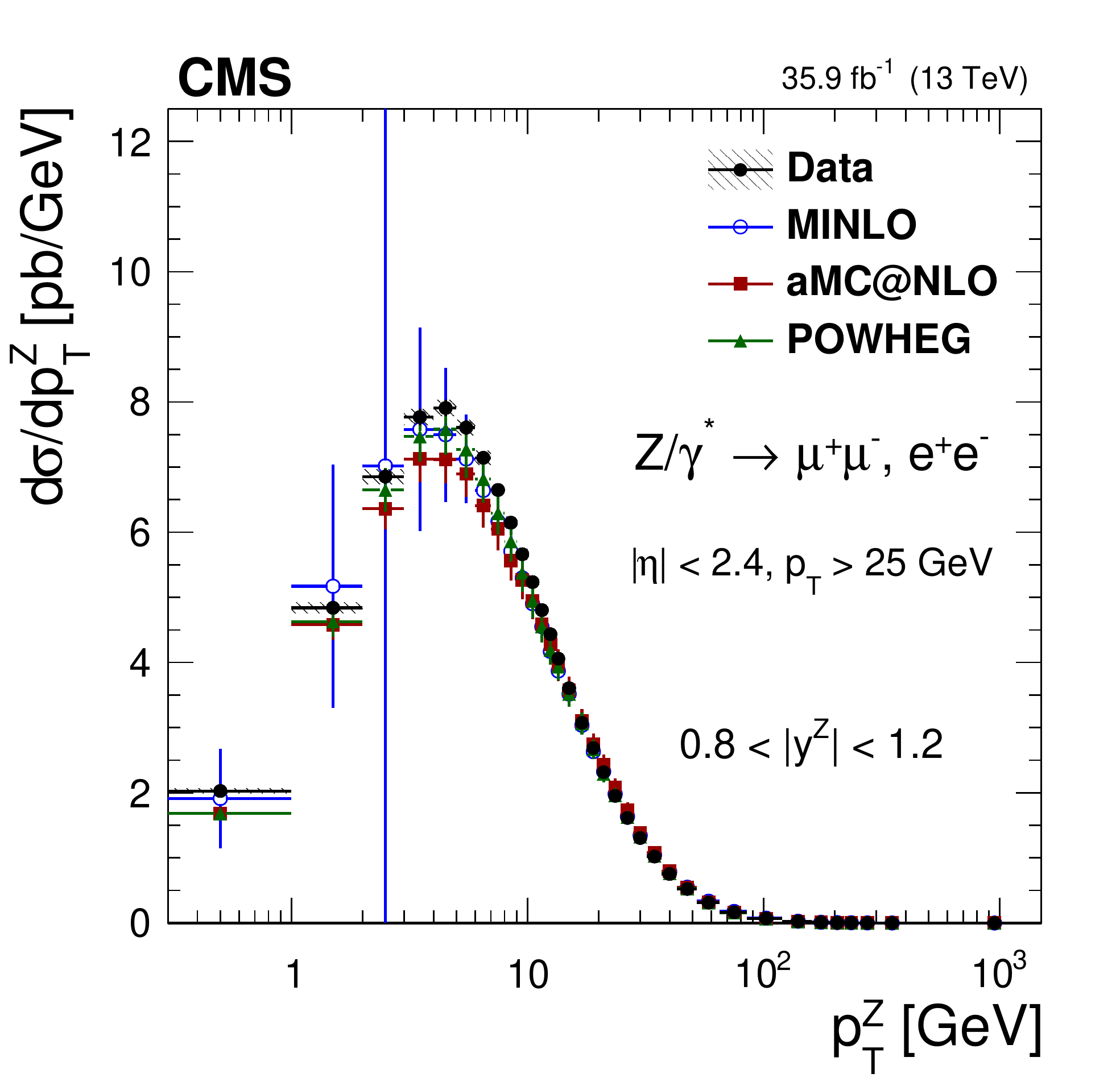}
        \includegraphics[width=0.45\textwidth]{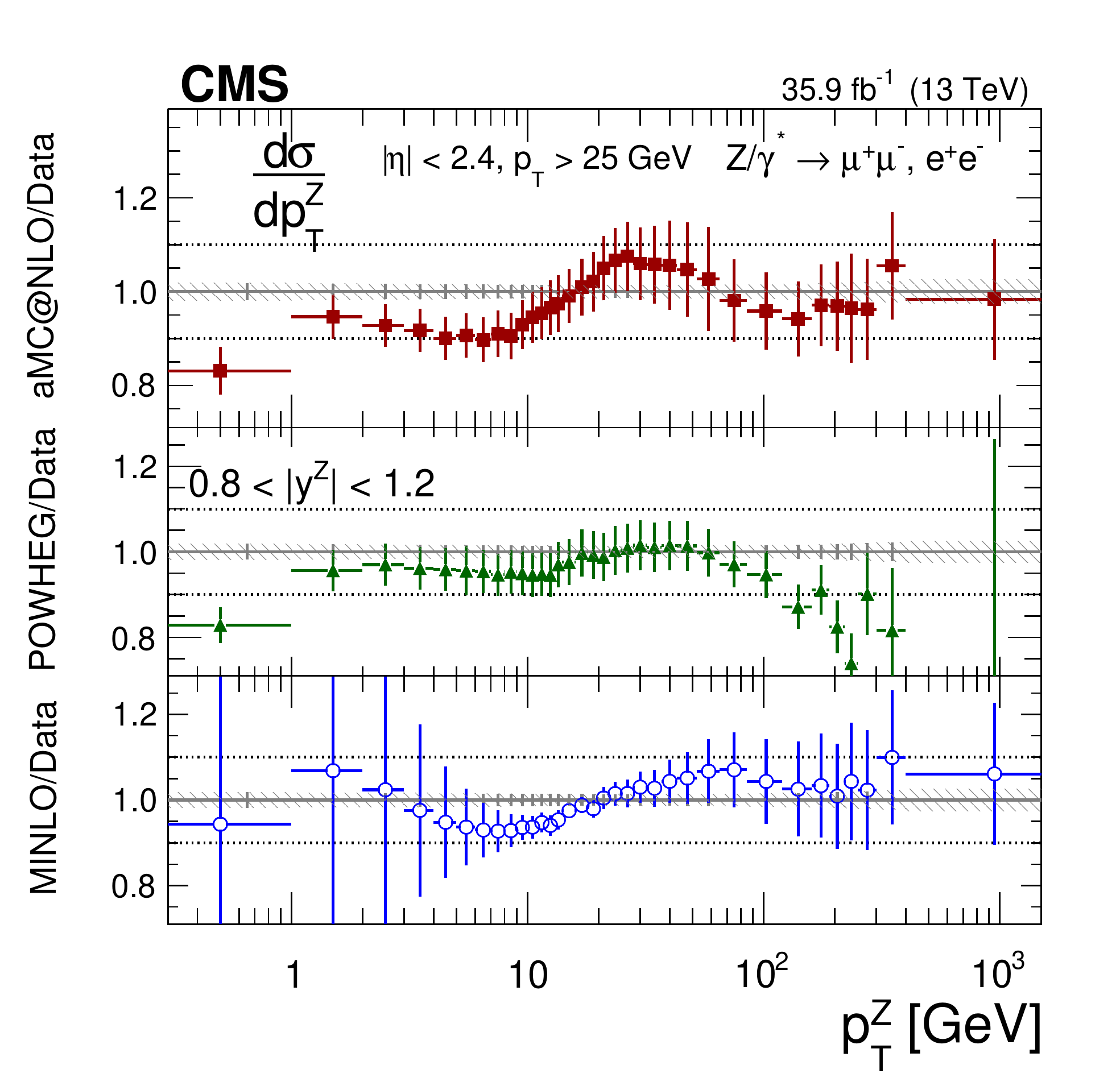}
	\caption{The measured absolute cross sections (left) in bins of $\pt^{\cPZ}$ for the $0.8 < \abs{\rapidity^{\cPZ}} < 1.2$ region. The ratios of the predictions to the data are also shown (right). The shaded bands around the data points (black) correspond to the total experimental uncertainty. The measurement is compared to the predictions with \MGvATNLO (square red markers),  $\POWHEG$ (green triangles), and $\POWHEG$-\textsc{MINLO} (blue circles). The error bands around the predictions correspond to the combined statistical, PDF, and scale uncertainties.}
	\label{fig:zll_double2}
\end{figure}

\begin{figure}
	\centering
	\includegraphics[width=0.45\textwidth]{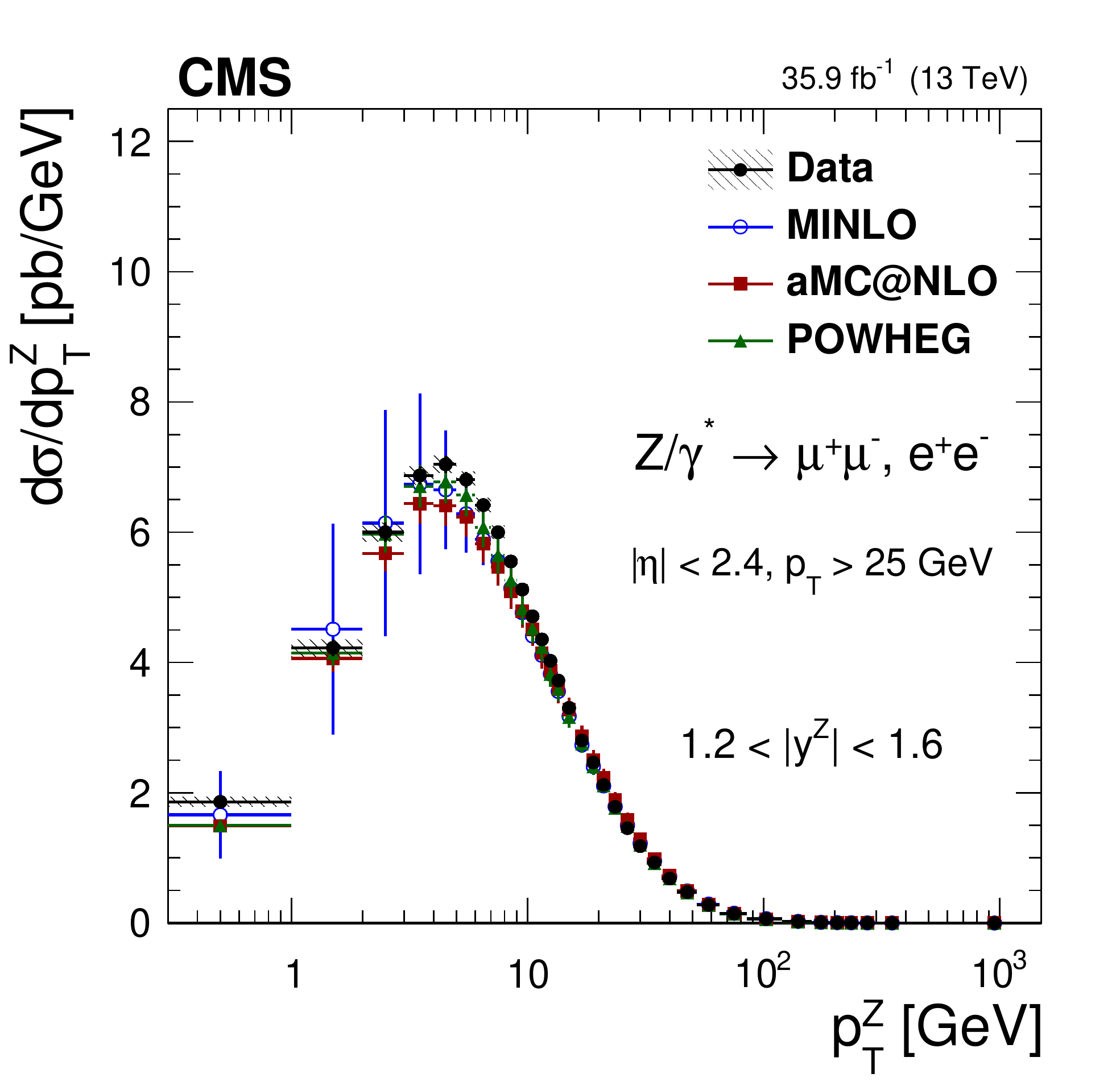}
        \includegraphics[width=0.45\textwidth]{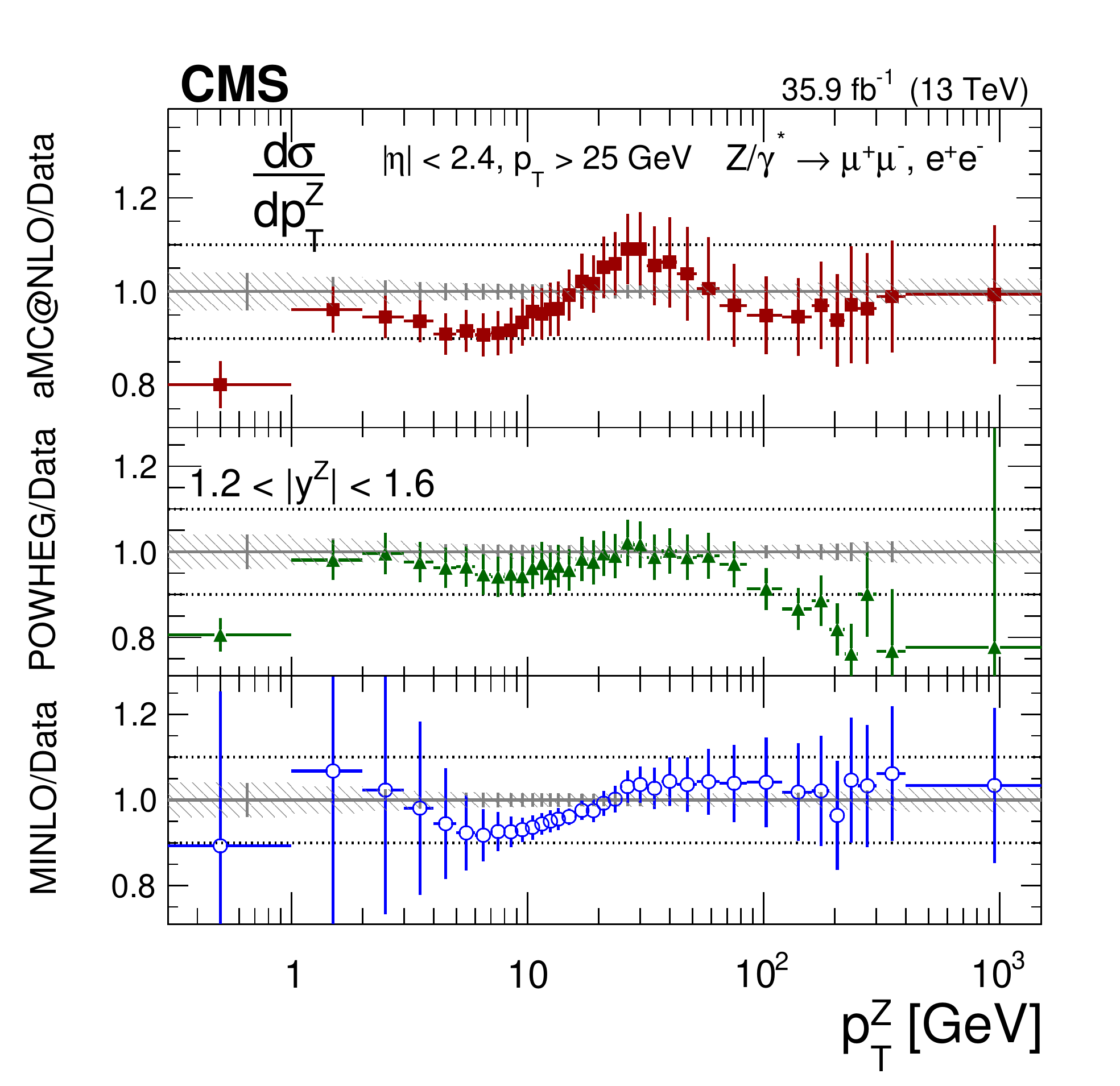}
	\caption{The measured absolute cross sections (left) in bins of $\pt^{\cPZ}$ for the $1.2 < \abs{\rapidity^{\cPZ}} < 1.6$ region. The ratios of the predictions to the data are also shown (right). The shaded bands around the data points (black) correspond to the total experimental uncertainty. The measurement is compared to the predictions with \MGvATNLO (square red markers),  $\POWHEG$ (green triangles), and $\POWHEG$-\textsc{MINLO} (blue circles). The error bands around the predictions correspond to the combined statistical, PDF, and scale uncertainties.}
	\label{fig:zll_double3}
\end{figure}

\begin{figure}
	\centering
	\includegraphics[width=0.45\textwidth]{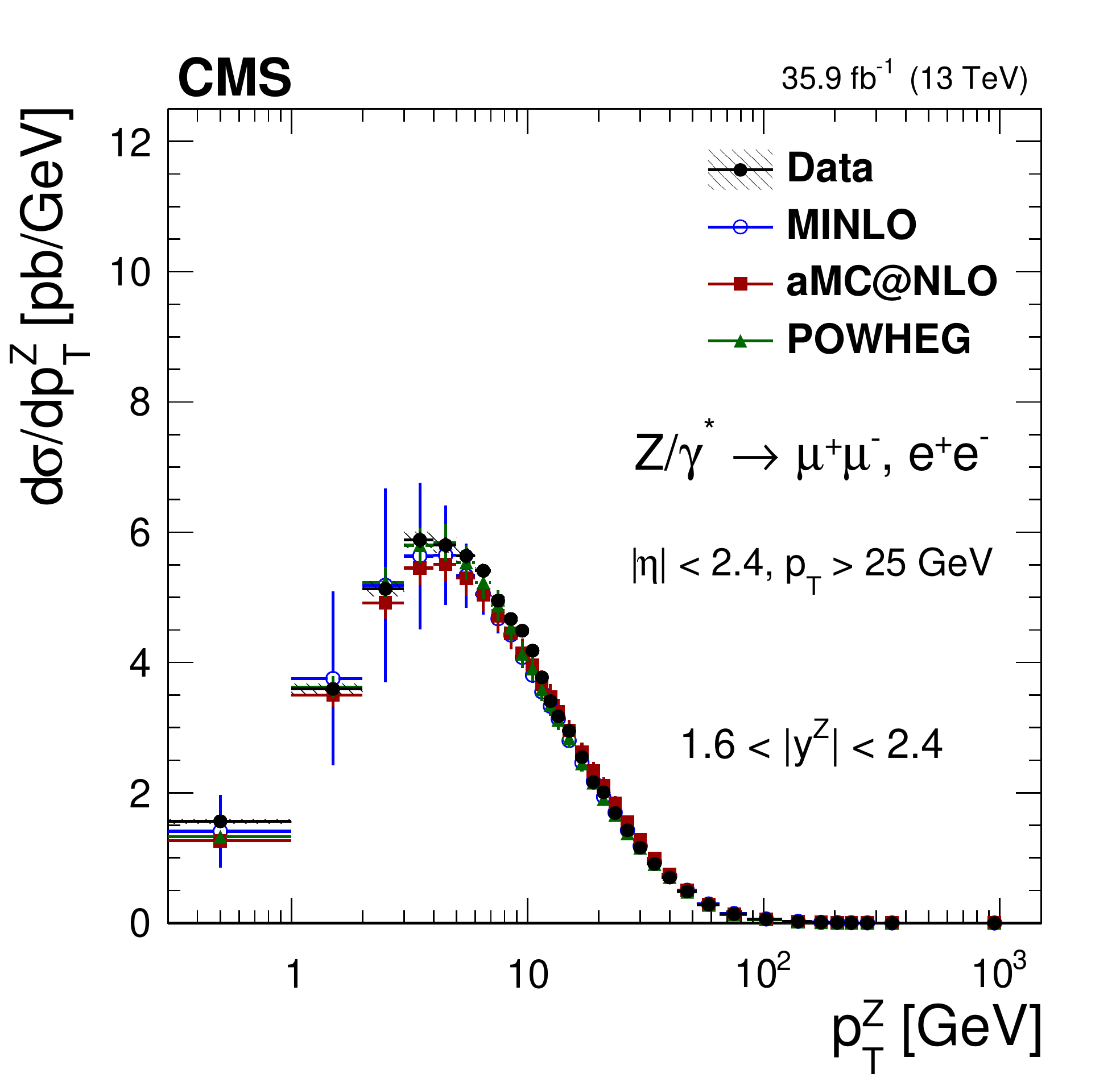}
        \includegraphics[width=0.45\textwidth]{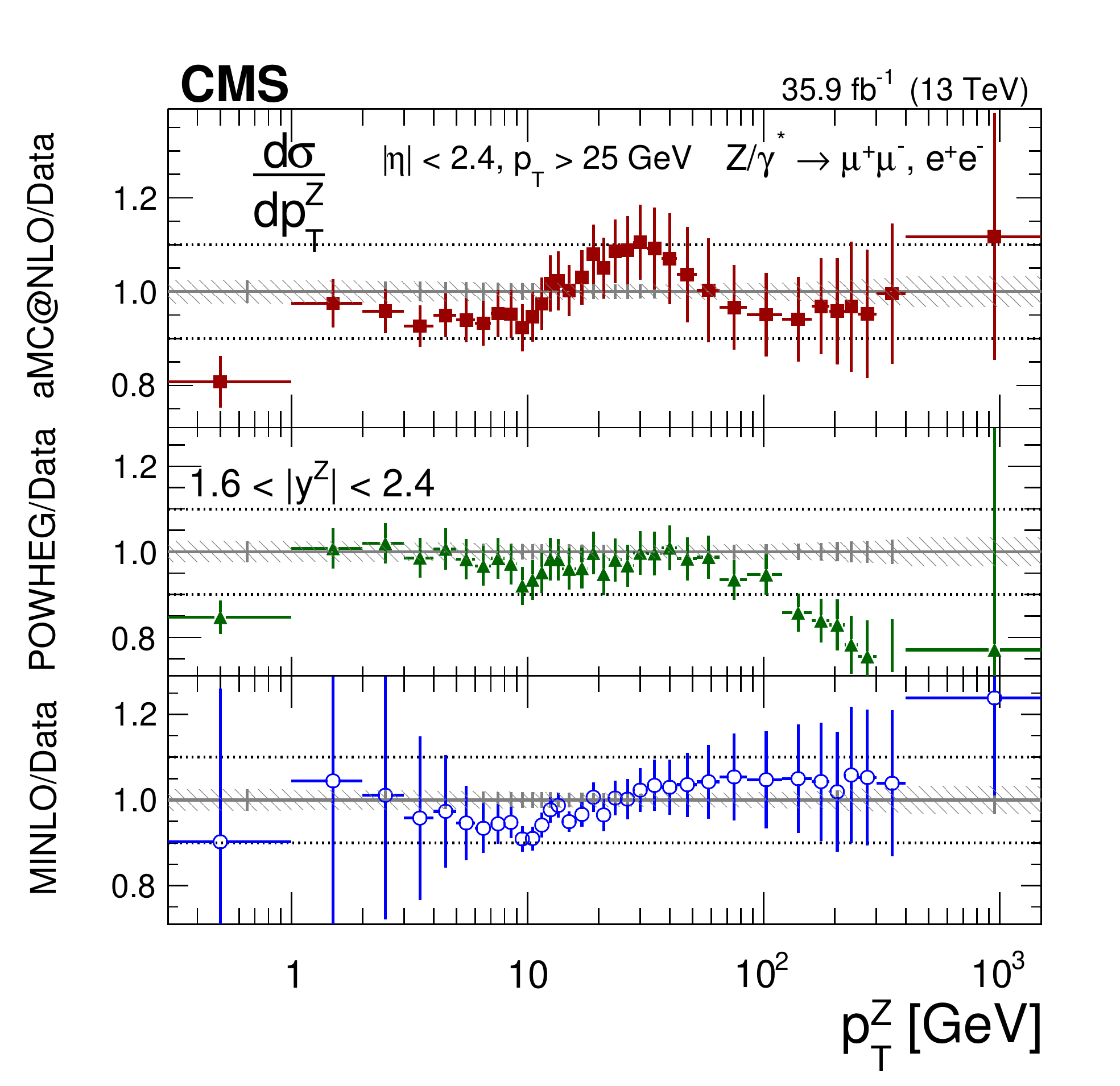}
	\caption{The measured absolute cross sections (left) in bins of $\pt^{\cPZ}$ for the $1.6 < \abs{\rapidity^{\cPZ}} < 2.4$ region. The ratios of the predictions to the data are also shown (right). The shaded bands around the data points (black) correspond to the total experimental uncertainty. The measurement is compared to the predictions with \MGvATNLO (square red markers),  $\POWHEG$ (green triangles), and $\POWHEG$-\textsc{MINLO} (blue circles). The error bands around the predictions correspond to the combined statistical, PDF, and scale uncertainties.}
	\label{fig:zll_double4}
\end{figure}

\begin{figure}
	\centering
	\includegraphics[width=0.45\textwidth]{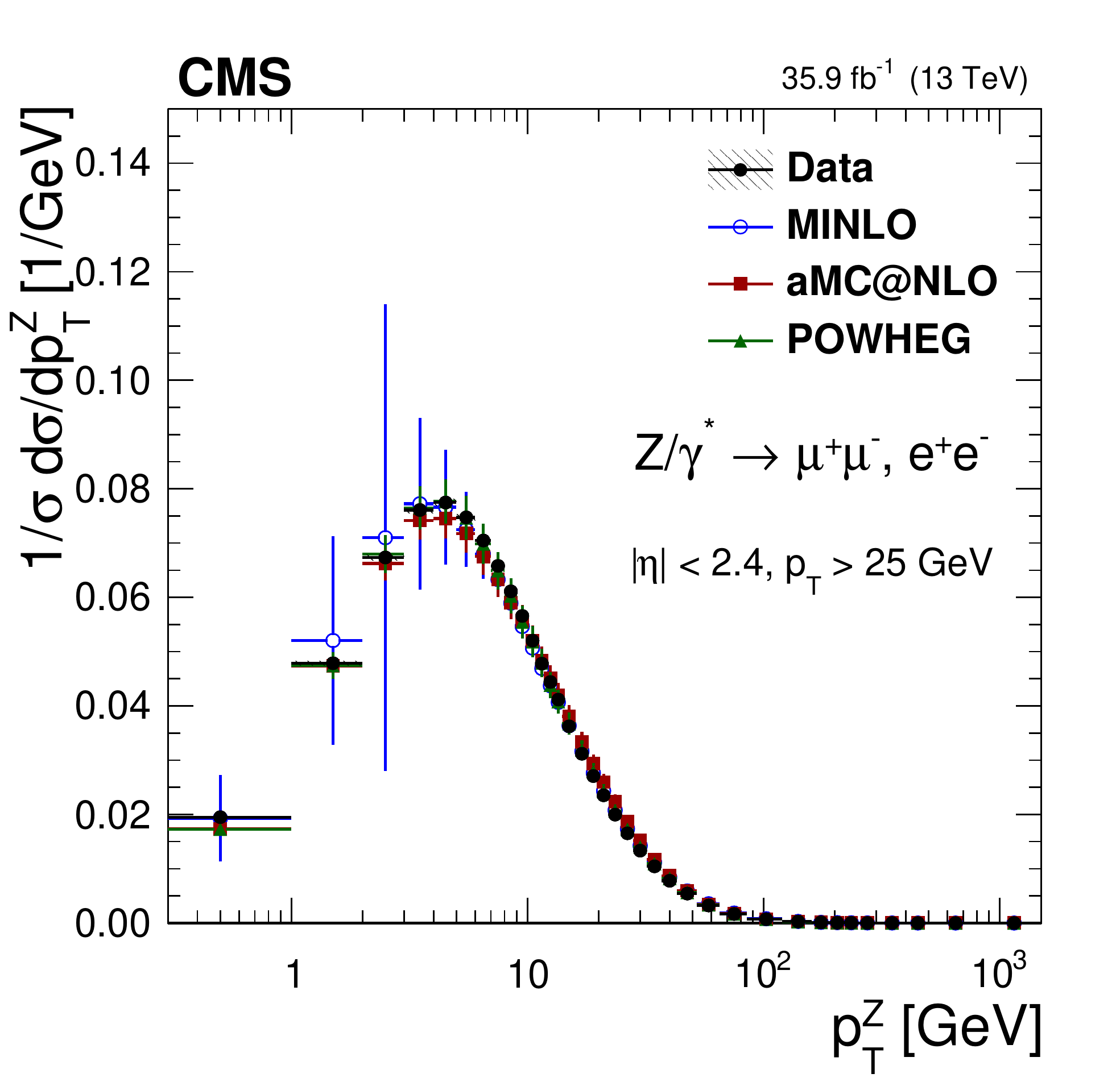}
        \includegraphics[width=0.45\textwidth]{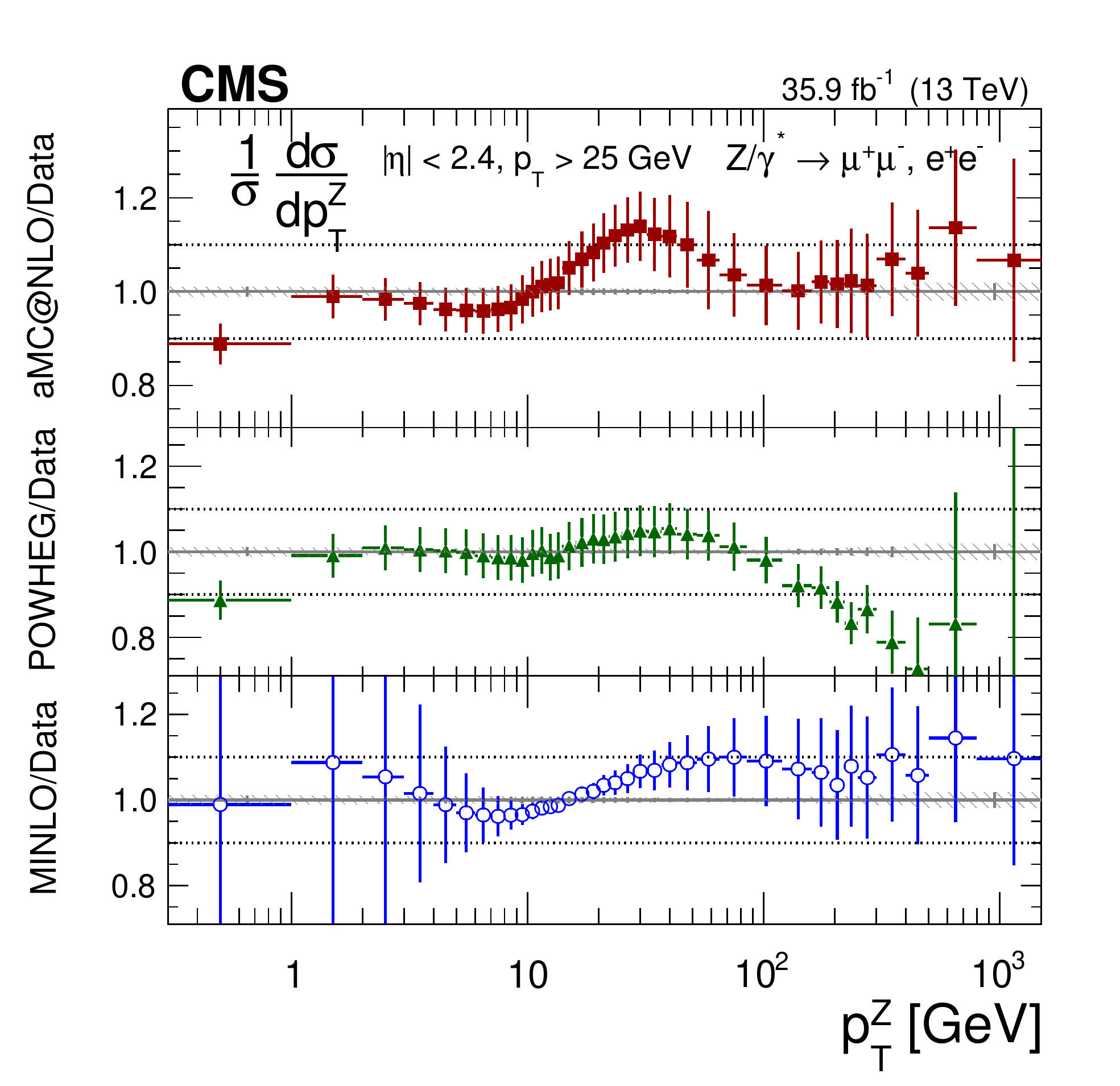}
	\includegraphics[width=0.45\textwidth]{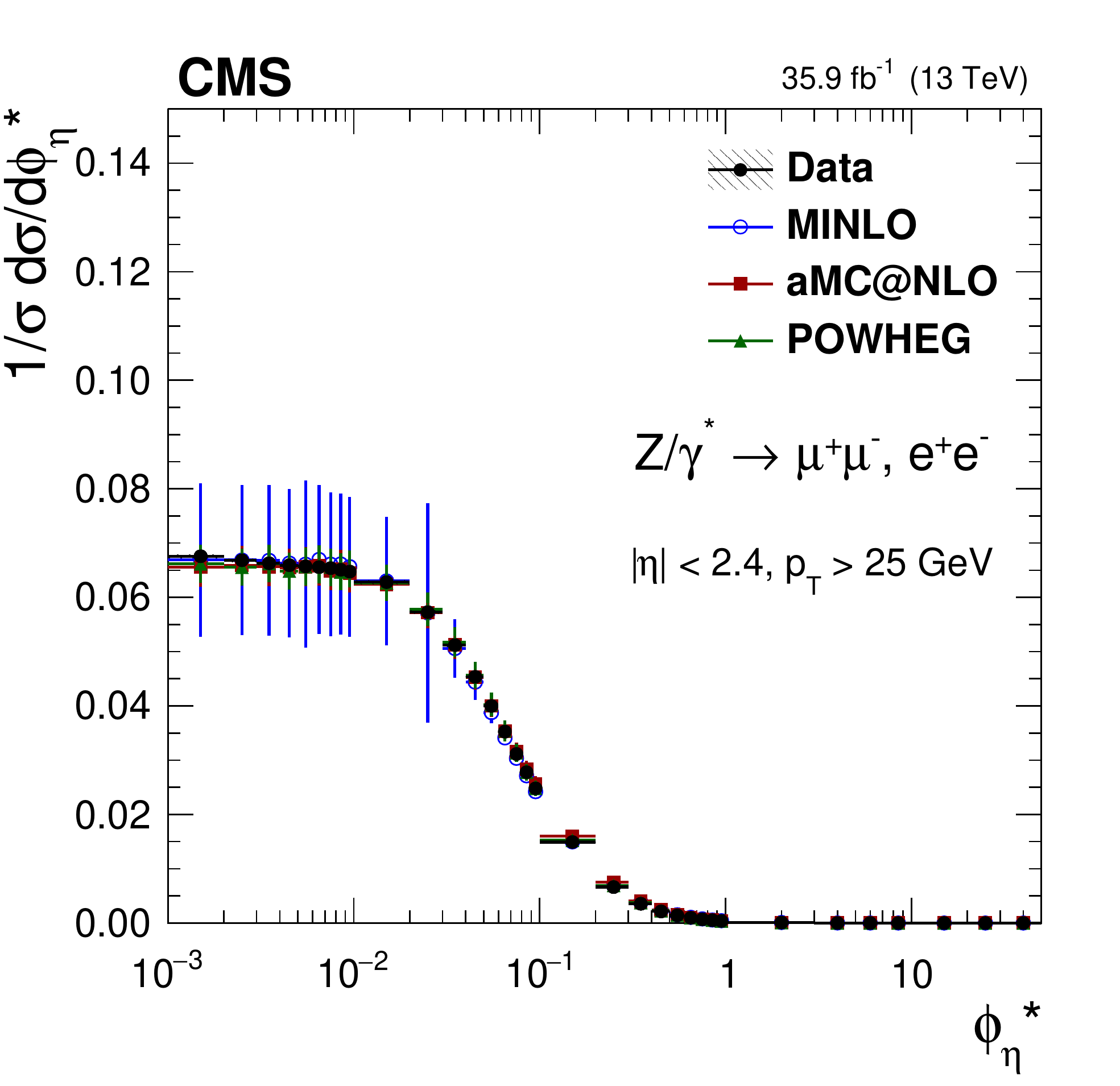}
	\includegraphics[width=0.45\textwidth]{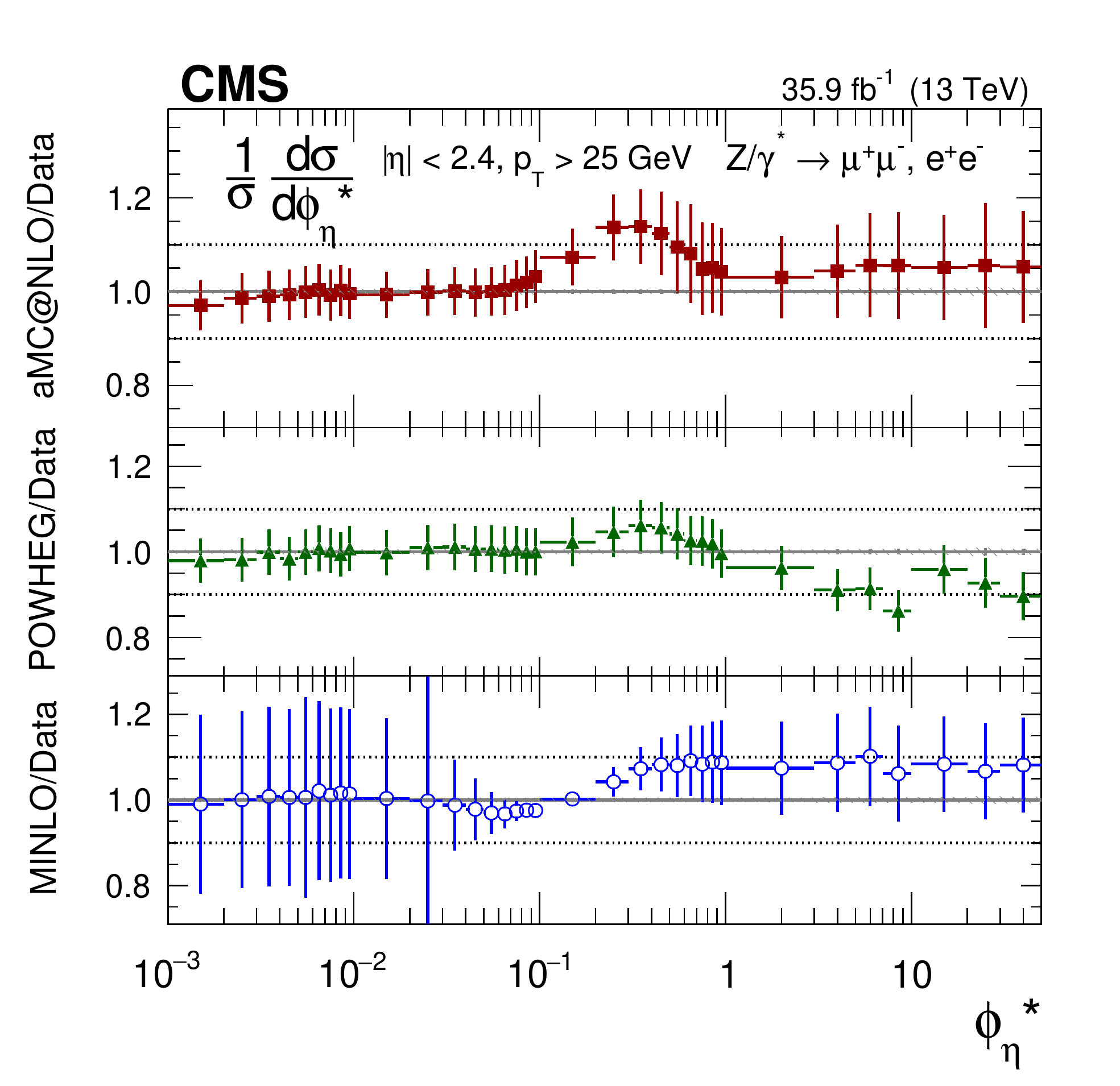}
        \includegraphics[width=0.45\textwidth]{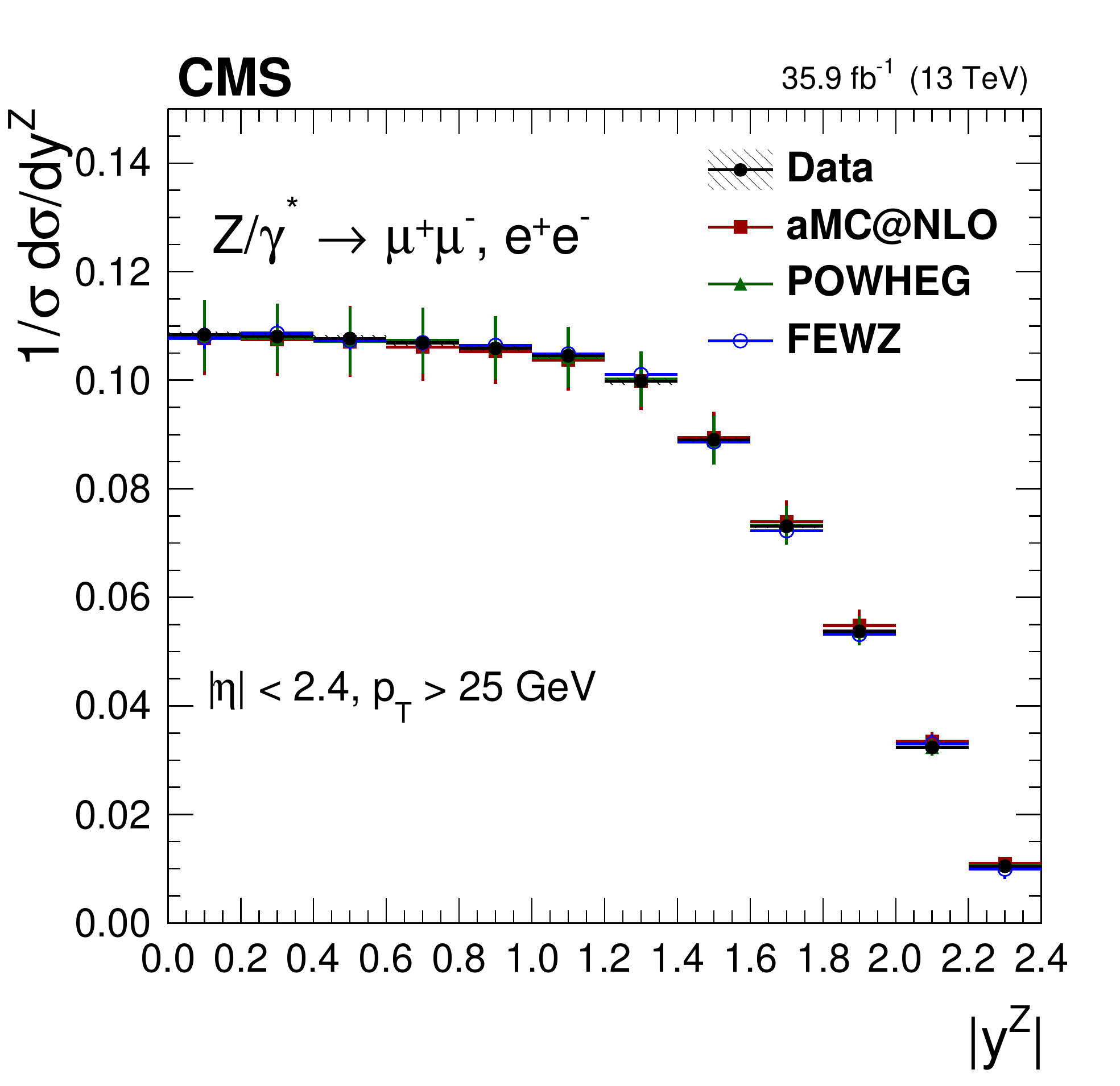}
        \includegraphics[width=0.45\textwidth]{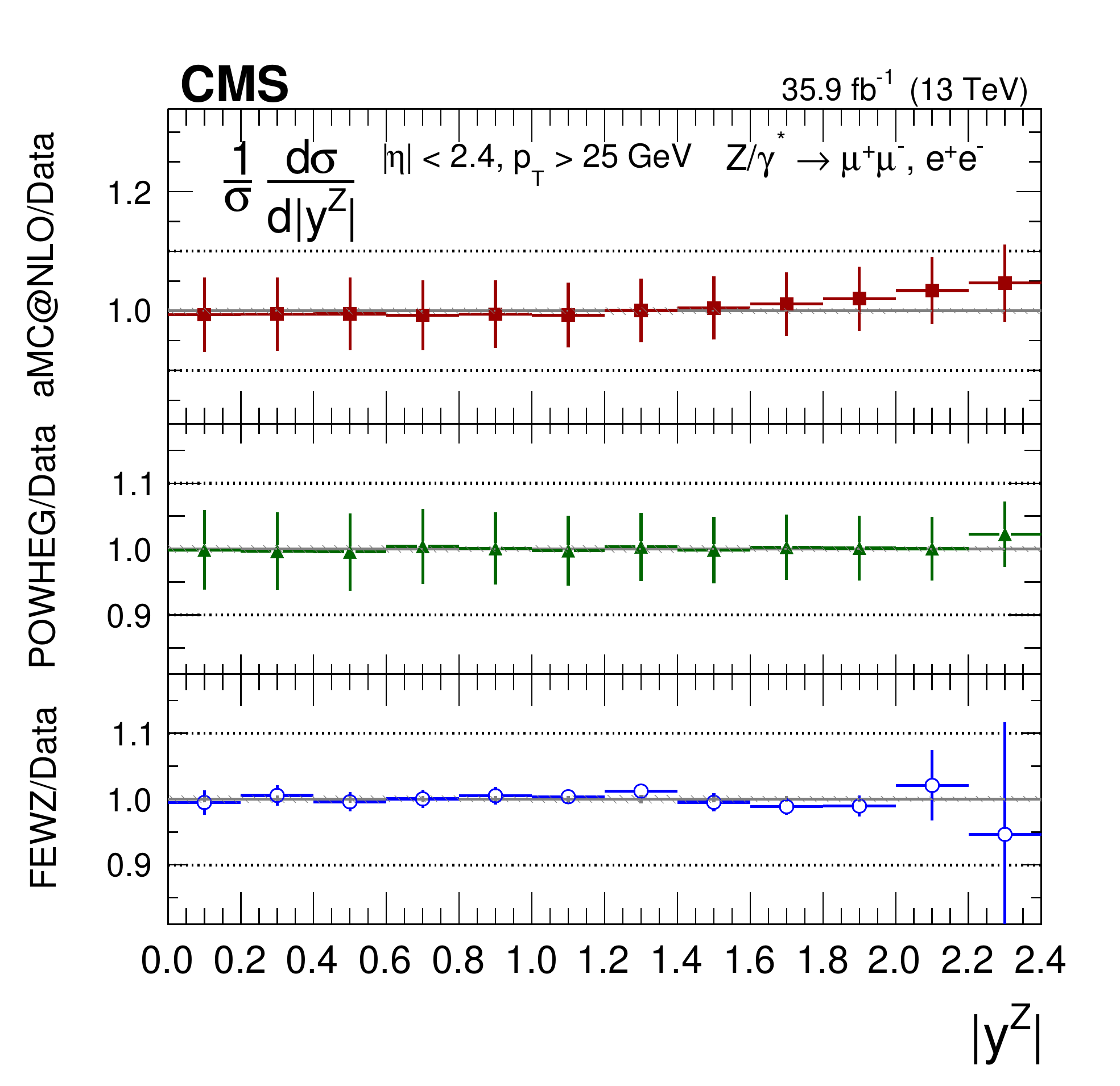}
	\caption{The measured normalized cross sections (left) in bins of $\pt^{\cPZ}$ (upper), $\phiStar$ (middle), and $\abs{\rapidity^{\cPZ}}$ (lower) for the combined measurement. The ratios of the predictions to the data are also shown (right). The shaded bands around the data points (black) correspond to the total experimental uncertainty. The $\pt^{\cPZ}$  and $\phiStar$ measurements are compared to the predictions with \MGvATNLO (square red markers), $\POWHEG$ (green triangles), and $\POWHEG$-\textsc{MINLO} (blue circles). The $\abs{\rapidity^{\cPZ}}$ measurement is compared to the predictions with \MGvATNLO (square red markers), $\POWHEG$ (green triangles), and $\FEWZ$ (blue circles). The error bars around the predictions correspond to the combined statistical, PDF, and scale uncertainties.}
	\label{fig:cross_norm}
\end{figure}

\begin{figure}
	\centering
	\includegraphics[width=0.45\textwidth]{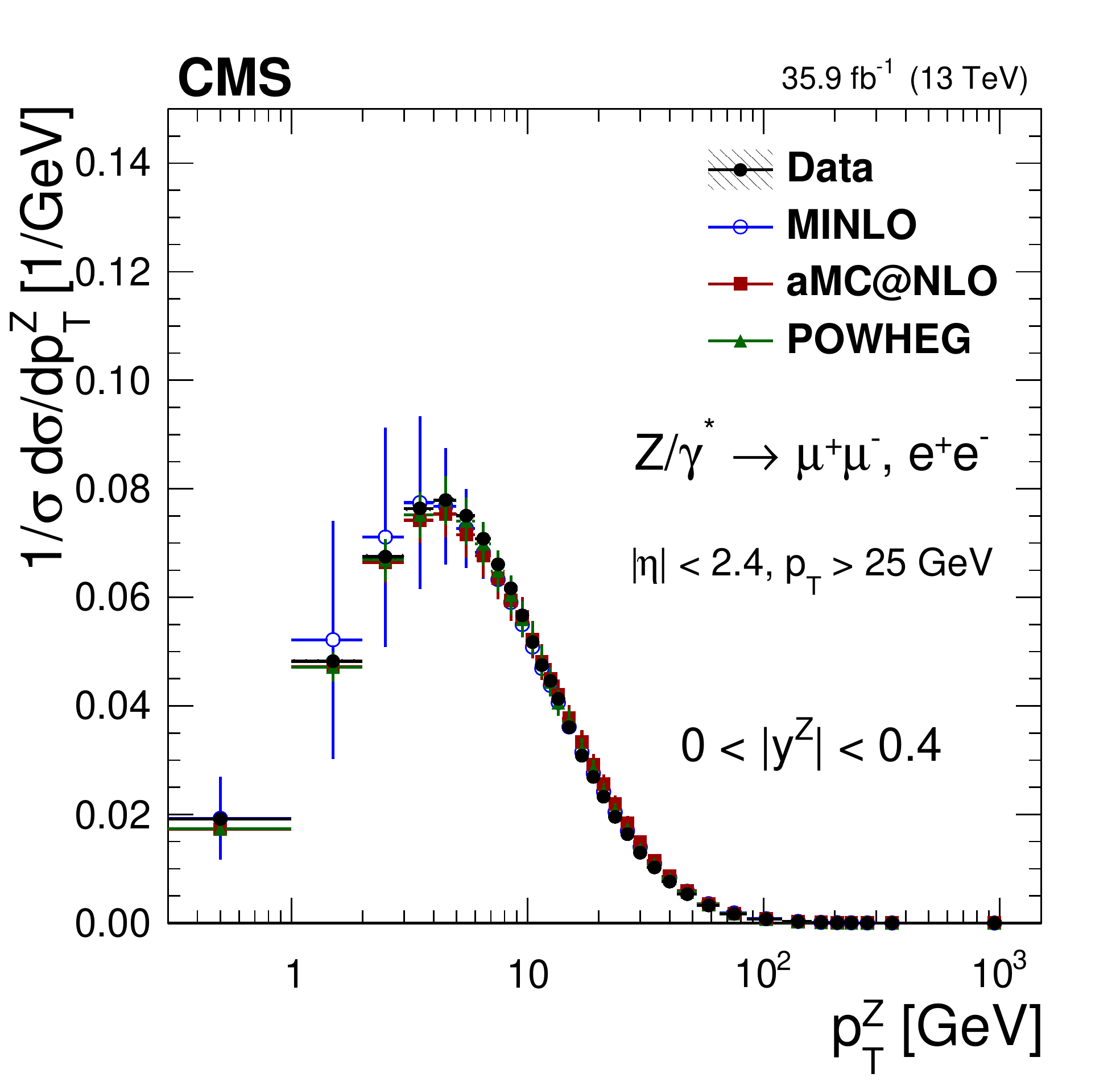}
        \includegraphics[width=0.45\textwidth]{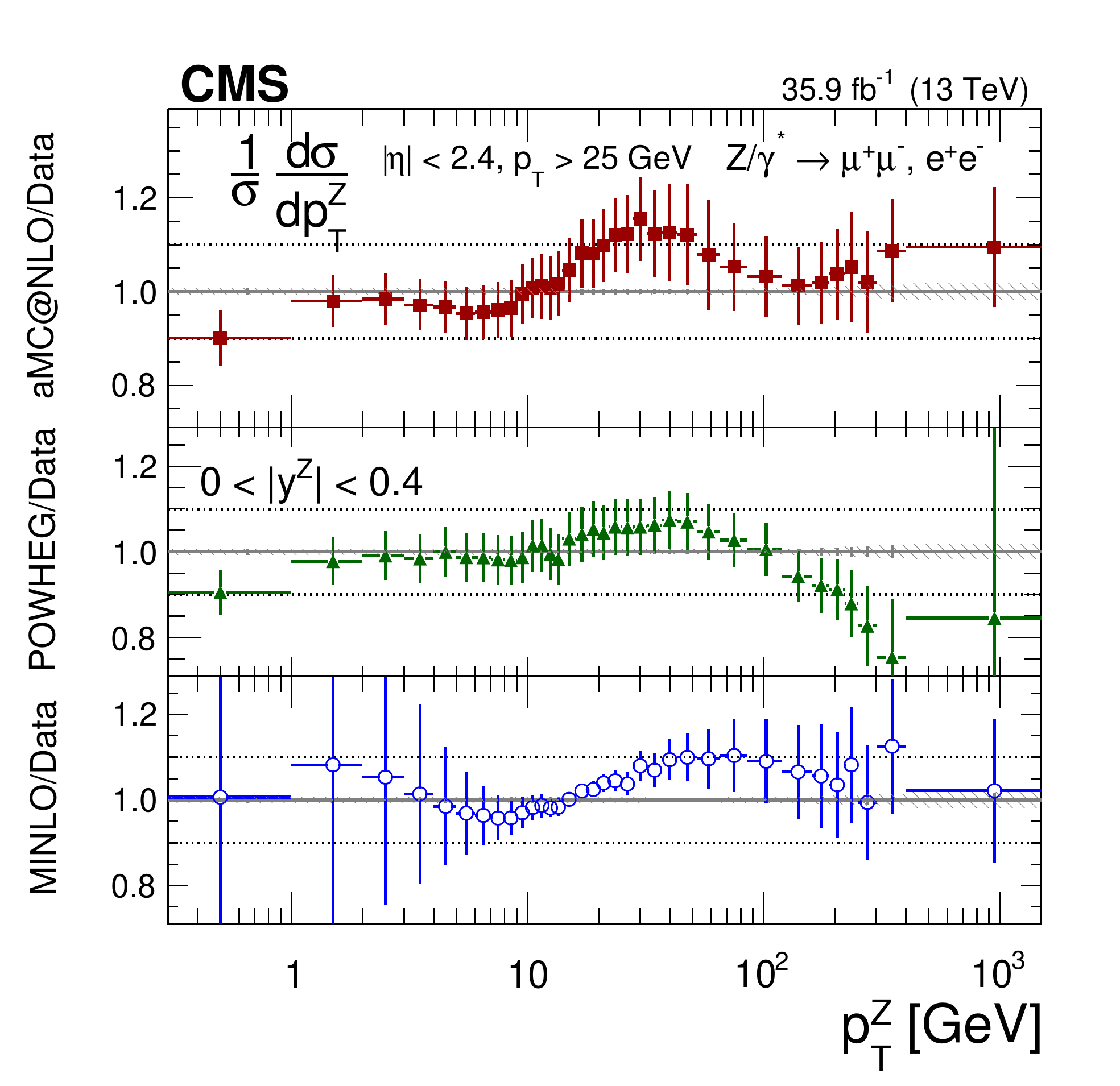}
	\caption{The measured normalized cross sections (left) in bins of $\pt^{\cPZ}$ for the $0.0 < \abs{\rapidity^{\cPZ}} < 0.4$ region. The ratios of the predictions to the data are also shown (right). The shaded bands around the data points (black) correspond to the total experimental uncertainty. The measurement is compared to the predictions with \MGvATNLO (square red markers),  $\POWHEG$ (green triangles), and $\POWHEG$-\textsc{MINLO} (blue circles). The error bands around the predictions correspond to the combined statistical, PDF, and scale uncertainties.}
	\label{fig:zll_norm0}
\end{figure}

\begin{figure}
	\centering
	\includegraphics[width=0.45\textwidth]{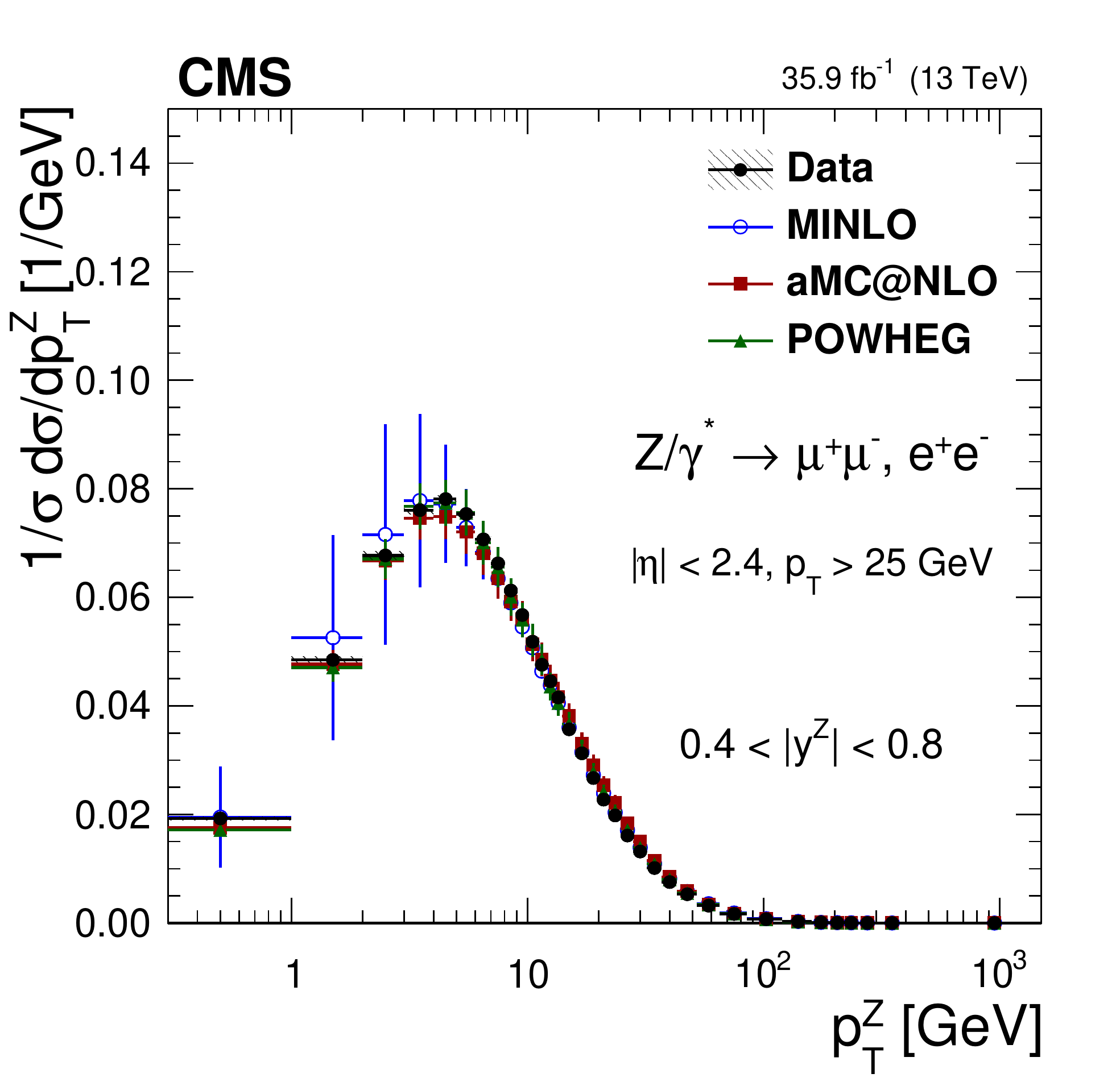}
        \includegraphics[width=0.45\textwidth]{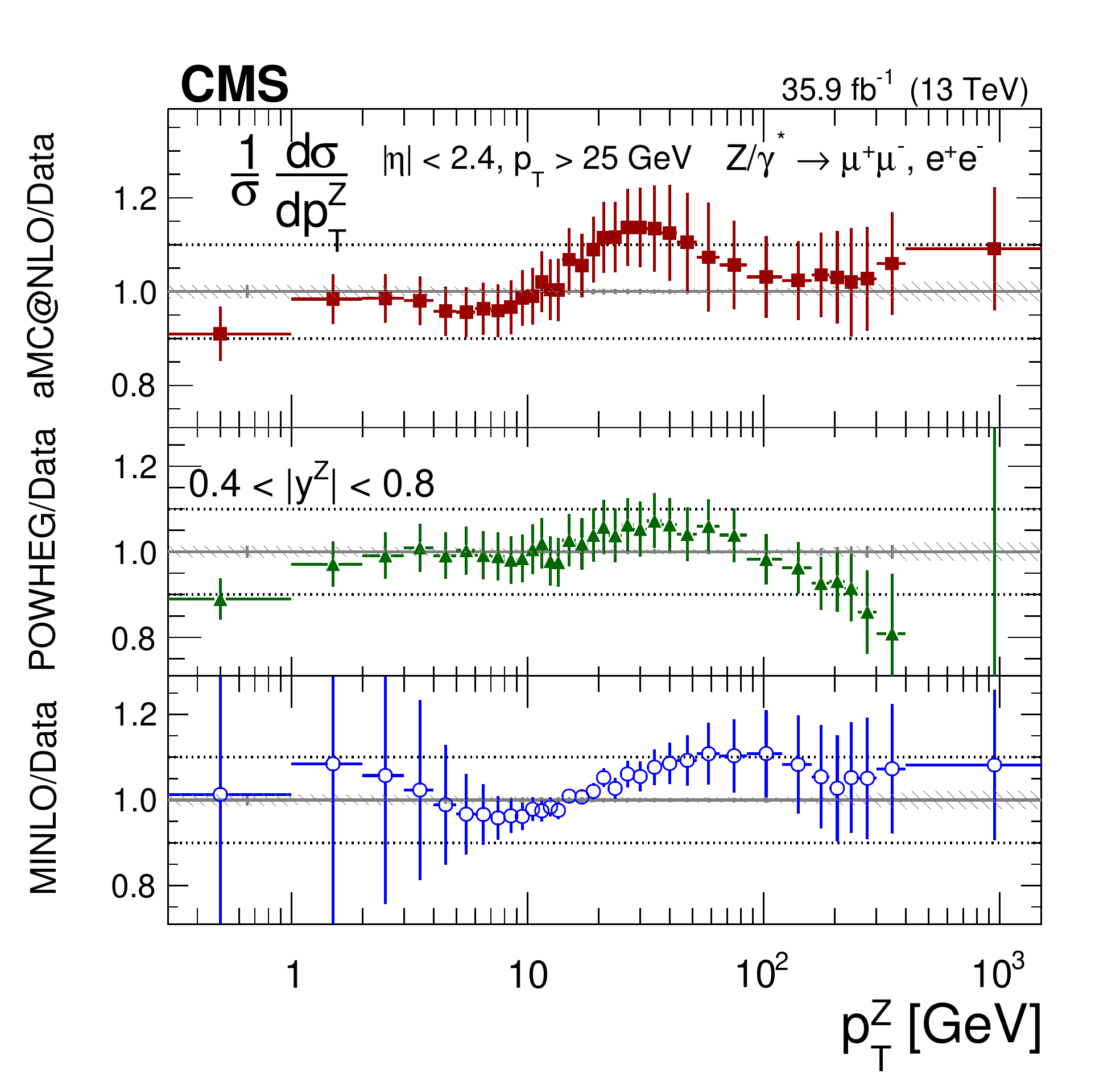}
	\caption{The measured normalized cross sections (left) in bins of $\pt^{\cPZ}$ for the $0.4 < \abs{\rapidity^{\cPZ}} < 0.8$ region. The ratios of the predictions to the data are also shown (right). The shaded bands around the data points (black) correspond to the total experimental uncertainty. The measurement is compared to the predictions with \MGvATNLO (square red markers),  $\POWHEG$ (green triangles), and $\POWHEG$-\textsc{MINLO} (blue circles). The error bands around the predictions correspond to the combined statistical, PDF, and scale uncertainties.}
	\label{fig:zll_norm1}
\end{figure}

\begin{figure}
	\centering
	\includegraphics[width=0.45\textwidth]{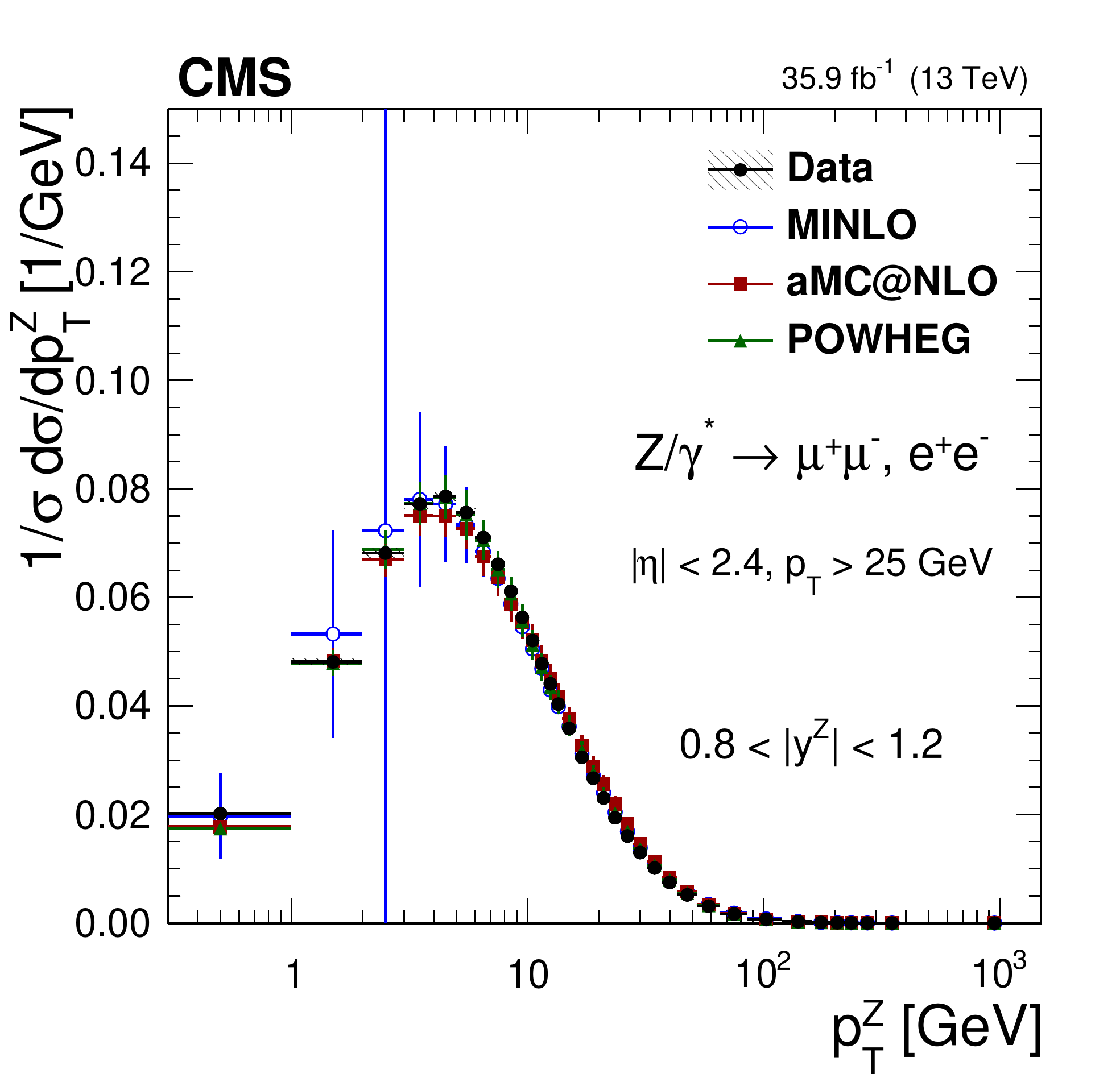}
        \includegraphics[width=0.45\textwidth]{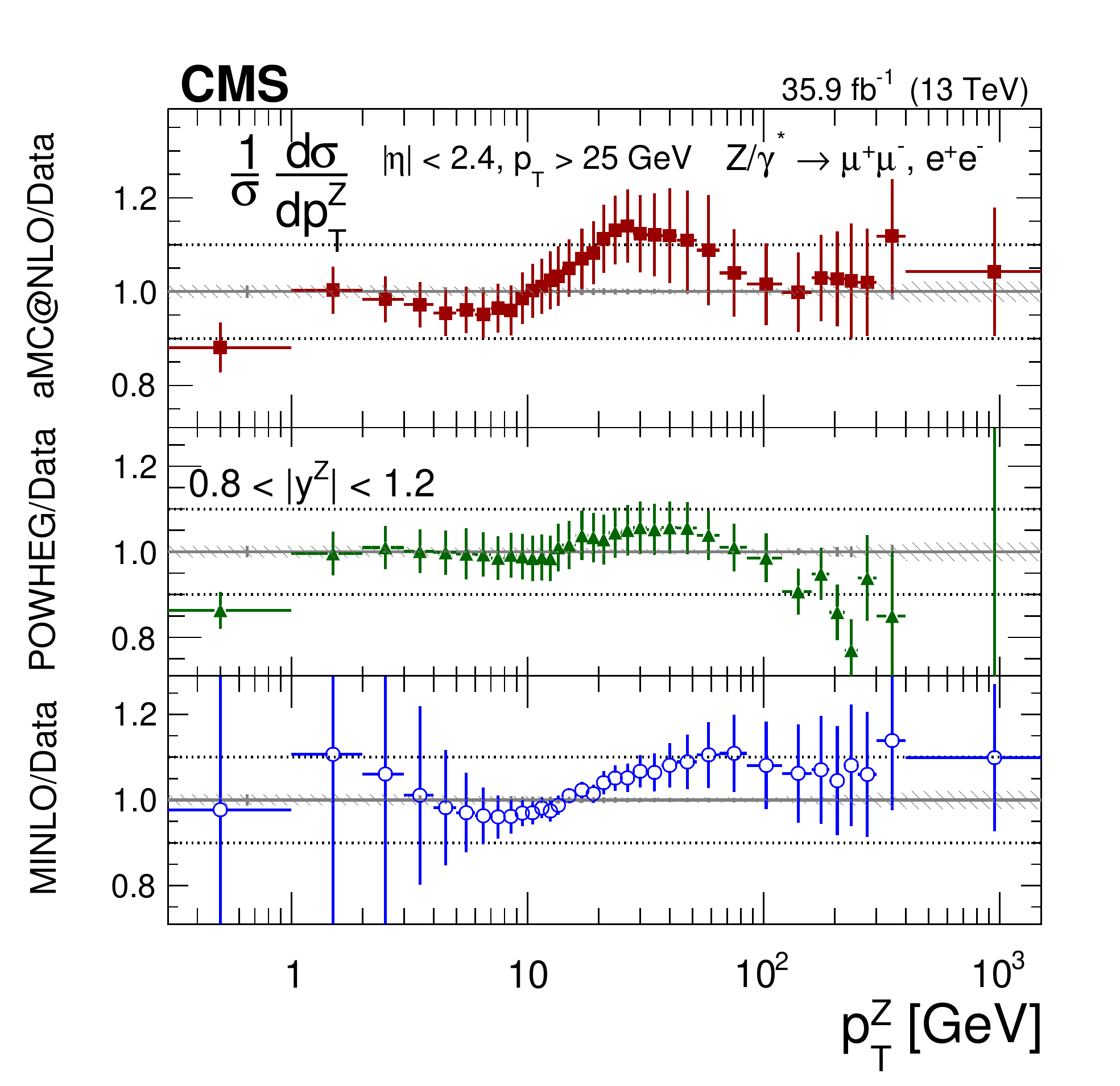}
	\caption{The measured normalized cross sections (left) in bins of $\pt^{\cPZ}$ for the $0.8 < \abs{\rapidity^{\cPZ}} < 1.2$ region. The ratios of the predictions to the data are also shown (right). The shaded bands around the data points (black) correspond to the total experimental uncertainty. The measurement is compared to the predictions with \MGvATNLO (square red markers),  $\POWHEG$ (green triangles), and $\POWHEG$-\textsc{MINLO} (blue circles). The error bands around the predictions correspond to the combined statistical, PDF, and scale uncertainties.}
	\label{fig:zll_norm2}
\end{figure}

\begin{figure}
	\centering
	\includegraphics[width=0.45\textwidth]{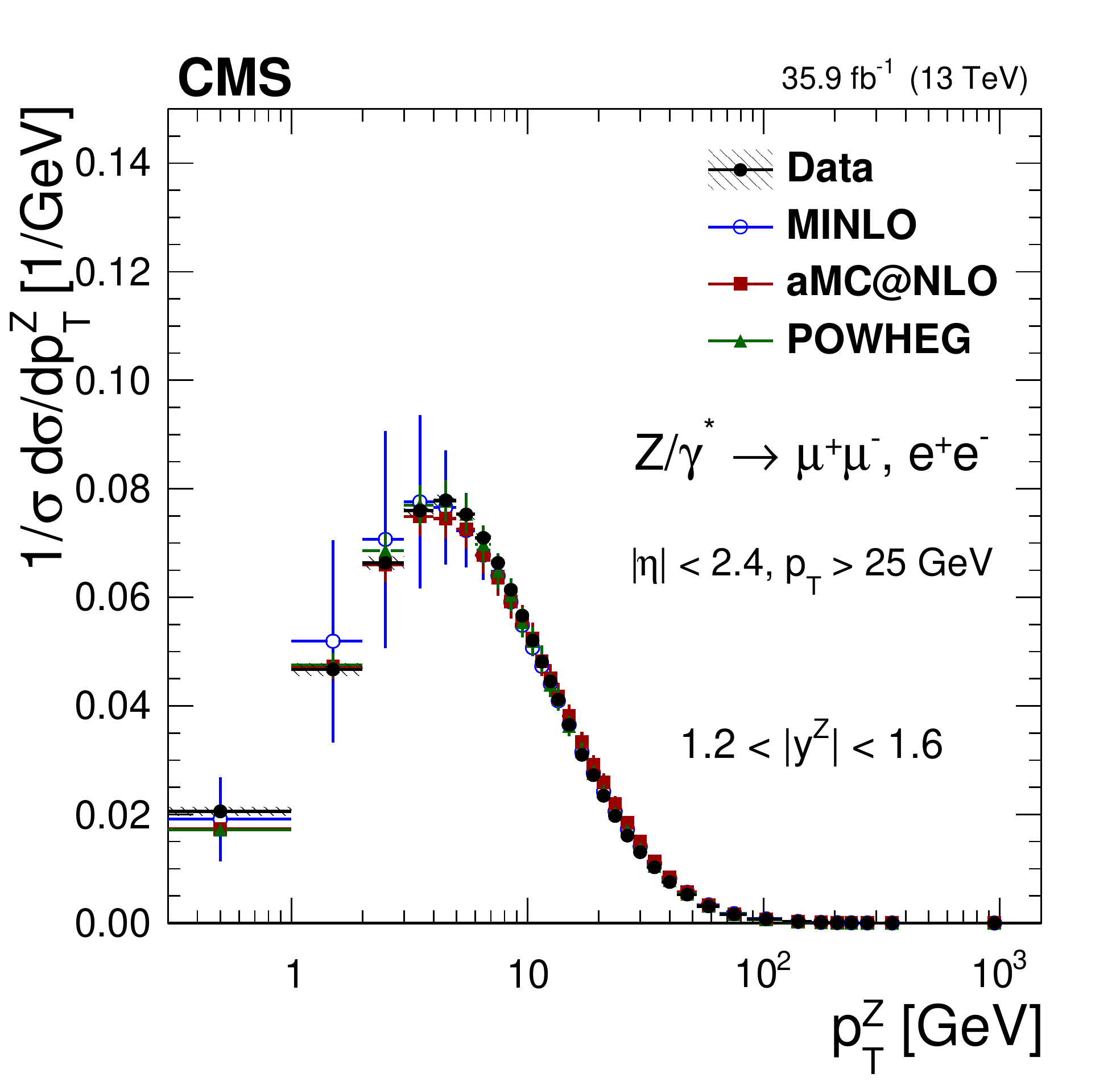}
        \includegraphics[width=0.45\textwidth]{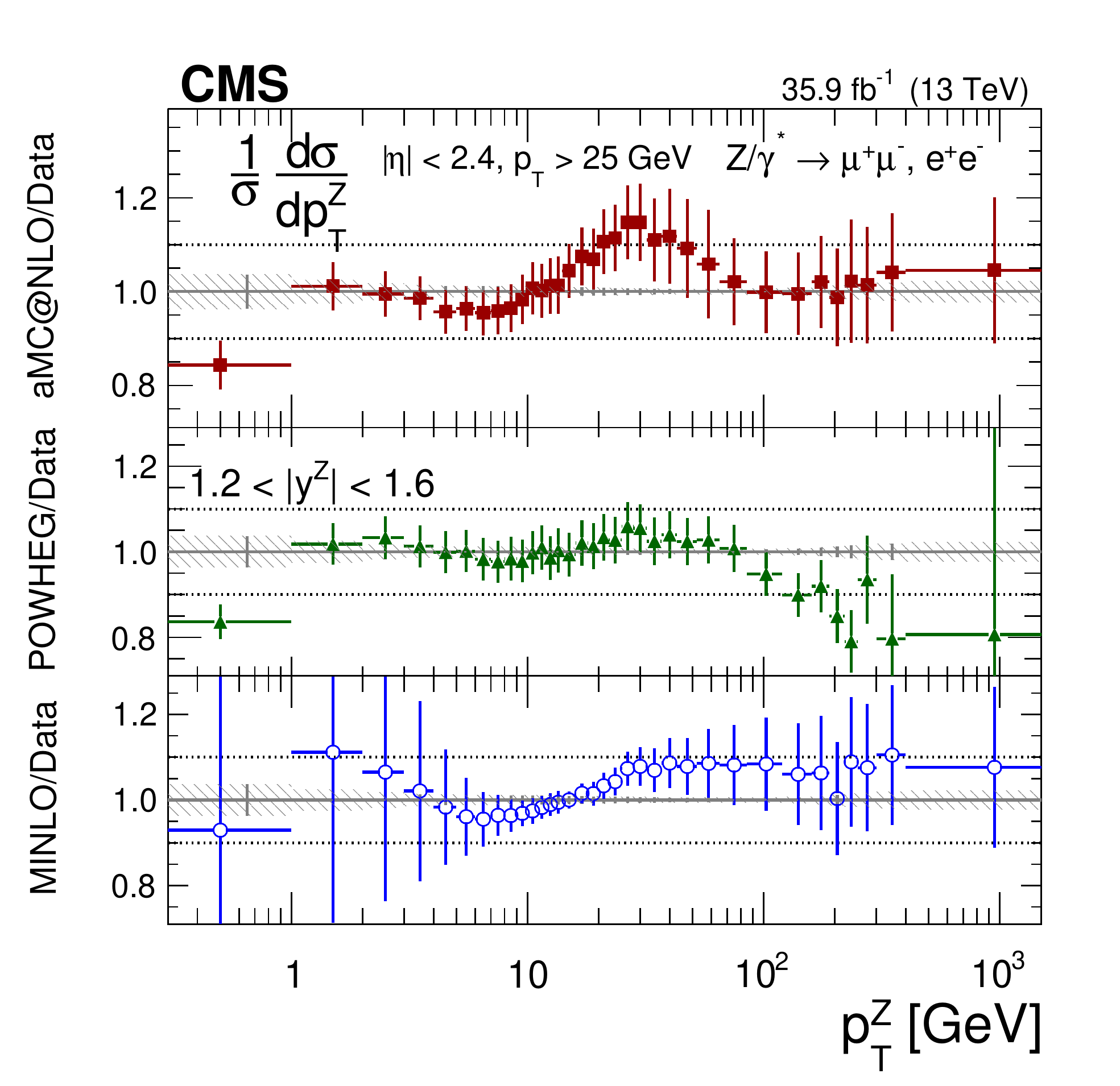}
	\caption{The measured normalized cross sections (left) in bins of $\pt^{\cPZ}$ for the $1.2 < \abs{\rapidity^{\cPZ}} < 1.6$ region. The ratios of the predictions to the data are also shown (right). The shaded bands around the data points (black) correspond to the total experimental uncertainty. The measurement is compared to the predictions with \MGvATNLO (square red markers),  $\POWHEG$ (green triangles), and $\POWHEG$-\textsc{MINLO} (blue circles). The error bands around the predictions correspond to the combined statistical, PDF, and scale uncertainties.}
	\label{fig:zll_norm3}
\end{figure}

\begin{figure}
	\centering
	\includegraphics[width=0.45\textwidth]{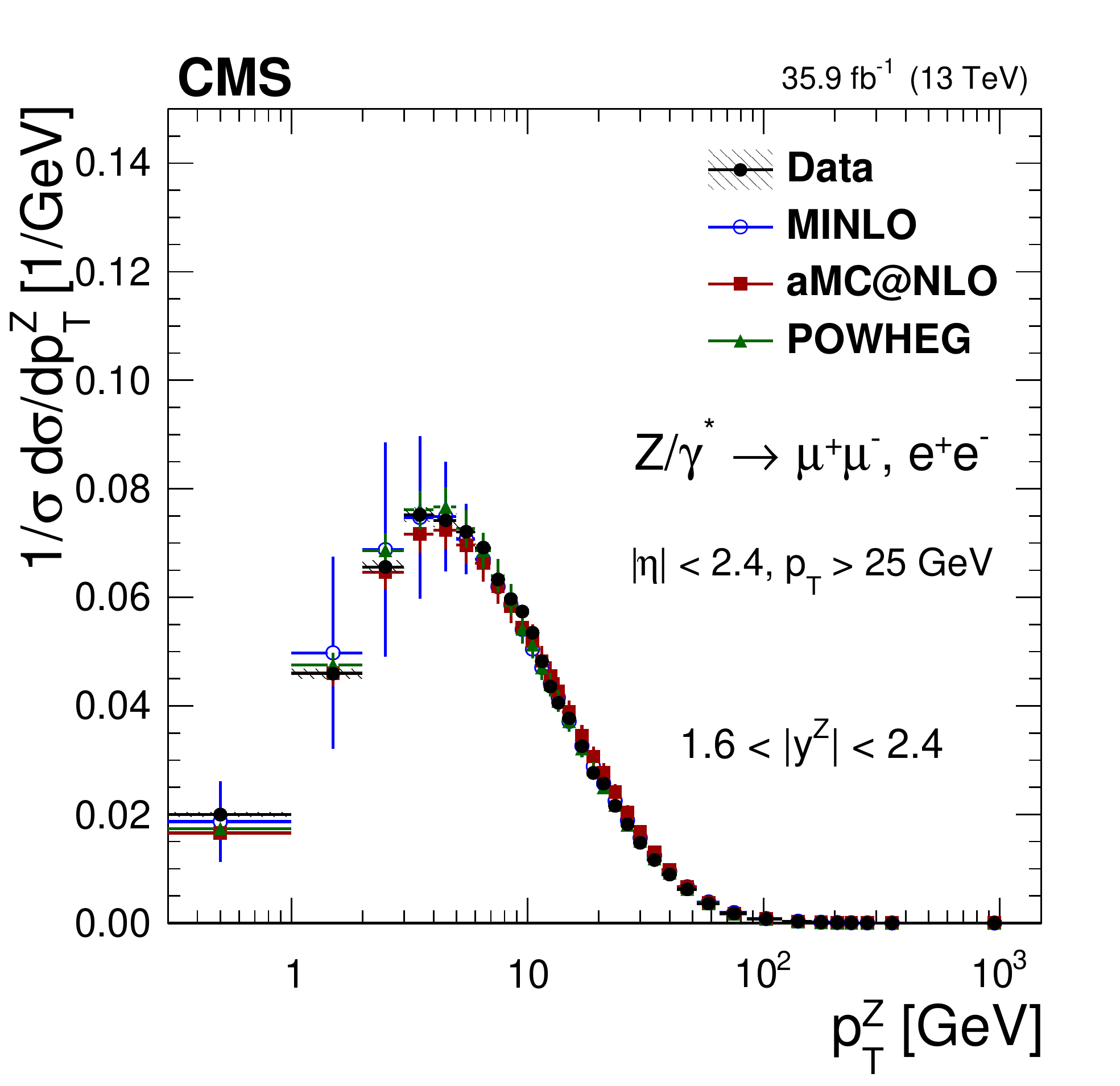}
        \includegraphics[width=0.45\textwidth]{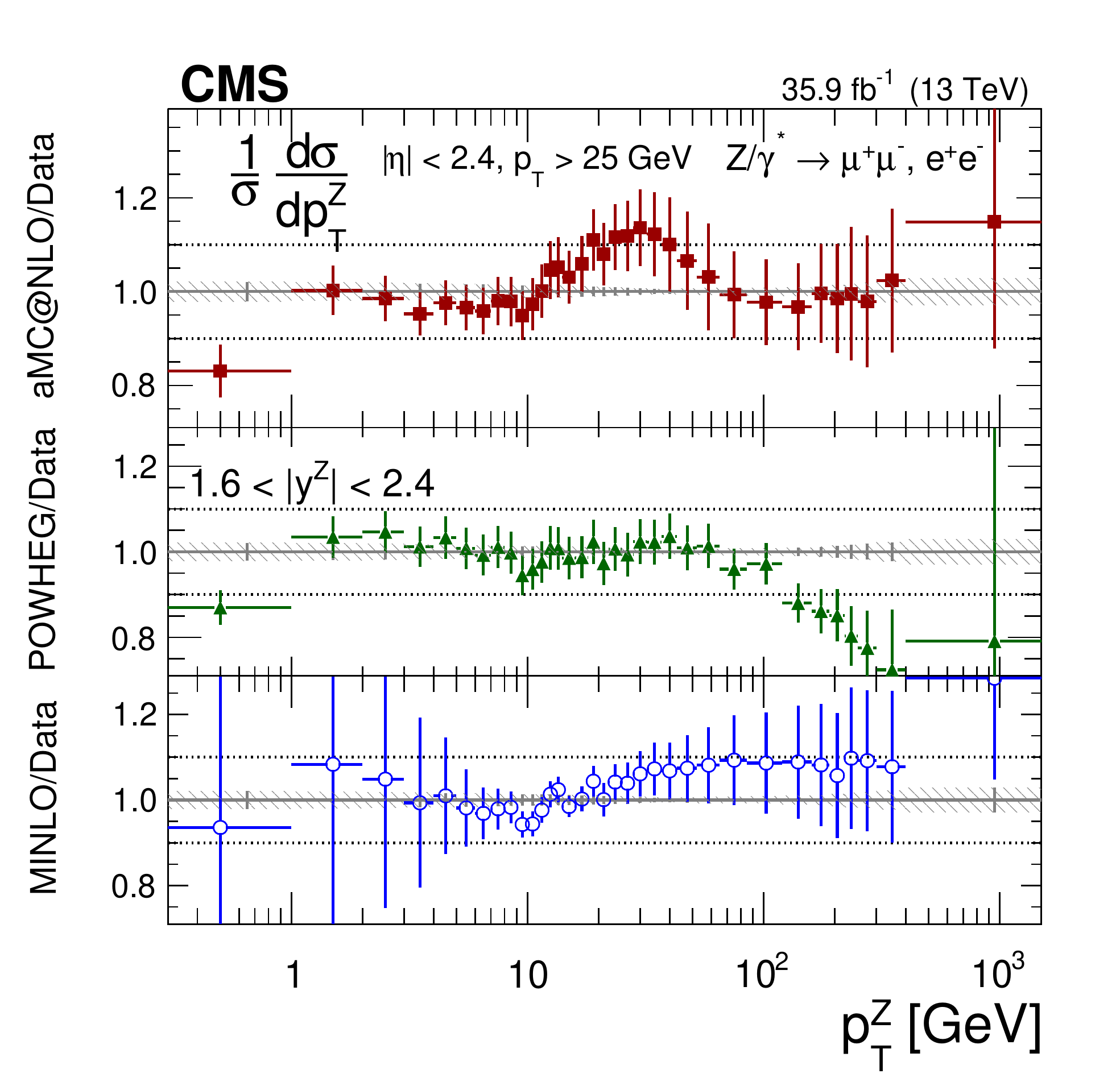}
	\caption{The measured normalized cross sections (left) in bins of $\pt^{\cPZ}$ for the $1.6 < \abs{\rapidity^{\cPZ}} < 2.4$ region. The ratios of the predictions to the data are also shown (right). The shaded bands around the data points (black) correspond to the total experimental uncertainty. The measurement is compared to the predictions with \MGvATNLO (square red markers),  $\POWHEG$ (green triangles), and $\POWHEG$-\textsc{MINLO} (blue circles). The error bands around the predictions correspond to the combined statistical, PDF, and scale uncertainties.}
	\label{fig:zll_norm4}
\end{figure}

Summaries of the absolute double-differential cross section measurements in
$\pt^{\cPZ}$ and $\abs{\rapidity^{\cPZ}}$ are shown in Figs.~\ref{fig:zll_double0}--\ref{fig:zll_double4}.
The normalized cross section measurements in bins of $\pt^{\cPZ}$, $\phiStar$, and $\abs{\rapidity^{\cPZ}}$
are shown in Fig.~\ref{fig:cross_norm}. The measured normalized cross section  uncertainties are
smaller than 0.5\% for $\phiStar<0.5$ and for $\pt^{\cPZ} < 50\GeV$. Summaries of the normalized
double-differential cross section measurements in $\pt^{\cPZ}$ and $\abs{\rapidity^{\cPZ}}$ are shown in Figs.~\ref{fig:zll_norm0}--\ref{fig:zll_norm4}. The cross sections are individually normalized in each $\abs{\rapidity^{\cPZ}}$ region. The measurements are compared to the predictions using parton shower
modeling with \MGvATNLO, $\POWHEG$, and $\POWHEG$-\textsc{MINLO}.
The predictions are consistent with the measurements within the theoretical uncertainties, 
although there is a trend of discrepancy of about 10\% 
in the range $20 < \pt^{\cPZ} < 60 \GeV$.

\section{Summary}
Measurements are reported of the differential cross sections for $\cPZ$ bosons produced in proton-proton collisions at $\sqrt{s}$ = 13\TeV and decaying to muons and electrons. The data set used corresponds to an integrated luminosity of  $35.9\fbinv$. Distributions of the transverse momentum \pt, the angular variable $\phi^{*}$, and the rapidity of
lepton pairs are measured. The results are corrected for detector effects and
compared to various theoretical predictions. The measurements provide sensitive
tests of theoretical predictions using fixed-order, resummed, and parton shower
calculations. The uncertainties in the normalized cross section measurements are
smaller than 0.5\% for $\phiStar < 0.5$ and for $\pt^{\cPZ} < 50\GeV$.
\clearpage
\begin{acknowledgments}
We congratulate our colleagues in the CERN accelerator departments for the excellent performance of the LHC and thank the technical and administrative staffs at CERN and at other CMS institutes for their contributions to the success of the CMS effort. In addition, we gratefully acknowledge the computing centers and personnel of the Worldwide LHC Computing Grid for delivering so effectively the computing infrastructure essential to our analyses. Finally, we acknowledge the enduring support for the construction and operation of the LHC and the CMS detector provided by the following funding agencies: BMBWF and FWF (Austria); FNRS and FWO (Belgium); CNPq, CAPES, FAPERJ, FAPERGS, and FAPESP (Brazil); MES (Bulgaria); CERN; CAS, MoST, and NSFC (China); COLCIENCIAS (Colombia); MSES and CSF (Croatia); RPF (Cyprus); SENESCYT (Ecuador); MoER, ERC IUT, PUT and ERDF (Estonia); Academy of Finland, MEC, and HIP (Finland); CEA and CNRS/IN2P3 (France); BMBF, DFG, and HGF (Germany); GSRT (Greece); NKFIA (Hungary); DAE and DST (India); IPM (Iran); SFI (Ireland); INFN (Italy); MSIP and NRF (Republic of Korea); MES (Latvia); LAS (Lithuania); MOE and UM (Malaysia); BUAP, CINVESTAV, CONACYT, LNS, SEP, and UASLP-FAI (Mexico); MOS (Montenegro); MBIE (New Zealand); PAEC (Pakistan); MSHE and NSC (Poland); FCT (Portugal); JINR (Dubna); MON, RosAtom, RAS, RFBR, and NRC KI (Russia); MESTD (Serbia); SEIDI, CPAN, PCTI, and FEDER (Spain); MOSTR (Sri Lanka); Swiss Funding Agencies (Switzerland); MST (Taipei); ThEPCenter, IPST, STAR, and NSTDA (Thailand); TUBITAK and TAEK (Turkey); NASU and SFFR (Ukraine); STFC (United Kingdom); DOE and NSF (USA).

\hyphenation{Rachada-pisek} Individuals have received support from the Marie-Curie program and the European Research Council and Horizon 2020 Grant, contract Nos.\ 675440, 752730, and 765710 (European Union); the Leventis Foundation; the A.P.\ Sloan Foundation; the Alexander von Humboldt Foundation; the Belgian Federal Science Policy Office; the Fonds pour la Formation \`a la Recherche dans l'Industrie et dans l'Agriculture (FRIA-Belgium); the Agentschap voor Innovatie door Wetenschap en Technologie (IWT-Belgium); the F.R.S.-FNRS and FWO (Belgium) under the ``Excellence of Science -- EOS" -- be.h project n.\ 30820817; the Beijing Municipal Science \& Technology Commission, No. Z181100004218003; the Ministry of Education, Youth and Sports (MEYS) of the Czech Republic; the Lend\"ulet (``Momentum") Program and the J\'anos Bolyai Research Scholarship of the Hungarian Academy of Sciences, the New National Excellence Program \'UNKP, the NKFIA research grants 123842, 123959, 124845, 124850, 125105, 128713, 128786, and 129058 (Hungary); the Council of Science and Industrial Research, India; the HOMING PLUS program of the Foundation for Polish Science, cofinanced from European Union, Regional Development Fund, the Mobility Plus program of the Ministry of Science and Higher Education, the National Science Center (Poland), contracts Harmonia 2014/14/M/ST2/00428, Opus 2014/13/B/ST2/02543, 2014/15/B/ST2/03998, and 2015/19/B/ST2/02861, Sonata-bis 2012/07/E/ST2/01406; the National Priorities Research Program by Qatar National Research Fund; the Ministry of Science and Education, grant no. 3.2989.2017 (Russia); the Programa Estatal de Fomento de la Investigaci{\'o}n Cient{\'i}fica y T{\'e}cnica de Excelencia Mar\'{\i}a de Maeztu, grant MDM-2015-0509 and the Programa Severo Ochoa del Principado de Asturias; the Thalis and Aristeia programs cofinanced by EU-ESF and the Greek NSRF; the Rachadapisek Sompot Fund for Postdoctoral Fellowship, Chulalongkorn University and the Chulalongkorn Academic into Its 2nd Century Project Advancement Project (Thailand); the Welch Foundation, contract C-1845; and the Weston Havens Foundation (USA). \end{acknowledgments}

\bibliography{auto_generated}

\providecommand{\href}[2]{#2}\begingroup\raggedright\begin{thebibliography}{10}%
\makeatletter
\providecommand{\hrefCMSnoop }[0]{\@secondoftwo}%
\makeatother
\providecommand{\doi}{\texttt{doi:}\begingroup \urlstyle{tt}\Url}

\bibitem{Melnikov:2006kv}
\hrefCMSnoop {}{K.~Melnikov and F.~Petriello, ``{Electroweak gauge boson
  production at hadron colliders through
  $\mathcal{O}(\alpha_{\mathrm{s}}^2)$}'',} \textit{ Phys. Rev. D} \textbf{ 74}
  (2006) 114017,
  \href{http://dx.doi.org/10.1103/PhysRevD.74.114017}{\doi{10.1103/PhysRevD.74.114017}},
\href{http://www.arXiv.org/abs/hep-ph/0609070}{\texttt{arXiv:hep-ph/0609070}}.

\bibitem{Catani:2009sm}
S.~Catani\hrefCMSnoop {}{ {et~al.}, ``{Vector boson production at hadron
  colliders: a fully exclusive QCD calculation at NNLO}'',} \textit{ Phys. Rev.
  Lett.} \textbf{ 103} (2009) 082001,
  \href{http://dx.doi.org/10.1103/PhysRevLett.103.082001}{\doi{10.1103/PhysRevLett.103.082001}},
\href{http://www.arXiv.org/abs/0903.2120}{\texttt{arXiv:0903.2120}}.

\bibitem{Ridder:2015dxa}
A.~Gehrmann-De~Ridder\hrefCMSnoop {}{ {et~al.}, ``{Precise QCD predictions for
  the production of a Z boson in association with a hadronic jet}'',} \textit{
  Phys. Rev. Lett.} \textbf{ 117} (2016) 022001,
  \href{http://dx.doi.org/10.1103/PhysRevLett.117.022001}{\doi{10.1103/PhysRevLett.117.022001}},
\href{http://www.arXiv.org/abs/1507.02850}{\texttt{arXiv:1507.02850}}.

\bibitem{Boughezal:2015ded}
R.~Boughezal\hrefCMSnoop {}{ {et~al.}, ``{Z-boson production in association
  with a jet at next-to-next-to-leading order in perturbative QCD}'',} \textit{
  Phys. Rev. Lett.} \textbf{ 116} (2016) 152001,
  \href{http://dx.doi.org/10.1103/PhysRevLett.116.152001}{\doi{10.1103/PhysRevLett.116.152001}},
\href{http://www.arXiv.org/abs/1512.01291}{\texttt{arXiv:1512.01291}}.

\bibitem{Boughezal:2015dva}
\hrefCMSnoop {}{R.~Boughezal, C.~Focke, X.~Liu, and F.~Petriello, ``{$W$-boson
  production in association with a jet at next-to-next-to-leading order in
  perturbative QCD}'',} \textit{ Phys. Rev. Lett.} \textbf{ 115} (2015) 062002,
  \href{http://dx.doi.org/10.1103/PhysRevLett.115.062002}{\doi{10.1103/PhysRevLett.115.062002}},
\href{http://www.arXiv.org/abs/1504.02131}{\texttt{arXiv:1504.02131}}.

\bibitem{Dittmaier:2014qza}
\hrefCMSnoop {}{S.~Dittmaier, A.~Huss, and C.~Schwinn, ``{Mixed QCD-electroweak
  $\mathcal{O}(\alpha_s\alpha)$ corrections to Drell-Yan processes in the
  resonance region: pole approximation and non-factorizable corrections}'',}
  \textit{ Nucl. Phys. B} \textbf{ 885} (2014) 318,
  \href{http://dx.doi.org/10.1016/j.nuclphysb.2014.05.027}{\doi{10.1016/j.nuclphysb.2014.05.027}},
\href{http://www.arXiv.org/abs/1403.3216}{\texttt{arXiv:1403.3216}}.

\bibitem{Lindert:2017olm}
\hrefCMSnoop {}{J.~M. Lindert {et~al.}, ``{Precise predictions for V+jets dark
  matter backgrounds}'',} \textit{ Eur. Phys. J. C} \textbf{ 77} (2017) 829,
  \href{http://dx.doi.org/10.1140/epjc/s10052-017-5389-1}{\doi{10.1140/epjc/s10052-017-5389-1}},
\href{http://www.arXiv.org/abs/1705.04664}{\texttt{arXiv:1705.04664}}.

\bibitem{Collins:1984kg}
\hrefCMSnoop {}{J.~C. Collins, D.~E. Soper, and G.~F. Sterman, ``{Transverse
  momentum distribution in Drell-Yan pair and W and Z boson production}'',}
  \textit{ Nucl. Phys. B} \textbf{ 250} (1985)
\href{http://dx.doi.org/10.1016/0550-3213(85)90479-1}{\doi{10.1016/0550-3213(85)90479-1}}.

\bibitem{Balazs:1995nz}
\hrefCMSnoop {}{C.~Balazs, J.-W. Qiu, and C.~P. Yuan, ``{Effects of QCD
  resummation on distributions of leptons from the decay of electroweak vector
  bosons}'',} \textit{ Phys. Lett. B} \textbf{ 355} (1995) 548,
  \href{http://dx.doi.org/10.1016/0370-2693(95)00726-2}{\doi{10.1016/0370-2693(95)00726-2}},
\href{http://www.arXiv.org/abs/hep-ph/9505203}{\texttt{arXiv:hep-ph/9505203}}.

\bibitem{Catani:2015vma}
\hrefCMSnoop {}{S.~Catani, D.~de~Florian, G.~Ferrera, and M.~Grazzini,
  ``{Vector boson production at hadron colliders: transverse-momentum
  resummation and leptonic decay}'',} \textit{ JHEP} \textbf{ 12} (2015) 047,
  \href{http://dx.doi.org/10.1007/JHEP12(2015)047}{\doi{10.1007/JHEP12(2015)047}},
\href{http://www.arXiv.org/abs/1507.06937}{\texttt{arXiv:1507.06937}}.

\bibitem{Sjostrand:2014zea}
T.~Sj{\"o}strand\hrefCMSnoop {}{ {et~al.}, ``{An introduction to PYTHIA
  8.2}'',} \textit{ Comput. Phys. Commun.} \textbf{ 191} (2015) 159,
  \href{http://dx.doi.org/10.1016/j.cpc.2015.01.024}{\doi{10.1016/j.cpc.2015.01.024}},
\href{http://www.arXiv.org/abs/1410.3012}{\texttt{arXiv:1410.3012}}.

\bibitem{Gleisberg:2008ta}
T.~Gleisberg\hrefCMSnoop {}{ {et~al.}, ``{Event generation with SHERPA 1.1}'',}
  \textit{ JHEP} \textbf{ 02} (2009) 007,
  \href{http://dx.doi.org/10.1088/1126-6708/2009/02/007}{\doi{10.1088/1126-6708/2009/02/007}},
\href{http://www.arXiv.org/abs/0811.4622}{\texttt{arXiv:0811.4622}}.

\bibitem{Bahr:2008pv}
M.~B{\"a}hr\hrefCMSnoop {}{ {et~al.}, ``Herwig++ physics and manual'',}
  \textit{ Eur. Phys. J. C} \textbf{ 58} (2008) 639,
  \href{http://dx.doi.org/10.1140/epjc/s10052-008-0798-9}{\doi{10.1140/epjc/s10052-008-0798-9}},
\href{http://www.arXiv.org/abs/0803.0883}{\texttt{arXiv:0803.0883}}.

\bibitem{Nason:2004rx}
\hrefCMSnoop {}{P.~Nason, ``A new method for combining {NLO QCD} with shower
  {Monte Carlo} algorithms'',} \textit{ JHEP} \textbf{ 11} (2004) 040,
  \href{http://dx.doi.org/10.1088/1126-6708/2004/11/040}{\doi{10.1088/1126-6708/2004/11/040}},
\href{http://www.arXiv.org/abs/hep-ph/0409146}{\texttt{arXiv:hep-ph/0409146}}.

\bibitem{Frixione:2002ik}
\hrefCMSnoop {}{S.~Frixione and B.~R. Webber, ``{Matching NLO QCD computations
  and parton shower simulations}'',} \textit{ JHEP} \textbf{ 06} (2002) 029,
  \href{http://dx.doi.org/10.1088/1126-6708/2002/06/029}{\doi{10.1088/1126-6708/2002/06/029}},
\href{http://www.arXiv.org/abs/hep-ph/0204244}{\texttt{arXiv:hep-ph/0204244}}.

\bibitem{Alioli:2010xd}
\hrefCMSnoop {}{S.~Alioli, P.~Nason, C.~Oleari, and E.~Re, ``{A general
  framework for implementing NLO calculations in shower Monte Carlo programs:
  the POWHEG BOX}'',} \textit{ JHEP} \textbf{ 06} (2010) 043,
  \href{http://dx.doi.org/10.1007/JHEP06(2010)043}{\doi{10.1007/JHEP06(2010)043}},
\href{http://www.arXiv.org/abs/1002.2581}{\texttt{arXiv:1002.2581}}.

\bibitem{Alwall:2014hca}
J.~Alwall\hrefCMSnoop {}{ {et~al.}, ``{The automated computation of tree-level
  and next-to-leading order differential cross sections, and their matching to
  parton shower simulations}'',} \textit{ JHEP} \textbf{ 07} (2014) 079,
  \href{http://dx.doi.org/10.1007/JHEP07(2014)079}{\doi{10.1007/JHEP07(2014)079}},
\href{http://www.arXiv.org/abs/1405.0301}{\texttt{arXiv:1405.0301}}.

\bibitem{Angeles-Martinez:2015sea}
\hrefCMSnoop {}{R.~Angeles-Martinez {et~al.}, ``{Transverse momentum dependent
  (TMD) parton distribution functions: status and prospects}'',} \textit{ Acta
  Phys. Polon. B} \textbf{ 46} (2015) 2501,
  \href{http://dx.doi.org/10.5506/APhysPolB.46.2501}{\doi{10.5506/APhysPolB.46.2501}},
\href{http://www.arXiv.org/abs/1507.05267}{\texttt{arXiv:1507.05267}}.

\bibitem{ATLAS_ZpT7TeV}
\hrefCMSnoop {}{{ATLAS Collaboration}, ``{Measurement of the transverse
  momentum distribution of Z$/\PGg^*$ bosons in proton-proton collisions at
  $\sqrt{s} = 7\TeV$ with the ATLAS detector}'',} \textit{ Phys. Lett. B}
  \textbf{ 705} (2011) 415,
  \href{http://dx.doi.org/10.1016/j.physletb.2011.10.018}{\doi{10.1016/j.physletb.2011.10.018}},
\href{http://www.arXiv.org/abs/1107.2381}{\texttt{arXiv:1107.2381}}.

\bibitem{ATLAS_ZptEta7TeV}
\hrefCMSnoop {}{{ATLAS Collaboration}, ``{Measurement of the Z$/\PGg^*$ boson
  transverse momentum distribution in $\Pp\Pp$ collisions at $\sqrt{s} = 7\TeV$
  with the ATLAS detector}'',} \textit{ JHEP} \textbf{ 09} (2014) 145,
  \href{http://dx.doi.org/10.1007/JHEP09(2014)145}{\doi{10.1007/JHEP09(2014)145}},
\href{http://www.arXiv.org/abs/1406.3660}{\texttt{arXiv:1406.3660}}.

\bibitem{Aad:2015auj}
\hrefCMSnoop {}{{ATLAS Collaboration}, ``{Measurement of the transverse
  momentum and $\phi ^*_{\eta }$ distributions of Drell-Yan lepton pairs in
  proton-proton collisions at $\sqrt{s}=8$ TeV with the ATLAS detector}'',}
  \textit{ Eur. Phys. J. C} \textbf{ 76} (2016) 291,
  \href{http://dx.doi.org/10.1140/epjc/s10052-016-4070-4}{\doi{10.1140/epjc/s10052-016-4070-4}},
\href{http://www.arXiv.org/abs/1512.02192}{\texttt{arXiv:1512.02192}}.

\bibitem{Aaboud:2016btc}
\hrefCMSnoop {}{{ATLAS Collaboration}, ``{Precision measurement and
  interpretation of inclusive $W^+$ , $W^-$ and $Z/\gamma ^*$ production cross
  sections with the ATLAS detector}'',} \textit{ Eur. Phys. J. C} \textbf{ 77}
  (2017) 367,
  \href{http://dx.doi.org/10.1140/epjc/s10052-017-4911-9}{\doi{10.1140/epjc/s10052-017-4911-9}},
\href{http://www.arXiv.org/abs/1612.03016}{\texttt{arXiv:1612.03016}}.

\bibitem{Sirunyan:2018owv}
\hrefCMSnoop {}{{CMS Collaboration}, ``{Measurement of the differential
  Drell-Yan cross section in proton-proton collisions at $\sqrt{s} =$ 13
  TeV}'',} (2018).
  \href{http://www.arXiv.org/abs/1812.10529}{\texttt{arXiv:1812.10529}}.
Submitted to \textsc{JHEP}.

\bibitem{CMS_ZpT7TeV}
\hrefCMSnoop {}{{CMS Collaboration}, ``{Measurement of the rapidity and
  transverse momentum distributions of Z bosons in $\Pp\Pp$ collisions at
  $\sqrt{s} = 7\TeV$}'',} \textit{ Phys. Rev. D} \textbf{ 85} (2012) 032002,
  \href{http://dx.doi.org/10.1103/PhysRevD.85.032002}{\doi{10.1103/PhysRevD.85.032002}},
  \href{http://www.arXiv.org/abs/1110.4973}{\texttt{arXiv:1110.4973}}.

\bibitem{CMS_ZpT8TeV}
\hrefCMSnoop {}{{CMS Collaboration}, ``{Measurement of the Z boson differential
  cross section in transverse momentum and rapidity in proton-proton collisions
  at 8\TeV}'',} \textit{ Phys. Lett. B} \textbf{ 749} (2015) 187,
  \href{http://dx.doi.org/10.1016/j.physletb.2015.07.065}{\doi{10.1016/j.physletb.2015.07.065}},
\href{http://www.arXiv.org/abs/1504.03511}{\texttt{arXiv:1504.03511}}.

\bibitem{CMS:2014jea}
\hrefCMSnoop {}{{CMS Collaboration}, ``{Measurements of differential and
  double-differential Drell-Yan cross sections in proton-proton collisions at
  $\sqrt{s}= 8$ ~TeV}'',} \textit{ Eur. Phys. J. C} \textbf{ 75} (2015) 147,
  \href{http://dx.doi.org/10.1140/epjc/s10052-015-3364-2}{\doi{10.1140/epjc/s10052-015-3364-2}},
\href{http://www.arXiv.org/abs/1412.1115}{\texttt{arXiv:1412.1115}}.

\bibitem{Khachatryan:2016nbe}
\hrefCMSnoop {}{{CMS Collaboration}, ``{Measurement of the transverse momentum
  spectra of weak vector bosons produced in proton-proton collisions at $
  \sqrt{s}=8 $ TeV}'',} \textit{ JHEP} \textbf{ 02} (2017) 096,
  \href{http://dx.doi.org/10.1007/JHEP02(2017)096}{\doi{10.1007/JHEP02(2017)096}},
\href{http://www.arXiv.org/abs/1606.05864}{\texttt{arXiv:1606.05864}}.

\bibitem{LHCb_WZ7TeV}
\hrefCMSnoop {}{{LHCb Collaboration}, ``{Inclusive $\PW$ and Z production in
  the forward region at $\sqrt{s} = 7\TeV$}'',} \textit{ JHEP} \textbf{ 06}
  (2012) 058,
  \href{http://dx.doi.org/10.1007/JHEP06(2012)058}{\doi{10.1007/JHEP06(2012)058}},
  \href{http://www.arXiv.org/abs/1204.1620}{\texttt{arXiv:1204.1620}}.

\bibitem{LHCb_Zee7TeV}
\hrefCMSnoop {}{{LHCb Collaboration}, ``{Measurement of the cross-section for
  Z$\rightarrow \rm{e^+e^-}$ production in pp collisions at $\sqrt{s} =
  7\TeV$}'',} \textit{ JHEP} \textbf{ 02} (2013) 106,
  \href{http://dx.doi.org/10.1007/JHEP02(2013)106}{\doi{10.1007/JHEP02(2013)106}},
  \href{http://www.arXiv.org/abs/1212.4620}{\texttt{arXiv:1212.4620}}.

\bibitem{LHCb_ZpT7TeV}
\hrefCMSnoop {}{{LHCb Collaboration}, ``{Measurement of the forward Z boson
  production cross-section in pp collisions at $\sqrt{s} = 7\TeV$}'',} \textit{
  JHEP} \textbf{ 08} (2015) 039,
  \href{http://dx.doi.org/10.1007/JHEP08(2015)039}{\doi{10.1007/JHEP08(2015)039}},
  \href{http://www.arXiv.org/abs/1505.07024}{\texttt{arXiv:1505.07024}}.

\bibitem{LHCb_WZ8TeV}
\hrefCMSnoop {}{{LHCb Collaboration}, ``{Measurement of forward W and Z boson
  production in pp collisions at $\sqrt{s} = 8\TeV$}'',} \textit{ JHEP}
  \textbf{ 01} (2016) 155,
  \href{http://dx.doi.org/10.1007/JHEP01(2016)155}{\doi{10.1007/JHEP01(2016)155}},
  \href{http://www.arXiv.org/abs/1511.08039}{\texttt{arXiv:1511.08039}}.

\bibitem{Aaij:2016mgv}
\hrefCMSnoop {}{{LHCb Collaboration}, ``{Measurement of the forward Z boson
  production cross-section in pp collisions at $\sqrt{s} = 13$ TeV}'',}
  \textit{ JHEP} \textbf{ 09} (2016) 136,
  \href{http://dx.doi.org/10.1007/JHEP09(2016)136}{\doi{10.1007/JHEP09(2016)136}},
\href{http://www.arXiv.org/abs/1607.06495}{\texttt{arXiv:1607.06495}}.

\bibitem{Affolder:1999jh}
\hrefCMSnoop {}{{CDF} Collaboration, ``{The transverse momentum and total cross
  section of $\rm{ e^+e^-}$ pairs in the Z boson region from $\Pp\bar{\Pp}$
  collisions at $\sqrt{s} = 1.8$ TeV}'',} \textit{ Phys. Rev. Lett.} \textbf{
  84} (2000) 845,
  \href{http://dx.doi.org/10.1103/PhysRevLett.84.845}{\doi{10.1103/PhysRevLett.84.845}},
\href{http://www.arXiv.org/abs/hep-ex/0001021}{\texttt{arXiv:hep-ex/0001021}}.

\bibitem{Abbott:1999yd}
\hrefCMSnoop {}{{D0} Collaboration, ``{Differential production cross section of
  Z bosons as a function of transverse momentum at $\sqrt{s} = 1.8$ TeV}'',}
  \textit{ Phys. Rev. Lett.} \textbf{ 84} (2000) 2792,
  \href{http://dx.doi.org/10.1103/PhysRevLett.84.2792}{\doi{10.1103/PhysRevLett.84.2792}},
\href{http://www.arXiv.org/abs/hep-ex/9909020}{\texttt{arXiv:hep-ex/9909020}}.

\bibitem{TevatronWZ:D0PhysRevLett2008_100}
\hrefCMSnoop {}{{D0} Collaboration, ``{Measurement of the shape of the boson
  transverse momentum distribution in $\Pp \bar{\Pp} \to \PZ / \gamma^{*} \to
  \Pe^+ \Pe^- + X$ events produced at $\sqrt{s}=1.96~\TeV$}'',} \textit{ Phys.
  Rev. Lett.} \textbf{ 100} (2008) 102002,
  \href{http://dx.doi.org/10.1103/PhysRevLett.100.102002}{\doi{10.1103/PhysRevLett.100.102002}},
\href{http://www.arXiv.org/abs/0712.0803}{\texttt{arXiv:0712.0803}}.

\bibitem{TevatronWZ:D0PhysLettB2010_693}
\hrefCMSnoop {}{{D0} Collaboration, ``{Measurement of the normalized
  $\PZ/\gamma^* \to \mu^+\mu^-$ transverse momentum distribution in
  $\Pp\bar{\Pp}$ collisions at $\sqrt{s}=1.96$ TeV}'',} \textit{ Phys. Lett. B}
  \textbf{ 693} (2010) 522,
  \href{http://dx.doi.org/10.1016/j.physletb.2010.09.012}{\doi{10.1016/j.physletb.2010.09.012}},
\href{http://www.arXiv.org/abs/1006.0618}{\texttt{arXiv:1006.0618}}.

\bibitem{TevatronWZ:D0PhysRevLett2011_106}
\hrefCMSnoop {}{{D0} Collaboration, ``{Precise study of the $\PZ/\gamma^*$
  boson transverse momentum distribution in $\Pp\bar{\Pp}$ collisions using a
  novel technique}'',} \textit{ Phys. Rev. Lett.} \textbf{ 106} (2011) 122001,
  \href{http://dx.doi.org/10.1103/PhysRevLett.106.122001}{\doi{10.1103/PhysRevLett.106.122001}},
\href{http://www.arXiv.org/abs/1010.0262}{\texttt{arXiv:1010.0262}}.

\bibitem{Banfi:2010cf}
A.~Banfi\hrefCMSnoop {}{ {et~al.}, ``Optimisation of variables for studying
  dilepton transverse momentum distributions at hadron colliders'',} \textit{
  Eur. Phys. J. C} \textbf{ 71} (2011) 1600,
  \href{http://dx.doi.org/10.1140/epjc/s10052-011-1600-y}{\doi{10.1140/epjc/s10052-011-1600-y}},
\href{http://www.arXiv.org/abs/1009.1580}{\texttt{arXiv:1009.1580}}.

\bibitem{Banfi:2012du}
\hrefCMSnoop {}{A.~Banfi, M.~Dasgupta, S.~Marzani, and L.~Tomlinson,
  ``{Predictions for Drell-Yan $\phi^*$ and $\mathrm{Q_T}$ observables at the
  LHC}'',} \textit{ Phys. Lett. B} \textbf{ 715} (2012) 152,
  \href{http://dx.doi.org/10.1016/j.physletb.2012.07.035}{\doi{10.1016/j.physletb.2012.07.035}},
\href{http://www.arXiv.org/abs/1205.4760}{\texttt{arXiv:1205.4760}}.

\bibitem{Sirunyan:2017igm}
\hrefCMSnoop {}{{CMS Collaboration}, ``{Measurement of differential cross
  sections in the kinematic angular variable $\phi^*$ for inclusive Z boson
  production in pp collisions at $\sqrt{s}=$ 8 TeV}'',} \textit{ JHEP} \textbf{
  03} (2018) 172,
  \href{http://dx.doi.org/10.1007/JHEP03(2018)172}{\doi{10.1007/JHEP03(2018)172}},
\href{http://www.arXiv.org/abs/1710.07955}{\texttt{arXiv:1710.07955}}.

\bibitem{Chatrchyan:2008zzk}
\hrefCMSnoop {}{{CMS Collaboration}, ``The {CMS} experiment at the {CERN}
  {LHC}'',} \textit{ JINST} \textbf{ 3} (2008) S08004,
  \href{http://dx.doi.org/10.1088/1748-0221/3/08/S08004}{\doi{10.1088/1748-0221/3/08/S08004}}.

\bibitem{Khachatryan:2016bia}
\hrefCMSnoop {}{{CMS Collaboration}, ``{The CMS trigger system}'',} \textit{
  JINST} \textbf{ 12} (2017) P01020,
  \href{http://dx.doi.org/10.1088/1748-0221/12/01/P01020}{\doi{10.1088/1748-0221/12/01/P01020}},
\href{http://www.arXiv.org/abs/1609.02366}{\texttt{arXiv:1609.02366}}.

\bibitem{Agostinelli:2002hh}
\hrefCMSnoop {}{{GEANT4} Collaboration, ``{\GEANTfour}---a simulation
  toolkit'',} \textit{ Nucl. Instrum. Meth. A} \textbf{ 506} (2003) 250,
  \href{http://dx.doi.org/10.1016/S0168-9002(03)01368-8}{\doi{10.1016/S0168-9002(03)01368-8}}.

\bibitem{Alioli:2008gx}
\hrefCMSnoop {}{S.~Alioli, P.~Nason, C.~Oleari, and E.~Re, ``{NLO} vector boson
  production matched with shower in {POWHEG}'',} \textit{ JHEP} \textbf{ 07}
  (2008) 060,
  \href{http://dx.doi.org/10.1088/1126-6708/2008/07/060}{\doi{10.1088/1126-6708/2008/07/060}},
\href{http://www.arXiv.org/abs/0805.4802}{\texttt{arXiv:0805.4802}}.

\bibitem{MCFM}
\hrefCMSnoop {}{J.~M. Campbell and R.~K. Ellis, ``{MCFM for the Tevatron and
  the LHC}'',} \textit{ Nucl. Phys. Proc. Suppl.} \textbf{ 205-206} (2010) 10,
  \href{http://dx.doi.org/10.1016/j.nuclphysbps.2010.08.011}{\doi{10.1016/j.nuclphysbps.2010.08.011}},
\href{http://www.arXiv.org/abs/1007.3492}{\texttt{arXiv:1007.3492}}.

\bibitem{Skands:2014pea}
\hrefCMSnoop {}{P.~Skands, S.~Carrazza, and J.~Rojo, ``Tuning {PYTHIA} 8.1: the
  {Monash} 2013 tune'',} \textit{ Eur. Phys. J. C} \textbf{ 74} (2014) 3024,
  \href{http://dx.doi.org/10.1140/epjc/s10052-014-3024-y}{\doi{10.1140/epjc/s10052-014-3024-y}},
  \href{http://www.arXiv.org/abs/1404.5630}{\texttt{arXiv:1404.5630}}.

\bibitem{Khachatryan:2015pea}
\hrefCMSnoop {}{{CMS Collaboration}, ``Event generator tunes obtained from
  underlying event and multiparton scattering measurements'',} \textit{ Eur.
  Phys. J. C} \textbf{ 76} (2016) 155,
  \href{http://dx.doi.org/10.1140/epjc/s10052-016-3988-x}{\doi{10.1140/epjc/s10052-016-3988-x}},
\href{http://www.arXiv.org/abs/1512.00815}{\texttt{arXiv:1512.00815}}.

\bibitem{Ball:2014uwa}
\hrefCMSnoop {}{{NNPDF} Collaboration, ``{Parton distributions for the LHC Run
  II}'',} \textit{ JHEP} \textbf{ 04} (2015) 040,
  \href{http://dx.doi.org/10.1007/JHEP04(2015)040}{\doi{10.1007/JHEP04(2015)040}},
\href{http://www.arXiv.org/abs/1410.8849}{\texttt{arXiv:1410.8849}}.

\bibitem{Sirunyan:2017ulk}
\hrefCMSnoop {}{{CMS Collaboration}, ``{Particle-flow reconstruction and global
  event description with the CMS detector}'',} \textit{ JINST} \textbf{ 12}
  (2017) P10003,
  \href{http://dx.doi.org/10.1088/1748-0221/12/10/P10003}{\doi{10.1088/1748-0221/12/10/P10003}},
\href{http://www.arXiv.org/abs/1706.04965}{\texttt{arXiv:1706.04965}}.

\bibitem{Cacciari:2008gp}
\hrefCMSnoop {}{M.~Cacciari, G.~P. Salam, and G.~Soyez, ``{The anti-\kt jet
  clustering algorithm}'',} \textit{ JHEP} \textbf{ 04} (2008) 063,
  \href{http://dx.doi.org/10.1088/1126-6708/2008/04/063}{\doi{10.1088/1126-6708/2008/04/063}},
\href{http://www.arXiv.org/abs/0802.1189}{\texttt{arXiv:0802.1189}}.

\bibitem{Cacciari:2011ma}
\hrefCMSnoop {}{M.~Cacciari, G.~P. Salam, and G.~Soyez, ``{FastJet} user
  manual'',} \textit{ Eur. Phys. J. C} \textbf{ 72} (2012) 1896,
  \href{http://dx.doi.org/10.1140/epjc/s10052-012-1896-2}{\doi{10.1140/epjc/s10052-012-1896-2}},
  \href{http://www.arXiv.org/abs/1111.6097}{\texttt{arXiv:1111.6097}}.

\bibitem{Chatrchyan:2012xi}
\hrefCMSnoop {}{{CMS Collaboration}, ``{Performance of CMS muon reconstruction
  in pp collision events at $\sqrt{s} = 7\TeV$}'',} \textit{ JINST} \textbf{ 7}
  (2012) P10002,
  \href{http://dx.doi.org/10.1088/1748-0221/7/10/P10002}{\doi{10.1088/1748-0221/7/10/P10002}},
\href{http://www.arXiv.org/abs/1206.4071}{\texttt{arXiv:1206.4071}}.

\bibitem{Sirunyan:2018fpa}
\hrefCMSnoop {}{{CMS Collaboration}, ``{Performance of the CMS muon detector
  and muon reconstruction with proton-proton collisions at $\sqrt{s}=$ 13
  TeV}'',} \textit{ JINST} \textbf{ 13} (2018) P06015,
  \href{http://dx.doi.org/10.1088/1748-0221/13/06/P06015}{\doi{10.1088/1748-0221/13/06/P06015}},
\href{http://www.arXiv.org/abs/1804.04528}{\texttt{arXiv:1804.04528}}.

\bibitem{Khachatryan:2015hwa}
\hrefCMSnoop {}{{CMS Collaboration}, ``{Performance of electron reconstruction
  and selection with the CMS detector in proton-proton collisions at $\sqrt{s}
  = 8\TeV$}'',} \textit{ JINST} \textbf{ 10} (2015) P06005,
  \href{http://dx.doi.org/10.1088/1748-0221/10/06/P06005}{\doi{10.1088/1748-0221/10/06/P06005}},
\href{http://www.arXiv.org/abs/1502.02701}{\texttt{arXiv:1502.02701}}.

\bibitem{TRK-11-001}
\hrefCMSnoop {}{{CMS Collaboration}, ``{Description and performance of track
  and primary-vertex reconstruction with the CMS tracker}'',} \textit{ JINST}
  \textbf{ 9} (2014) P10009,
  \href{http://dx.doi.org/10.1088/1748-0221/9/10/P10009}{\doi{10.1088/1748-0221/9/10/P10009}},
\href{http://www.arXiv.org/abs/1405.6569}{\texttt{arXiv:1405.6569}}.

\bibitem{Tanabashi:2018oca}
\hrefCMSnoop {}{{Particle Data Group}, M.~Tanabashi {et~al.}, ``Review of
  particle physics'',} \textit{ Phys. Rev. D} \textbf{ 98} (2018) 030001,
  \href{http://dx.doi.org/10.1103/PhysRevD.98.030001}{\doi{10.1103/PhysRevD.98.030001}}.

\bibitem{Chatrchyan:2014tja}
\hrefCMSnoop {}{{CMS Collaboration}, ``{Search for invisible decays of Higgs
  bosons in the vector boson fusion and associated ZH production modes}'',}
  \textit{ Eur. Phys. J. C} \textbf{ 74} (2014) 2980,
  \href{http://dx.doi.org/10.1140/epjc/s10052-014-2980-6}{\doi{10.1140/epjc/s10052-014-2980-6}},
\href{http://www.arXiv.org/abs/1404.1344}{\texttt{arXiv:1404.1344}}.

\bibitem{CMS:2011aa}
\hrefCMSnoop {}{{CMS Collaboration}, ``{Measurement of the Drell--Yan cross
  section in $\Pp\Pp$ collisions at $\sqrt{s} = 7$~TeV}'',} \textit{ JHEP}
  \textbf{ 10} (2011) 132,
  \href{http://dx.doi.org/10.1007/JHEP10(2011)132}{\doi{10.1007/JHEP10(2011)132}},
\href{http://www.arXiv.org/abs/1107.4789}{\texttt{arXiv:1107.4789}}.

\bibitem{Bodek:2012id}
A.~Bodek\hrefCMSnoop {}{ {et~al.}, ``Extracting muon momentum scale corrections
  for hadron collider experiments'',} \textit{ Eur. Phys. J. C} \textbf{ 72}
  (2012) 2194,
  \href{http://dx.doi.org/10.1140/epjc/s10052-012-2194-8}{\doi{10.1140/epjc/s10052-012-2194-8}},
\href{http://www.arXiv.org/abs/1208.3710}{\texttt{arXiv:1208.3710}}.

\bibitem{Schmitt:2012kp}
\hrefCMSnoop {}{S.~Schmitt, ``{TUnfold: an algorithm for correcting migration
  effects in high energy physics}'',} \textit{ JINST} \textbf{ 7} (2012)
  T10003,
  \href{http://dx.doi.org/10.1088/1748-0221/7/10/T10003}{\doi{10.1088/1748-0221/7/10/T10003}},
\href{http://www.arXiv.org/abs/1205.6201}{\texttt{arXiv:1205.6201}}.

\bibitem{Schmitt:2016orm}
\hrefCMSnoop {}{S.~Schmitt, ``{Data unfolding methods in high energy
  physics}'',} \textit{ Eur. Phys. J. Web Conf.} \textbf{ 137} (2017) 11008,
  \href{http://dx.doi.org/10.1051/epjconf/201713711008}{\doi{10.1051/epjconf/201713711008}},
\href{http://www.arXiv.org/abs/1611.01927}{\texttt{arXiv:1611.01927}}.

\bibitem{LUM-17-001}
\href {http://cds.cern.ch/record/2257069}{{CMS Collaboration}, ``{CMS}
  luminosity measurements for the 2016 data taking period'',} CMS Physics
  Analysis Summary CMS-PAS-LUM-17-001, 2017.

\bibitem{Golonka:2005pn}
\hrefCMSnoop {}{P.~Golonka and Z.~Was, ``{PHOTOS Monte Carlo: A precision tool
  for QED corrections in $Z$ and $W$ decays}'',} \textit{ Eur. Phys. J. C}
  \textbf{ 45} (2006) 97,
  \href{http://dx.doi.org/10.1140/epjc/s2005-02396-4}{\doi{10.1140/epjc/s2005-02396-4}},
\href{http://www.arXiv.org/abs/hep-ph/0506026}{\texttt{arXiv:hep-ph/0506026}}.

\bibitem{Butterworth:2015oua}
\hrefCMSnoop {}{J.~Butterworth {et~al.}, ``{PDF4LHC recommendations for LHC Run
  II}'',} \textit{ J. Phys. G} \textbf{ 43} (2016) 023001,
  \href{http://dx.doi.org/10.1088/0954-3899/43/2/023001}{\doi{10.1088/0954-3899/43/2/023001}},
\href{http://www.arXiv.org/abs/1510.03865}{\texttt{arXiv:1510.03865}}.

\bibitem{Lai:2010vv}
H.-L. Lai\hrefCMSnoop {}{ {et~al.}, ``New parton distributions for collider
  physics'',} \textit{ Phys. Rev. D} \textbf{ 82} (2010) 074024,
  \href{http://dx.doi.org/10.1103/PhysRevD.82.074024}{\doi{10.1103/PhysRevD.82.074024}},
  \href{http://www.arXiv.org/abs/1007.2241}{\texttt{arXiv:1007.2241}}.

\bibitem{Martin:2009iq}
\hrefCMSnoop {}{A.~D. Martin, W.~J. Stirling, R.~S. Thorne, and G.~Watt,
  ``{Parton distributions for the LHC}'',} \textit{ Eur. Phys. J. C} \textbf{
  63} (2009) 189,
  \href{http://dx.doi.org/10.1140/epjc/s10052-009-1072-5}{\doi{10.1140/epjc/s10052-009-1072-5}},
  \href{http://www.arXiv.org/abs/0901.0002}{\texttt{arXiv:0901.0002}}.

\bibitem{Ball:2011mu}
R.~D. Ball\hrefCMSnoop {}{ {et~al.}, ``{Impact of heavy quark masses on parton
  distributions and LHC phenomenology}'',} \textit{ Nucl. Phys. B} \textbf{
  849} (2011) 296,
  \href{http://dx.doi.org/10.1016/j.nuclphysb.2011.03.021}{\doi{10.1016/j.nuclphysb.2011.03.021}},
\href{http://www.arXiv.org/abs/1101.1300}{\texttt{arXiv:1101.1300}}.

\bibitem{FEWZ}
\hrefCMSnoop {}{K.~Melnikov and F.~Petriello, ``{The $W$ boson production cross
  section at the LHC through $O(\alpha^2_s)$}'',} \textit{ Phys. Rev. Lett.}
  \textbf{ 96} (2006) 231803,
  \href{http://dx.doi.org/10.1103/PhysRevLett.96.231803}{\doi{10.1103/PhysRevLett.96.231803}},
\href{http://www.arXiv.org/abs/hep-ph/0603182}{\texttt{arXiv:hep-ph/0603182}}.

\bibitem{Gavin:2010az}
\hrefCMSnoop {}{R.~Gavin, Y.~Li, F.~Petriello, and S.~Quackenbush, ``{FEWZ 2.0:
  A code for hadronic Z production at next-to-next-to-leading order}'',}
  \textit{ Comput. Phys. Commun.} \textbf{ 182} (2011) 2388,
  \href{http://dx.doi.org/10.1016/j.cpc.2011.06.008}{\doi{10.1016/j.cpc.2011.06.008}},
\href{http://www.arXiv.org/abs/1011.3540}{\texttt{arXiv:1011.3540}}.

\bibitem{Gavin:2012sy}
\hrefCMSnoop {}{R.~Gavin, Y.~Li, F.~Petriello, and S.~Quackenbush, ``{W physics
  at the LHC with FEWZ 2.1}'',} \textit{ Comput. Phys. Commun.} \textbf{ 184}
  (2013) 208,
  \href{http://dx.doi.org/10.1016/j.cpc.2012.09.005}{\doi{10.1016/j.cpc.2012.09.005}},
\href{http://www.arXiv.org/abs/1201.5896}{\texttt{arXiv:1201.5896}}.

\bibitem{Li:2012wna}
\hrefCMSnoop {}{Y.~Li and F.~Petriello, ``{Combining QCD and electroweak
  corrections to dilepton production in FEWZ}'',} \textit{ Phys. Rev. B}
  \textbf{ 86} (2012) 094034,
  \href{http://dx.doi.org/10.1103/PhysRevD.86.094034}{\doi{10.1103/PhysRevD.86.094034}},
\href{http://www.arXiv.org/abs/1208.5967}{\texttt{arXiv:1208.5967}}.

\bibitem{Ball:2017nwa}
\hrefCMSnoop {}{{NNPDF} Collaboration, ``{Parton distributions from
  high-precision collider data}'',} \textit{ Eur. Phys. J. C} \textbf{ 77}
  (2017) 663,
  \href{http://dx.doi.org/10.1140/epjc/s10052-017-5199-5}{\doi{10.1140/epjc/s10052-017-5199-5}},
\href{http://www.arXiv.org/abs/1706.00428}{\texttt{arXiv:1706.00428}}.

\bibitem{Frederix:2012ps}
\hrefCMSnoop {}{R.~Frederix and S.~Frixione, ``{Merging meets matching in
  MC@NLO}'',} \textit{ JHEP} \textbf{ 12} (2012) 061,
  \href{http://dx.doi.org/10.1007/JHEP12(2012)061}{\doi{10.1007/JHEP12(2012)061}},
\href{http://www.arXiv.org/abs/1209.6215}{\texttt{arXiv:1209.6215}}.

\bibitem{Hamilton:2012rf}
\hrefCMSnoop {}{K.~Hamilton, P.~Nason, C.~Oleari, and G.~Zanderighi, ``{Merging
  H/W/Z + 0 and 1 jet at NLO with no merging scale: a path to parton shower +
  NNLO matching}'',} \textit{ JHEP} \textbf{ 05} (2013) 082,
  \href{http://dx.doi.org/10.1007/JHEP05(2013)082}{\doi{10.1007/JHEP05(2013)082}},
\href{http://www.arXiv.org/abs/1212.4504}{\texttt{arXiv:1212.4504}}.

\bibitem{Ladinsky:1993zn}
\hrefCMSnoop {}{G.~A. Ladinsky and C.~P. Yuan, ``The nonperturbative regime in
  {QCD} resummation for gauge boson production at hadron colliders'',} \textit{
  Phys. Rev. D} \textbf{ 50} (1994) R4239,
  \href{http://dx.doi.org/10.1103/PhysRevD.50.R4239}{\doi{10.1103/PhysRevD.50.R4239}},
\href{http://www.arXiv.org/abs/hep-ph/9311341}{\texttt{arXiv:hep-ph/9311341}}.

\bibitem{Balazs:1997xd}
\hrefCMSnoop {}{C.~Balazs and C.~P. Yuan, ``{Soft gluon effects on lepton pairs
  at hadron colliders}'',} \textit{ Phys. Rev. D} \textbf{ 56} (1997) 5558,
  \href{http://dx.doi.org/10.1103/PhysRevD.56.5558}{\doi{10.1103/PhysRevD.56.5558}},
\href{http://www.arXiv.org/abs/hep-ph/9704258}{\texttt{arXiv:hep-ph/9704258}}.

\bibitem{Landry:2002ix}
\hrefCMSnoop {}{F.~Landry, R.~Brock, P.~M. Nadolsky, and C.~P. Yuan,
  ``{Tevatron Run-1 $Z$ boson data and Collins-Soper-Sterman resummation
  formalism}'',} \textit{ Phys. Rev. D} \textbf{ 67} (2003) 073016,
  \href{http://dx.doi.org/10.1103/PhysRevD.67.073016}{\doi{10.1103/PhysRevD.67.073016}},
\href{http://www.arXiv.org/abs/hep-ph/0212159}{\texttt{arXiv:hep-ph/0212159}}.

\bibitem{Alioli:2015toa}
S.~Alioli\hrefCMSnoop {}{ {et~al.}, ``{Drell-Yan production at NNLL'+NNLO
  matched to parton showers}'',} \textit{ Phys. Rev. D} \textbf{ 92} (2015)
  094020,
  \href{http://dx.doi.org/10.1103/PhysRevD.92.094020}{\doi{10.1103/PhysRevD.92.094020}},
\href{http://www.arXiv.org/abs/1508.01475}{\texttt{arXiv:1508.01475}}.

\bibitem{Martinez:2018jxt}
A.~Bermudez~Martinez\hrefCMSnoop {}{ {et~al.}, ``{Collinear and TMD parton
  densities from fits to precision DIS measurements in the parton branching
  method}'',} \textit{ Phys. Rev. D} \textbf{ 99} (2019) 074008,
  \href{http://dx.doi.org/10.1103/PhysRevD.99.074008}{\doi{10.1103/PhysRevD.99.074008}},
\href{http://www.arXiv.org/abs/1804.11152}{\texttt{arXiv:1804.11152}}.

\bibitem{Hautmann:2017fcj}
F.~Hautmann\hrefCMSnoop {}{ {et~al.}, ``{Collinear and TMD quark and gluon
  densities from parton branching solution of QCD evolution equations}'',}
  \textit{ JHEP} \textbf{ 01} (2018) 070,
  \href{http://dx.doi.org/10.1007/JHEP01(2018)070}{\doi{10.1007/JHEP01(2018)070}},
\href{http://www.arXiv.org/abs/1708.03279}{\texttt{arXiv:1708.03279}}.

\bibitem{Hautmann:2017xtx}
F.~Hautmann\hrefCMSnoop {}{ {et~al.}, ``{Soft-gluon resolution scale in QCD
  evolution equations}'',} \textit{ Phys. Lett. B} \textbf{ 772} (2017) 446,
  \href{http://dx.doi.org/10.1016/j.physletb.2017.07.005}{\doi{10.1016/j.physletb.2017.07.005}},
\href{http://www.arXiv.org/abs/1704.01757}{\texttt{arXiv:1704.01757}}.

\bibitem{Martinez:2019mwt}
A.~B. Martinez\hrefCMSnoop {}{ {et~al.}, ``Production of {Z} bosons in the
  parton branching method'',} \textit{ Phys. Rev. D} \textbf{ 100} (2019)
  074027,
  \href{http://dx.doi.org/10.1103/PhysRevD.100.074027}{\doi{10.1103/PhysRevD.100.074027}},
\href{http://www.arXiv.org/abs/1906.00919}{\texttt{arXiv:1906.00919}}.

\bibitem{Dulat:2015mca}
S.~Dulat\hrefCMSnoop {}{ {et~al.}, ``{New parton distribution functions from a
  global analysis of quantum chromodynamics}'',} \textit{ Phys. Rev. D}
  \textbf{ 93} (2016) 033006,
  \href{http://dx.doi.org/10.1103/PhysRevD.93.033006}{\doi{10.1103/PhysRevD.93.033006}},
\href{http://www.arXiv.org/abs/1506.07443}{\texttt{arXiv:1506.07443}}.

\end{thebibliography}\endgroup
\cleardoublepage \appendix\section{The CMS Collaboration \label{app:collab}}\begin{sloppypar}\hyphenpenalty=5000\widowpenalty=500\clubpenalty=5000\vskip\cmsinstskip
\textbf{Yerevan Physics Institute, Yerevan, Armenia}\\*[0pt]
A.M.~Sirunyan$^{\textrm{\dag}}$, A.~Tumasyan
\vskip\cmsinstskip
\textbf{Institut f\"{u}r Hochenergiephysik, Wien, Austria}\\*[0pt]
W.~Adam, F.~Ambrogi, T.~Bergauer, J.~Brandstetter, M.~Dragicevic, J.~Er\"{o}, A.~Escalante~Del~Valle, M.~Flechl, R.~Fr\"{u}hwirth\cmsAuthorMark{1}, M.~Jeitler\cmsAuthorMark{1}, N.~Krammer, I.~Kr\"{a}tschmer, D.~Liko, T.~Madlener, I.~Mikulec, N.~Rad, J.~Schieck\cmsAuthorMark{1}, R.~Sch\"{o}fbeck, M.~Spanring, D.~Spitzbart, W.~Waltenberger, C.-E.~Wulz\cmsAuthorMark{1}, M.~Zarucki
% ...
\vskip\cmsinstskip
\dag: Deceased\\
1:  Also at Vienna University of Technology, Vienna, Austria\\
% ...
\end{sloppypar}
\end{document}